\documentclass[article]{aa}
\usepackage{natbib}
\usepackage{graphicx}
\usepackage{rotating}
\usepackage{times}
\usepackage{amsmath}
\usepackage{txfonts}
\usepackage{multirow}

\newcommand{\msun}{{\,\rm M}_{\odot}}
\newcommand{\lsun}{{\,\rm L}_{\odot}}

\newcommand{\kms}{\,{\rm km.s}^{-1}}

\newcommand{\nht}{\ifmmode {{\rm NH}_3} \else {NH{\bas 3}} \fi}
\newcommand{\tco}{\ifmmode {^{13}{\rm CO}} \else {$^{13}{\rm CO}$}\fi}
\newcommand{\dco}{\ifmmode {^{12}{\rm CO}} \else {$^{12}{\rm CO}$}\fi}
\newcommand{\cdo}{\ifmmode {{\rm C}^{18}{\rm O}} \else {${\rm C}^{18}{\rm O}$}\fi}

\newcommand{\htco}{\ifmmode {{\rm H}^{13}{\rm CO}^{+} } \else {${\rm H}^{13}
{\rm CO}^{+}$ }\fi}
\newcommand{\hco}{\ifmmode {{\rm H}^{12}{\rm CO}^{+} } \else {${\rm H}^{12}
{\rm CO}^{+}$ }\fi}
\newcommand{\juz}{\ifmmode {{\rm J}=1\rightarrow 0} \else
{J=1$\rightarrow$0}\fi}
\newcommand{\jdu}{\ifmmode {{\rm J}=2\rightarrow 1} \else
{J=2$\rightarrow$1}\fi}
\newcommand{\jtd}{\ifmmode {{\rm J}=3\rightarrow 2} \else
{J=3$\rightarrow$2} \fi}
\newcommand{\jcq}{\ifmmode {{\rm J}=5\!\rightarrow\!4} \else
{${\rm J}=5\!\rightarrow\!4$} \fi}
\newcommand{\as}{\ifmmode {^{\scriptscriptstyle\prime\prime}}
        \else $^{\scriptscriptstyle\prime\prime}$\fi}
\newcommand{\am}{\ifmmode {^{\scriptscriptstyle\prime}}
        \else $^{\scriptscriptstyle\prime}$\fi}
\newcommand{\hh}{\ifmmode {{\rm H}_2} \else {H$_2$} \fi}
\renewcommand{\hco}{\ifmmode {{\rm HCO}^+} \else {HCO$^+$} \fi}
\newcommand{\hhco}{\ifmmode {{\rm H}_2{\rm CO}} \else {H$_2$CO} \fi}
\newcommand{\ddco}{\ifmmode {{\rm D}_2{\rm CO}} \else {D$_2$CO} \fi}
\newcommand{\chhdoh}{\ifmmode {{\rm CH}_2{\rm DOH}^+} \else {CH$_2$DOH} \fi}
\newcommand{\chhhod}{\ifmmode {{\rm CH}_3{\rm OD}^+} \else {CH$_3$OD} \fi}
\newcommand{\chhhoh}{\ifmmode {{\rm CH}_3{\rm OH}^+} \else {CH$_3$OH} \fi}
\newcommand{\tchhhoh}{\ifmmode {^{13}{\rm CH}_3{\rm OH}^+} \else {$^{13}$CH$_3$OH} \fi}
\newcommand{\dcop}{\ifmmode {{\rm DCO}^+} \else {DCO$^+$} \fi}
\newcommand{\cchh}{\ifmmode {{\rm C}_2{\rm H}_2} \else {C$_2$H$_2$} \fi}

\newcommand{\hcccn}{\ifmmode {{\rm HC}_3{\rm N}} \else {HC$_3$N} \fi}

\begin{document}
\title{A sensitive survey for $^{13}$CO, CN, H$_2$CO and SO in the disks of T Tauri and Herbig Ae stars
\thanks{Based on observations carried out with the IRAM 30-m telescope.
 IRAM is supported by INSU/CNRS (France), MPG (Germany) and IGN (Spain).}
}
\author{
 S.~Guilloteau \inst{1,2},
 E.~Di Folco \inst{1,2},
 A.~Dutrey \inst{1,2},
 M.~Simon \inst{3},
% F.~Palla \inst{4},
 N.~Grosso \inst{4},
 V.Pi\'etu \inst{5}}
\institute{
Univ. Bordeaux, LAB, UMR 5804, F-33270, Floirac, France
\and{}
CNRS, LAB, UMR 5804, F-33270 Floirac, France\\
  \email{guilloteau@obs.u-bordeaux1.fr, difolco@obs.u-bordeaux1.fr, dutrey@obs.u-bordeaux1.fr}
 \and{}
 Department of Physics and Astronomy, Stony Brook University, Stony Brook, NY 11794-3800, USA
 \and{}
 Observatoire Astronomique de Strasbourg, Universit\'e de Strasbourg, CNRS, UMR 7550, 11 rue de l'Universit\'e, 67000 Strasbourg, France
 \and{}
 IRAM, 300 rue de la piscine, F-38406
 Saint Martin d'H\`eres, France
}
% \and{}
% Obs.di Arcetri, Firenze, Italy

\offprints{S.Guilloteau, \email{guilloteau@obs.u-bordeaux1.fr}}

\date{Received 28-Aug-2012, Accepted 8-Nov-2012} %
\authorrunning{Guilloteau et al.} %
\titlerunning{Molecules in T Tauri disks}

\abstract
% {} leave it empty if  necessary
{}
{We investigate the suitability of several molecular tracers, in particular CN N=2-1 line emission,
as a confusion-free probe of the kinematics of circumstellar disks.
}
{We use the IRAM 30-m telescope to perform a sensitive search for CN N=2-1 in 42 T Tauri or Herbig Ae systems located principally
in the Taurus-Auriga
region. $^{13}$CO J=2-1 is observed simultaneously to provide an indication of the level of confusion
with the surrounding molecular cloud. The bandpass also contains two transitions of ortho-H$_2$CO, one of
SO and the C$^{17}$O J=2-1 line which provide complementary information on the nature of the emission.
}
{While $^{13}$CO is in general dominated by residual emission from the cloud, CN exhibits a high disk detection
rate $> 50$\,\% in our sample.  We even report CN detection in stars for which interferometric
searches failed to detect $^{12}$CO, presumably because of obscuration by a foreground, optically thick, cloud.
Comparison between  CN and o-H$_2$CO or SO line profiles and intensities divide
the sample in two main categories. Sources with SO emission are bright and have strong H$_2$CO emission, leading in general to [H$_2$CO/CN]$ > 0.5$.
Furthermore, their line profiles, combined with a priori information on the objects, suggest that
the emission is coming from outflows or envelopes rather than from a circumstellar disk.  On the other hand, most sources have [H$_2$CO/CN]$ < 0.3$, no SO emission, and some of them exhibit clear double-peaked profiles characteristics of rotating disks. In this second category, CN is likely tracing the proto-planetary disks.
From the line flux and opacity derived from the hyperfine ratios, we constrain the outer radii
of the disks, which range from 300 to 600 AU.  The overall gas disk detection rate (including all
molecular tracers) is $\sim 68 \%$, and decreases for fainter continuum sources.}
{The current study demonstrates that gas disks, like dust disks,
are ubiquitous around young PMS stars in regions of isolated star formation, and that a large
fraction of them have $R_\mathrm{out} \gtrsim 300$ AU.
It also shows the potential of the CN N=2-1 transition to probe the kinematics of these disks, thereby
measuring the stellar mass, using high resolution observations with ALMA.}


%----------------------------------
\maketitle
%----------------------------------

%-------------------------------------------------------------------
\section{Introduction}

Masses of young stars are a fundamental property with impacts on
theories of the stars' formation, calculations of their evolution to the
main sequence, and the chronology of the planets that may be, or
will be, associated with them.  Dynamical techniques provide the only
absolutely reliable measurements of mass.  Because most binary orbits are
at least a few years long, the quickest way to accomplish a dynamical
measurement is by mapping the Keplerian rotation of the circumstellar
disk  of a young star \citep{Guilloteau+Dutrey_1998,Simon+etal_2000}.
These early studies, and especially the more recent
observations by  \citet{Schaefer+etal_2009}, were all carried out by
mm-wave interferometry  in the J=1-0 and J=2-1 transitions of
 {$^{12}$CO} and revealed two limitations of the method.  The first
is intrinsic:  Young stars with disks are generally located in
molecular clouds whose optically thick  {$^{12}$CO} emission may
mask the disk emission of the star.  The second limitation was
technical: it was difficult to find young stars with masses smaller than $0.5 \msun$
that have detectable gaseous circumstellar disks.
A possible solution to the first problem could be
observation in the lines of  {$^{13}$CO} or {C$^{18}$O}
because the optical depth of the molecular cloud would be less
in these transitions than in  {$^{12}$CO}. However, the disk emission
in the lines of  the CO isotopologues would of course be less
than that of  the {$^{12}$CO} lines.  We judged that, at the time of
our observations, the instrumentation available to us had insufficient
sensitivity to obtain mass measurements to the required precision,
$\le 5\%$.  The commissioning of the Atacama Large
 Millimeter/Submillimeter Array (ALMA) offers the promise  to remove
both obstacles to mass measurement of young low mass stars by
mapping the rotation of their disks.

Our goal in this work is to assess the applicability of several
molecular tracers for  confusion-free and  high sensitivity measurements
of disk kinematics using ALMA. After  {$^{12}$CO},  the strongest
lines  in disks are the rotational transitions of {$^{13}$CO} J=2-1 or 3-2, HCO$^+$
J=3-2 and  the hyperfine split lines of CN N=2-1 \citep{Dutrey+etal_1997,Oberg+etal_2010}.  However,
{$^{13}$CO}  line
emission can also be affected by confusion \citep[e.g.][for BP Tau]{Dutrey+etal_2003}.
The situation is less clear for HCO$^+$ J=3-2, but this transition can still have
substantial optical depth and be excited in clouds \citep[see for example CW Tau in][]{Salter+etal_2011};
furthermore, its relatively high frequency
requires better observing conditions.
Other previous searches for molecular tracers in disks were conducted with the single-dishes (principally the IRAM 30-m and JCMT) or mm/sub-mm arrays
(IRAM PdBI, SMA and CARMA). The former were limited in sensitivity \citep[e.g.][for a study of CN N=2-1 and HCO$^+$ J=3-2]{Salter+etal_2011}, while,
because of the need for sufficient UV coverage, the latter could only observe a few stars \citep[e.g.][]{Oberg+etal_2010}.

%
% Salter et al.'s (2011) study of CN (2-1) line emission was limited by sensitivity and Oberg et al.'s (2010) study of HCO$^+$
% emission was limited by the available UV-coverage to a few stars.
%

Recent advances in receiver and back-end technology at the IRAM-30m telescope
encouraged us to carry out a new assessment of molecular line
tracers in circumstellar disks, particularly the CN N=2-1 line hyperfine
split lines.  We report here our systematic search for CN N=2-1
emission in 40 young stars located in the Taurus/Auriga star forming
region (SFR).  Together with previous observations, 46 stars in this
SFR have now been observed  with high sensitivity in this transition,
a gain of a factor almost four in sample size.  \S \ref{sec:obs} describes our
sample, the observations, and their analysis. \S \ref{sec:results} presents our
results for the CO isotopologues, CN, H$_2$CO, and SO molecules.  \S \ref{sec:discussion} discusses
our assessment of CN as a disk tracer and comments on stars with
outflows and on general properties of the observed line emission.
\S \ref{sec:summary} summarizes our results.  An Appendix
shows all the observed spectra and provides
details for many of the stars.

%\input table-sources-s.tex

%  base-s
\begin{table*}[!ht]
\caption{Source sample}
\begin{tabular}{llllrrrrl}
\hline
Name & RA & Dec & Spectral & $L_*$ & $S_\nu$(1.3\,mm) & Incli. & A$_V$ & References \\
     & J2000.0 & J2000.0 & Type & $\lsun$ & (mJy) & ($^\circ$) & & \\
\hline
%
 %       FM Tau &
      FN Tau &
04:14:14.59 & 28:27:58.1 & M5 & 0.50 &   $< 18$ &  & 1.35 & a,A \\
      CW Tau &
04:14:17.0 & 28:10:56.51 & K3 & 0.68 &   96 &  & 2.19 & b,B \\
      CIDA-1 &
04:14:17.6 & 28:06:11 & M5.5 &  &   14 &  & - &  a,$\alpha$ \\
 %       FP Tau &
 %       CX Tau &
      CY Tau (*) &
04:17:33.729 & 28:20:46.86 & M1.5 & 0.40 &  111 & 28 $\pm$ 5 & 0.1 & c,1,B \\
 %   V410Anon13 &
      BP Tau &
04:19:15.834 & 29:06:26.98 & K7 & 0.65 &   58 & 33 $\pm$ 6 & 0.49 & c,1,B \\
      DE Tau &
04:21:55.6 & 27:55:05.55 & M1 & 1.14 &   36 &  & 1.56 & b,C,$\beta$ \\
      RY Tau &
04:21:57.42 & 28:26:35.6 & K1 & 6.59 &  229 & 66 $\pm$ 3 & 1.84 & b,2,B \\
       T Tau &
04:21:59.435 & 19:32:06.36 & K0 & 8.91 &  200 & 30 & 1.39 & c,3,C \\
 %      T Tau N &
 %      T Tau S &
   Haro 6-5B &
04:22:01.00 & 	26:57:35.5 & K5 & 0.05 &  134 & $~80$ & 9.96 & d,4,D,$\beta$ \\
 %       FS Tau &
      FT Tau &
04:23:39.188 & 24:56:14.28 & C & 0.38 &   73 & 23 $\pm$ 5 & & c,1,C \\
    DG Tau-B &
04:27:02.56 & 26:05:30.4 & C & $>0.02$ &  531 & 64 $\pm$ 2 & & c,1 \\
      DG Tau &
04:27:04.67 & 26:06:16.9 & K6 & 3.62 &  390 & 38 $\pm$ 2 & 3.32 & c,1,D,$\beta$ \\
   Haro 6-10 &
04:29:23.729 & 24:33:01.52 & K5 & 2.55 &   44 &  & & c,E \\
 %  Haro 6-10 N &
 %  Haro 6-10 S &
      IQ Tau &
04:29:51.56 & 	26:06:44.9	 & M0.5 & 0.53 &   60 & & 1.25 & a,B \\
 %        MHO-4 &
    LkHa 358 &
04:31:36.15 & 18:13:43.1 & K8 & 0.09 &   17 & 28 $\pm$ 9 & 0.13 & a,5,A,$\gamma$ \\ % VDA98
       HH 30 &
04:31:37.468 & 18:12:24.21 & M0 & 0.2 - 0.9 &   20 & 83 $\pm$ 2 & 2.96 & c,6,F,$\beta$ \\ % White 2004
      HL Tau &
04:31:38.413 & 18:13:57.55 & K7 & 1.53 &  819 & 45 $\pm$ 1 & 7.43 & c,1,D,$\beta$ \\
      HK Tau &
04:31:50.58 & 	24:24:17.9	 & M0.5 & 1.00 &   41 & $\sim 90$ & 3.41 & b,7,B \\
 %     HK Tau A &
 %     HK Tau B &
   Haro 6-13 &
04:32:15.419 & 24:28:59.47 & M0 & 2.11 &  114 & 40 $\pm$ 3 & 11.9 & c,5,A,$\beta$ \\
 %        MHO-5 &
      GG Tau &
04:32:30.34 & 17:31:40.5 & K7 & 0.64 &  593 & 37 $\pm$ 1 & 0.76 & b,8,B \\
 %    GG Tau Aa &
 %    GG Tau Ab &
 %       FZ Tau &
    UZ Tau E &
04:32:43.071 & 25:52:31.07 & M1 & 0.9-1.6 &  150 & 56 $\pm$ 2 & 1.49 & c,1,C+G \\
 %     UZ Tau W &
  04302+2247 &
04:33:16.2 & 22:53:20.0 & - &  &  130 & 88 $\pm$ 2 & & e,9 \\
      DL Tau &
04:33:39.077 & 25:20:38.10 & K7 & 1.16 &  204 & 43 $\pm$ 3 & 2.0 & c,1,H \\ %,$\delta$ \\
      DM Tau (\$) &
04:33:48.70 & 18:10:10.6 & M1 & 0.16 &  109 & 35 $\pm$ 1 & 0.31 & c,1,B,$\alpha$ \\ % Briceno 2002
      CI Tau (*) &
04:33:52.014 & 22:50:30.06 & K7 & 0.96 &  125 & 44 $\pm$ 3 & 1.77 & c,1,B \\
      AA Tau &
04:34:55.42 & 	24:28:53.1 & K7 & 0.66 &   73 & 70 $\pm$ 5 & 0.49 & b,10,C \\
 %       HO Tau &
      DN Tau &
04:35:27.38 & 	24:14:58.9	 & M0 & 0.68 &   89 &  & 1.89 & a,D,$\beta$ \\
 %       HP Tau &
 %    HP Tau/G3 &
 %    HP Tau/G2 &
 %       GM Tau &
      DO Tau &
04:38:28.59 & 	26:10:49.5	 & M0 & 1.29 &  136 &  & 2.64 & b,B \\
    HV Tau C &
04:38:35.31 & 	26:10:38.5	 & K6 & 0.60 &   40 & $\sim 90$ & 2.42 & b,11,B \\
 %       HV Tau &
     LkCa 15 (*) &
04:39:17.76 & 22:21:03.7 & K5 & 0.85 &  110 & 52 $\pm$ 1 & 0.62 & c,1,B \\
 %     ITG 33 A &
   Haro 6-33 &
04:41:38.827 & 25:56:26.68 & M0 & 0.76 &   34 & 52 $\pm$ 5 & 10.2 & c,5,D \\
      GO Tau (*) &
04:43:03.050 & 25:20:18.80 & M0 & 0.22 &   53 & 52 $\pm$ 1 & 1.18 & a,5,B\\
 %      CIDA-14 &
 %     RXJ04467 &
      DQ Tau &
04:46:53.064 & 17:00:00.09 & M0 & 0.91 &   83 &  & 0.97 & c,C \\
      DR Tau &
04:47:06.22 & 	16:58:42.8	 & K5 & 1.97 &  159 &  & 2.0  & b,B \\
      DS Tau &
04:47:48.6 & 29:25:10.96 & K5 & 0.67 &   25 &  & 0.31 & b,C \\
      UY Aur &
04:51:47.37 & 30:47:13.9 & M0 & 1.41 &   29 & 42 & 2.05 & b,12,B \\
      GM Aur (*) &
04:55:10.98 & 30:21:59.5 & K7 & 1.23 &  176 & 50 $\pm$ 1 & 0.14 & c,1,B \\
      AB Aur &
04:55:45.80 & 30:33:04.0 & B9 & 44.86 &  110 &  20-30 & 0.62 & f,13,I \\
      SU Aur &
04:55:59.4 & 30:34:01.39 & G2 & 9.29 &   $< 30$ & 62 & 0.90 & b,14,B \\
     MWC 480 &
04:58:46.27 & 29:50:37.0 & A2 & 11.50 &  289 & 37 $\pm$ 1 & 0.25 & c,1,G,$\gamma$ \\ VDA98
       CB 26 &
04:59:50.74 & 52:04:43.8 & - &  &  190 & $\sim 90$ & & g,15 \\
      CIDA-8 &
05:04:41.4 & 25:09:57 & M3.5 & 0.26 &    8 &  & 0.83 & a,A,$\alpha$ \\% A_J Briceno 2002
     CIDA-11 &
05:06:23.3 & 24:32:24 & M3.5 & 0.22 &    $< 4$ &  & 0.36 & a,A \\
      RW Aur &
 05:07:49.56 & 30:24:05.1 & K3 & 1.72 &   42 & 45 $\pm$ 5 & 0.32 & b,16,B \\
 %     RW Aur A &
 %     RW Aur B &
 %      CIDA-12 &
     MWC 758 &
05:30:27.51 & 25:19:58.4 & A3 & 11.00 &   55 & 21 & & c,17,J \\
      CQ Tau &
05:35:58.485 & 24:44:54.19 & A8/F2 & 12.00 &  162 & 29 $\pm$ 2 & 0.96 & c,1,J,$\gamma$ \\ %VDA98
   HD 163296 (\#) &
17:56:21.31 & -21:57:22.0 & A1 & 37.70 &  670 & 45 $\pm$ 4 & 0.25 & h,K,$\gamma$ \\ %VDA98
\hline
\end{tabular}
\tablefoot{Data for sources marked with (*) are for CN only, and
were obtained separately, (\$) no SO line, (\#) different SO line. \\
Flux references: (a) \citet{Schaefer+etal_2009} (b) \citet{Andrews+etal_2005}
(c) \citet{Guilloteau+etal_2011} (d) \citet{Osterloh+Beckwith_1995}
(e) Graeffe et al 2012, (f) \citet{Pietu+etal_2005}, (g) \citet{Launhardt+etal_2009}
(h) \citet{Qi+etal_2011}. \\
Inclination references: (1) \citet{Guilloteau+etal_2011}, (3) \citet{Isella+etal_2010a}
(3) \citet{Ratzka+etal_2009},  (4) \citet{Krist+etal_1998}, (5) \citet{Schaefer+etal_2009}
(6) \citet{Pety+etal_2006}, (7) \citet{McCabe+etal_2011}, (8) \citet{Guilloteau+etal_1999},
(9) Graeffe et al 2012, (10) \citet{Bouvier+etal_1999}, (11)\citet{Duchene+etal_2010},
(12) \citet{Duvert+etal_1998}, (13) \citet{Pietu+etal_2005}, (14) \citet{Akeson+etal_2002}
(15) \citet{Launhardt+etal_2009}, (16) \citet{Cabrit+etal_2006}, (17) \citet{Isella+etal_2010b},
(18) \citet{Isella+etal_2007}. \\
Luminosity references:
(A) \citet{Schaefer+etal_2009}, (B)  \citet{Bertout+etal_2007}, (C) \citet{Kenyon+Hartmann_1995},
(D) \citet{White+Hillenbrand_2004}
(E) \citet{Prato+etal_2009}, (F) \citet{Pety+etal_2006}, (G) \citet{Simon+etal_2000}, (H) \citet{White+Ghez_2001},
(I) \citet{Hernandez+etal_2004}, (J) \citet{Chapillon+etal_2008}, (K) \citet{Tilling+etal_2012}\\
Extinction references: \citet{Kenyon+Hartmann_1995}, except ($\alpha$) \cite{Briceno+etal_2002} ($A_J$ for CIDA-8),
($\beta$) \citet{White+Hillenbrand_2004},
($\gamma$)  \citet{vandenAncker+etal_1998} % , $\delta{White+Ghez_2001}
}
\label{tab:sources}

\end{table*}
%end input

\section{Observations and Data Analysis}
\label{sec:obs}

\subsection{Source Sample}
\label{sec:sub:sample}

We focus here principally on the Taurus Aurigae region, to which we add \object{CB26}, an embedded source
in an isolated Bok globule, and \object{HD 163296}, an isolated HAe star. Our study is thus only
relevant to regions of isolated star formation. Table \ref{tab:sources} lists the stars involved in this study; the 40 stars observed
anew at the IRAM 30-m radiotelescope appear without specific symbols.
This sub-sample was observed in an homogeneous way. It is completed by observations on a few sources (marked by specific symbols in the Table)
obtained independently, also with the
IRAM 30-m.
$^{13}$CO, CN and the ortho-H$_2$CO $3_{13} - 2_{12}$
transitions were observed towards the Herbig Ae star HD 163296, together with the SO $6_5-5_4$ line.
Data from \object{DM Tau} is from \citet{Dutrey+etal_1997}. The intensity reported for \object{LkCa 15} is from
a noisy IRAM 30-m spectrum, but is consistent with the interferometric data of \citet{Chapillon+etal_2012}.
\object{CI Tau}, \object{CY Tau} and \object{GO Tau} were detected with the IRAM 30-m, and later
imaged at the IRAM Plateau de Bure array \citep[][in prep]{Guilloteau+etal_2012}.
The overall sample covers a wide range of spectral types (from M5.5 to A0, taken from
\citet{Luhman+etal_2010}), luminosities ($0.2$ to $45 \lsun$), and 1.3 continuum flux densities
(from $< 4$ to $\sim 800$ mJy), but is not complete with respect to these quantities.
No specific account about extinction and location with respect to molecular clouds was made when
selecting sources. Sources with outflows were deliberately included. The epoch 2000 coordinates
are from interferometric mm continuum images \citep{Guilloteau+etal_2011} when
available, or SIMBAD otherwise.

\subsection{Observations}
\label{sec:sub:obs}
%\subsection{Sources}
Most observations were carried out with the IRAM 30-m telescope from Dec 4 to Dec 8 2011, under
excellent weather conditions (precipitable water vapor content below 1 mm). A few additional
sources were observed on the afternoon of May 1st, 2012 using the same instrumental setup.
The weather was good (water vapor content about
3 mm), but a few clouds occasionally produced anomalous refraction. The absolute calibration of this second
set is thus more uncertain.

The dual-polarization, sideband separating receivers were tuned to cover
approximately 7.8 GHz of bandwidth around 224.3 GHz in the upper sideband (USB) and 208.7 GHz in the lower sideband (LSB).
The 8 units of the FTS back-end (each of them being made of 3 adjacent sub-units) was covered more precisely 4 frequency ranges (with 2 polarization each) at
[220.325,224.365] and [224.055,228.105] GHz in USB, and
[204.655,208.695] and [208.375,212.425] GHz in LSB. The back-end provides a spectral resolution
of 195 kHz, which translates into 0.25 - 0.28 km.s$^{-1}$ in this frequency range.
The spectral setup includes several lines (listed in Table \ref{tab:lines}) which can be
relevant for disks, in particular all 19 hyperfine components of CN N=2-1.
The sideband separating mixers provide a mean rejection around 13 dB. However, the presence
of strong lines spread over the band indicate some variation across the band. A fortunate
case is the SO (5$_4$-4$_3$) transition at 206.176062 GHz, which appears at 226.572 GHz in the
image sideband and permits a control of the rejection near the CN N=2-1 lines, whose
strongest hyperfine components are at 226.874 and 226.657 GHz. Single sideband system temperatures
were in the range 170-190 K in December, and 170-250 K in May.

In December 2011, the exceptionally dry weather and clear sky produced rapid cooling
a few hours after sunset, which could not be compensated by the thermal regulation of
the telescope, and resulted in significant pointing changes. This also presumably
affected the telescope efficiency.  In May, anomalous refraction clearly occurred occasionally.
As a result, the absolute calibration is in general accurate to 10-15 \% but on several sources
is not better more than about 25 \%. However, since we control the receiver
sideband rejection, the relative intensities of the detected lines are expected to be
accurate to within 5 \% or better, noise being in general a more severe limitation.
Given the absolute calibration issues, we did not correct for the elevation dependent antenna
gain, and simply use a uniform conversion factor from antenna temperature (T$_A^*$) to flux
density of  9 Jy/K.

Since our targets should be unresolved by the IRAM 30-m, the observations were performed in symmetric
wobbler switching mode, with the two references $\pm 60''$ away at the same elevation. Together with the stable weather
in December, this mode provided flat baselines, allowing
clear detection of the continuum emission from the dust in most sources. Occasional baseline discontinuities between the 3 sub-units
(``platforming'') were visible in the spectra, but none of the spectral
lines we searched for were affected by this problem.

We observed the sample of  40 sources (see Table \ref{tab:sources}), for about 1 hour each.
For a typical linewidth of $2.6 \kms$\ (2 MHz), representative of our sources, this gives
a detection sensitivity of 0.08 Jy.$\kms$ ($1 \sigma$) on the integrated line flux.

\begin{figure}[!ht]
\centering
\includegraphics[width=7.5cm]{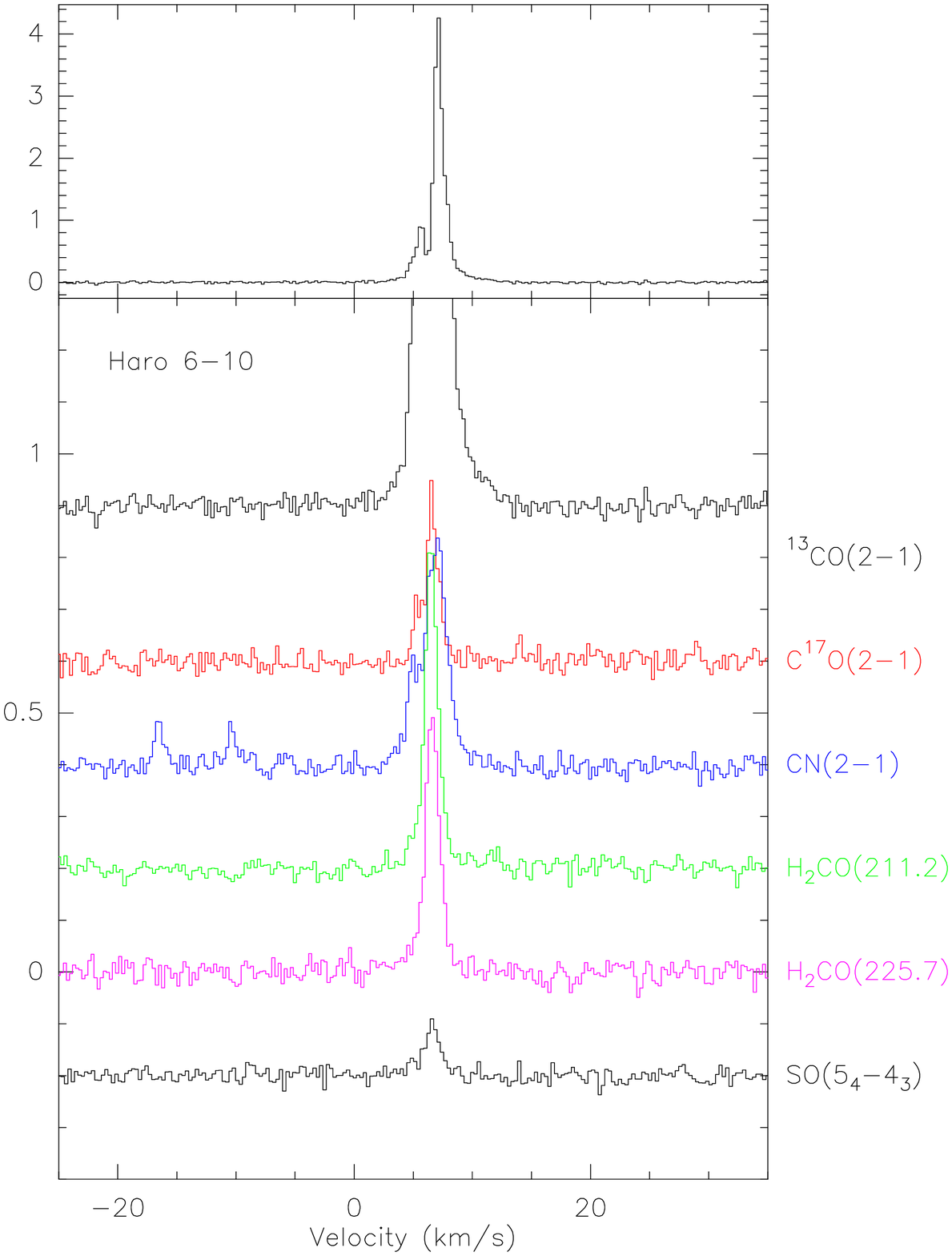}
\caption{Spectra of the observed transitions towards \object{Haro 6-10}.
Top panel: $^{13}$CO J=2-1 spectrum. Bottom panels, from
top to bottom: spectra of $^{13}$CO J=2-1, C$^{17}$O J=2-1,
CN N=2-1, ortho H$_2$CO $3_{13}-2_{12}$, ortho H$_2$CO $3_{12}-2_{11}$,
and SO $5_4-4_3$ presented on a common scale. The Y axis scale
is antenna temperature, T$_A^*$ (the conversion factor to
flux density is 9 Jy/K).}
\label{fig:haro6-10n}
\end{figure}

\begin{figure}[!ht]
\centering
\includegraphics[width=7.5cm]{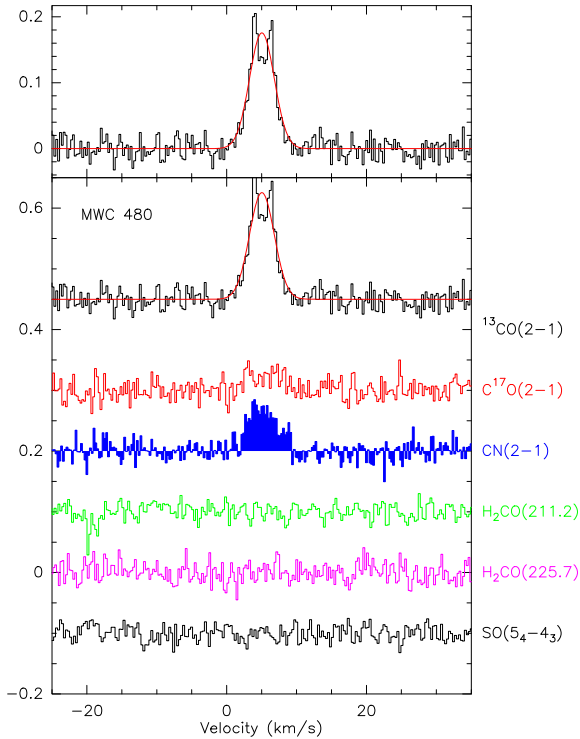}
\caption{As Fig.\ref{fig:haro6-10n} but towards \object{MWC 480}. The red curve
is a simple Gaussian fit to the disk emission in $^{13}$CO J=2-1.}
\label{fig:mwc_480}
\end{figure}

\subsection{Data Analysis}
\label{sec:sub:analysis}

We removed linear baselines  in a window $\pm$ 60 km.s$^{-1}$ wide
around each spectral line.  Spectra for all objects appear in Appendix \ref{app:spectra}.
Figures \ref{fig:haro6-10n} and \ref{fig:mwc_480}
show spectra for two objects: Haro 6-10 which exhibits bright lines in all molecules, and MWC\,480, an
isolated HAe star for which CN has already been imaged with the IRAM array \citep{Chapillon+etal_2012},
included in our sample as a reference object.
To derive line intensities,  we fitted simple Gaussian profiles to the observed lines
taking into account the hyperfine structure of the C$^{17}$O J=2-1 and
CN N=2-1 lines. We used the ``HFS (HyperFine Structure)'' method of the CLASS data
analysis program  in the GILDAS\footnote{See \texttt{http://www.iram.fr/IRAMFR/GILDAS} for more information
about GILDAS softwares} package for this purpose. This method assumes a common  excitation temperature
for all the hyperfine components, a likely situation at the high densities appearing in disks. For a Gaussian
line, the emerging profile is then
\begin{eqnarray}
T_A^*(v) & = & \eta \left( J_\nu(T_\mathrm{ex}) - J_\nu(T_\mathrm{bg}) \right) \times  \\ \nonumber
         &  & \Sigma_i \left(1-\exp\left(-R_i \tau \exp\left(-\left(\frac{v-V_\mathrm{sys}-v_i}{\delta V}\right)^2\right) \right) \right)
\end{eqnarray}
where $R_i$ are the relative intensities ($\Sigma_i R_i = 1$), $v_i$ the velocity of each hyperfine component
with respect to a common reference frequency, and $\tau$ the total line opacity (sum over all hyperfine components,
as the $R_i$ are normalized). $V_\mathrm{sys}$ is the source velocity,
$\delta V$ the line width at 1/e, and $\eta$ a beam filling factor. The HFS method uses
$X = \eta \tau (J_\nu(T_\mathrm{ex}) - J_\nu(T_\mathrm{bg})$,
$V_\mathrm{sys}$, $\delta V$ and $\tau$ as free parameters. The difference in saturation between the various
hyperfine components allows determination of $\tau$ provided sufficient signal to noise.
For CN, although all 19 hyperfine transitions were used for the fitting, the quoted
integrated line flux includes only the 3 blended hyperfine
components around 226.874 GHz and the two weak components at 226887.399 \&  226892.151 MHz
(which appear near V$_\mathrm{LSR}$ -11 and -17 $\kms$ in the brighter sources) \footnote{This is
essentially the flux in the J=5/2-3/2 fine structure line, as the only missing hyperfine component has
much smaller intensity}. Given the line ratios, the total line flux including all
hyperfine components would be 1.67 times larger in the optically thin limit.

\begin{table}[!h]
\caption{Frequencies of transitions}
\begin{tabular}{lcr}
 Frequency & Molecule and & Detection \\
 (MHz)  &  Transition     &  \\
 \hline
 206176.062 & SO ($5_4-4_3$) & Yes  \\
 209230.201 & HC$_3$N J=23-22 & No \\
 209419.15  & $^{13}$C$^{18}$O J=2-1 & cloud only \\
 211211.455 & ortho H$_2$CO $3_{13} - 2_{12}$ & Yes\\
% 211853.044 & $^{30}$SiO J=5-4 \\
 220398.688 & $^{13}$CO J=2-1 & Yes \\
 224714.385 & C$^{17}$O J=2-1 & Yes \\
 225697.781 & ortho H$_2$CO $3_{12} - 2_{11}$ & Yes \\
 226287.4185 & CN N=2-1 J=3/2-3/2, F=1/2-1/2 \\ % 0.0069 -4.1214 2    3.7866  2  265041234 2 0 2 1     1 0 2 1      CN, v=0,1
 226298.9427 & CN N=2-1 J=3/2-3/2, F=1/2-3/2 \\ % 0.0068 -4.2189 2    3.7862  2  265041234 2 0 2 1     1 0 2 2      CN, v=0,1
 226303.0372 & CN N=2-1 J=3/2-3/2, F=3/2-1/2 \\ % 0.0064 -4.2131 2    3.7866  4  265041234 2 0 2 2     1 0 2 1      CN, v=0,1
 226314.5400 & CN N=2-1 J=3/2-3/2, F=3/2-3/2 & Yes \\ % 0.0500 -3.8373 2    3.7862  4 -265041234 2 0 2 2     1 0 2 2      CN, v=0,1
 226332.4986 & CN N=2-1 J=3/2-3/2, F=3/2-5/2 \\ % 0.0056 -4.1747 2    3.7857  4  265041234 2 0 2 2     1 0 2 3      CN, v=0,1
 226341.9298 & CN N=2-1 J=3/2-3/2, F=5/2-3/2 \\ % 0.0054 -4.1578 2    3.7862  6  265041234 2 0 2 3     1 0 2 2      CN, v=0,1
 226359.8710 & CN N=2-1 J=3/2-3/2, F=5/2-5/2 & Yes \\ % 0.0500 -3.4508 2    3.7857  6 -265041234 2 0 2 3     1 0 2 3      CN, v=0,1
 226616.5714 & CN N=2-1 J=3/2-1/2, F=1/2-3/2 & Yes \\% 0.0053 -4.1042 2    3.7757  2  265041234 2 0 2 1     1 0 1 2      CN, v=0,1
 226632.1901 & CN N=2-1 J=3/2-1/2, F=3/2-3/2 & Yes \\ % 0.0035 -3.2044 2    3.7757  4  265041234 2 0 2 2     1 0 1 2      CN, v=0,1    0
 226659.5584 & CN N=2-1 J=3/2-1/2, F=5/2-3/2 & Yes \\ % 0.0026 -2.6815 2    3.7757  6  265041234 2 0 2 3     1 0 1 2      CN, v=0,1  +15
 226663.6928 & CN N=2-1 J=3/2-1/2, F=1/2-1/2 & Yes \\ % 0.0025 -3.2072 2    3.7741  2  265041234 2 0 2 1     1 0 1 1      CN, v=0,1   +7
 226679.3114 & CN N=2-1 J=3/2-1/2, F=3/2-1/2 & Yes \\ % 0.0031 -3.1122 2    3.7741  4  265041234 2 0 2 2     1 0 1 1      CN, v=0,1  -30
 226874.1908 & CN N=2-1 J=5/2-3/2, F=5/2-3/2 & Yes\\ % 0.0023 -2.6749 2    3.7862  6  265041234 2 0 3 3     1 0 2 2      CN, v=0,1     +8
 226874.7813 & CN N=2-1 J=5/2-3/2, F=7/2-5/2 & Yes \\ % 0.0030 -2.4751 2    3.7857  8  265041234 2 0 3 4     1 0 2 3      CN, v=0,1    +17
 226875.8960 & CN N=2-1 J=5/2-3/2, F=3/2-1/2 & Yes \\ % 0.0020 -2.9004 2    3.7866  4  265041234 2 0 3 2     1 0 2 1      CN, v=0,1      0
 226887.4202 & CN N=2-1 J=5/2-3/2, F=3/2-3/2 & Yes \\ % 0.0029 -3.3979 2    3.7862  4  265041234 2 0 3 2     1 0 2 2      CN, v=0,1    +21
 226892.1280 & CN N=2-1 J=5/2-3/2, F=5/2-5/2 & Yes \\ % 0.0027 -3.4004 2    3.7857  6  265041234 2 0 3 3     1 0 2 3      CN, v=0,1    -23
 226905.3574 & CN N=2-1 J=5/2-3/2, F=3/2-5/2 \\% 0.0044 -4.7824 2    3.7857  4  265041234 2 0 3 2     1 0 2 3      CN, v=0,1
 227191.8195 & CN N=2-1 J=5/2-1/2, F=5/2-3/2 \\% 0.0049 -7.4959 2    3.7757  6  265041234 2 0 3 3     1 0 1 2      CN, v=0,1 extremely weak...
\hline
\end{tabular}
\tablefoot{CN N=2-1 line frequencies were measured in laboratory by \citet{Skatrud+etal_1983};
we use here the fitted values from the CDMS Database \citep{CDMS_2001}.}
\label{tab:lines}
\end{table}
% end input

\begin{figure} %FIG.3
\includegraphics[width=\columnwidth]{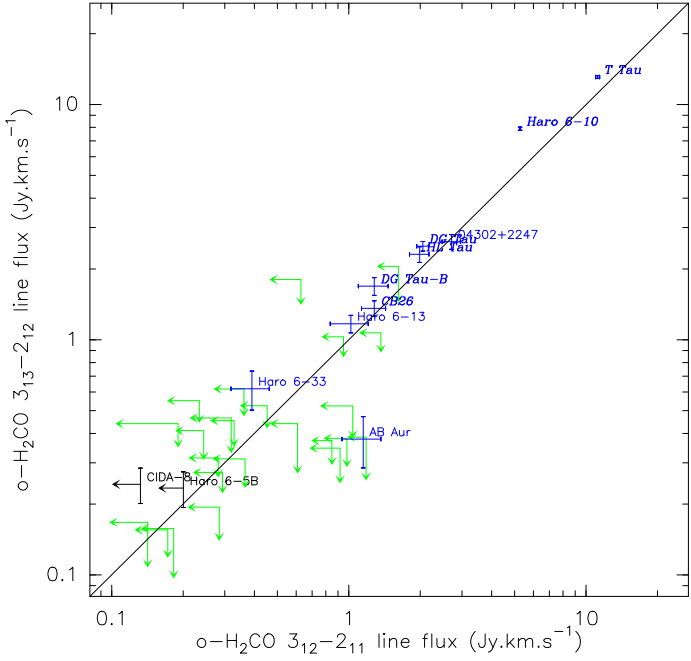}
\caption{Integrated line flux of the $3_{12}-2_{11}$ line of H$_2$CO vs
that of the $3_{13}-2_{12}$  transition.
Color codes indicate detected vs non-detected lines: blue for detection of both lines, black for detection of the $3_{13}-2_{12}$ line, green for non detections. The detection threshold is set at $4 \sigma$.
For detected lines, the errorbars are $1 \sigma$. For non-detected lines, the limit is the $2\sigma$ value, and
the arrow length is $1 \sigma$. Sources in italics are those with known molecular outflows.}
\label{fig:h2co-h2co}
\end{figure}

\begin{figure*} % FIG.4
\sidecaption
\includegraphics[width=12.0cm]{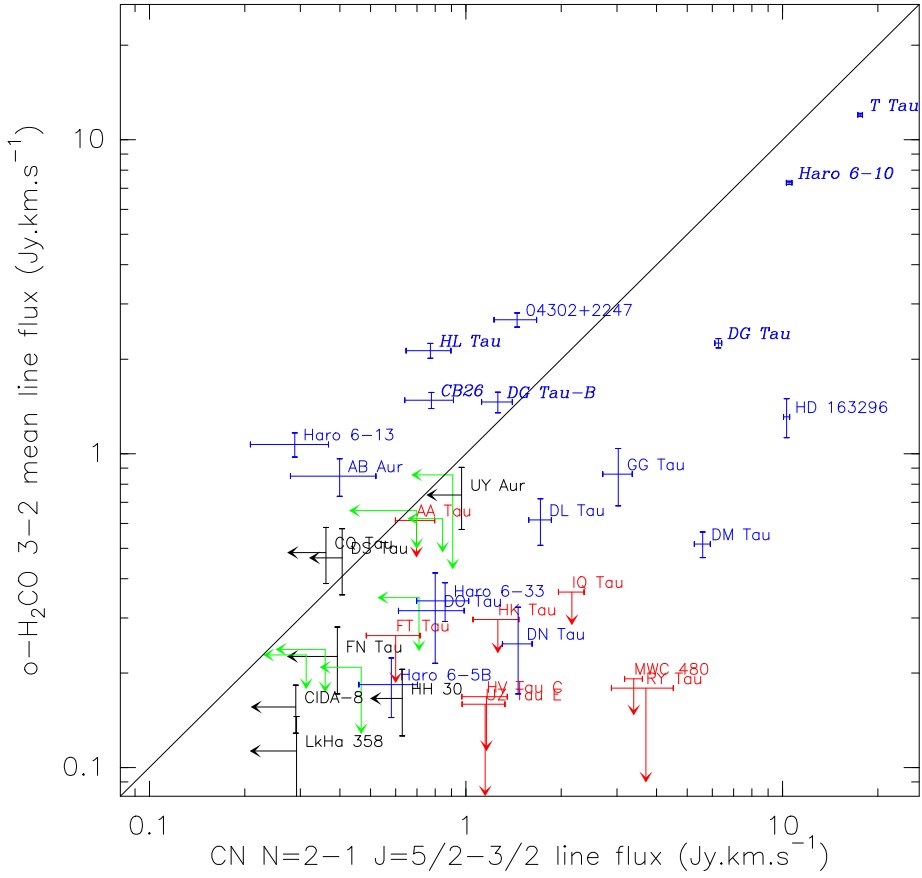}
\caption{Integrated line flux of H$_2$CO (combination of both transitions) vs CN N=2-1 (main group of
hyperfine components). Color codes indicate detected vs non-detected lines: blue for detection of both lines, red for detection
of the x-axis line only (here CN), black for detection
of the y-axis line only (here H$_2$CO), and green for non detections.
The detection threshold is set at $3 \sigma$.
For detected lines, the errorbars are $1 \sigma$. For non-detected lines, the limit is the $2\sigma$ value (best fit + $2 \sigma$),
and the arrow length is $1 \sigma$. Sources in italics are those with known molecular outflows.}
\label{fig:cn-h2co}
\end{figure*}

\begin{figure*} % FIG.5
\sidecaption
\includegraphics[width=12.0cm]{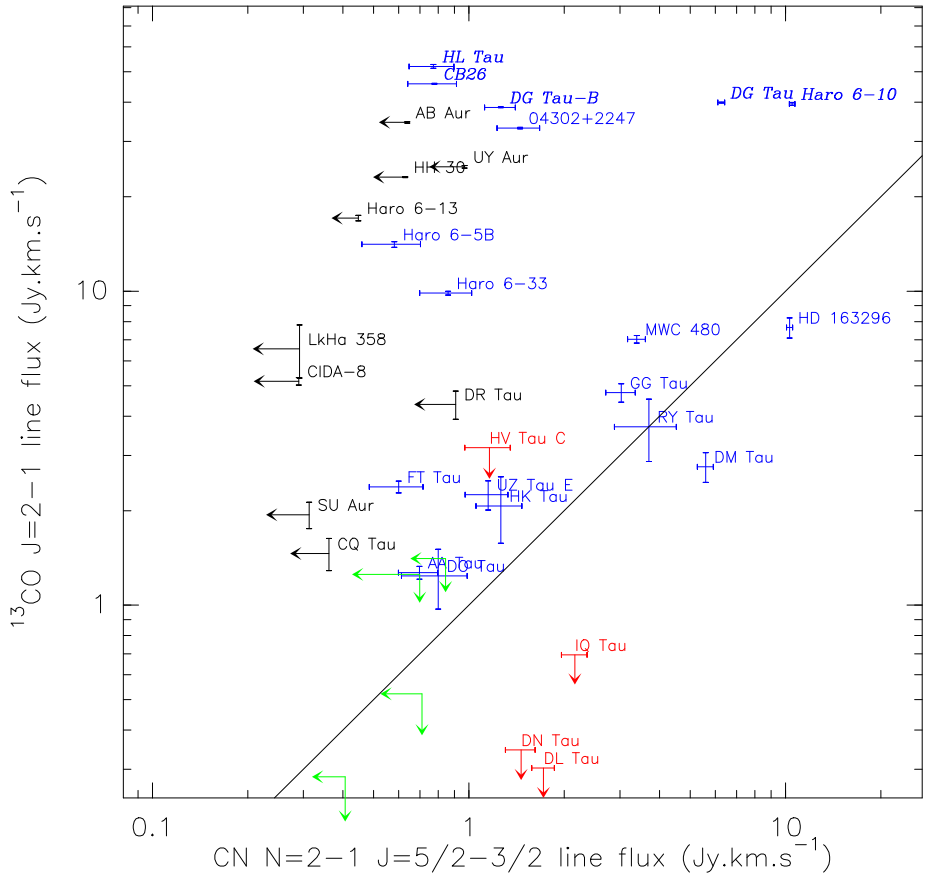}
\caption{As for {Fig.\ref{fig:cn-h2co}} but for $^{13}$CO J=2-1 (including contribution from
the cloud)  vs CN N=2-1. T Tau is out of scale here.}
\label{fig:13co-cn}
\end{figure*}

\begin{figure*} % FIG.6
\sidecaption
\includegraphics[width=12.0cm]{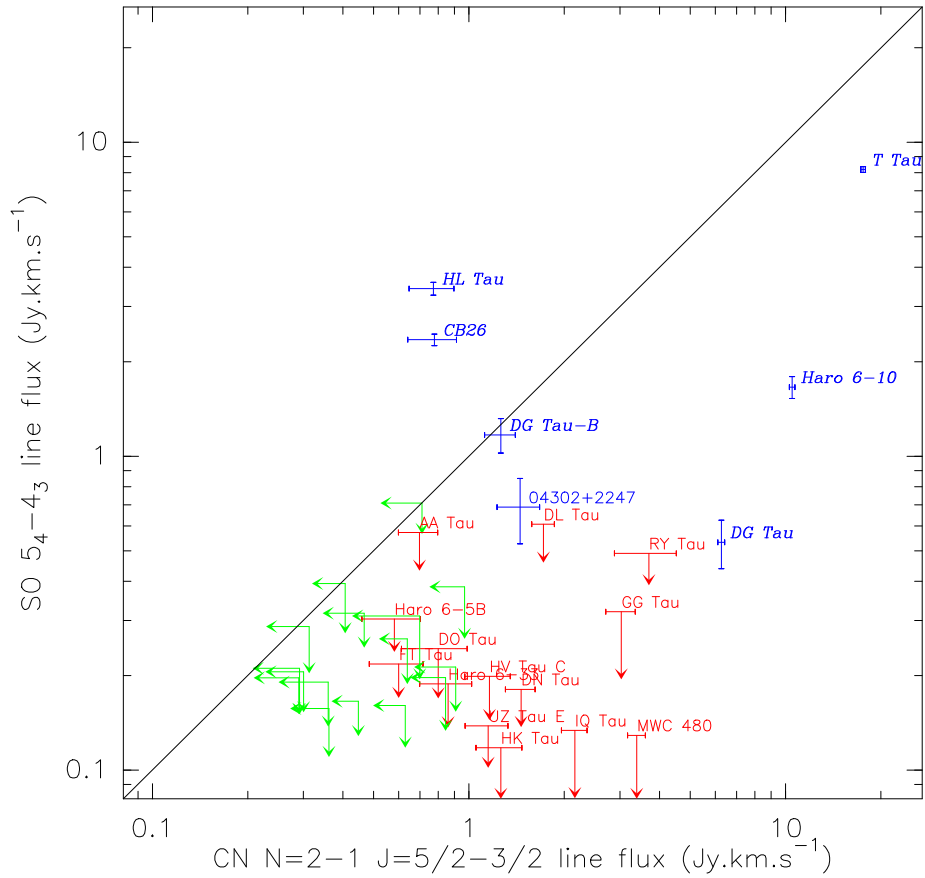}
\caption{As for {Fig.\ref{fig:cn-h2co}} but for SO vs CN N=2-1.}
\label{fig:so-cn}
\end{figure*}

\begin{figure*} % FIG.7
\sidecaption
\includegraphics[width=12.0cm]{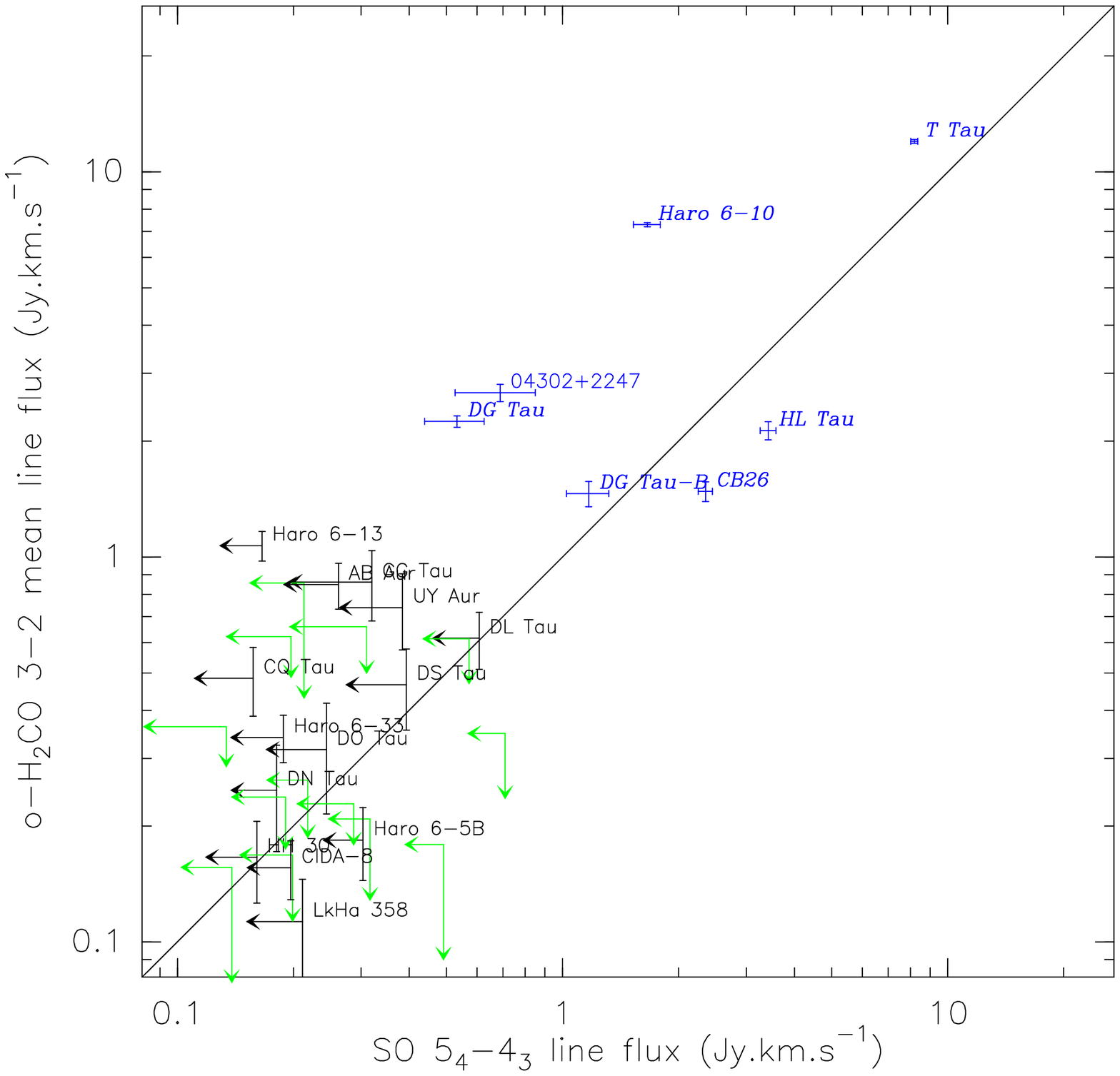}
\caption{As for {Fig.\ref{fig:cn-h2co}} but for H$_2$CO vs SO.}
\label{fig:h2co-so}
\end{figure*}

The two transitions of ortho-H$_2$CO exhibit a strong correlation (see Fig.\ref{fig:h2co-h2co})
because they arise in the same rotational levels  and differ only by the small K splitting,
with the 211.2 GHz line on average $20 \%$ stronger than the 225.2 GHz line.
We therefore computed an average spectrum and line intensity using
this ratio. For the several lines we analyzed,  Tables  \ref{tab:cn} and \ref{tab:h2co} list the derived line intensity,
 V$_{LSR}$ and line width, $\Delta V$.
Figures \ref{fig:cn-h2co} through \ref{fig:h2co-so} display relationships of the line intensities.
The color coding in these plots is as follows: blue designates
detection of both lines, red designates detection only of the line
plotted on the x-axis, and green detection only of the line plotted
along the y-axis, and black is for non detection of both lines.
Sources in italics are those with known molecular outflows.

\begin{figure*} % FIG.8
\sidecaption
\includegraphics[width=12.0cm]{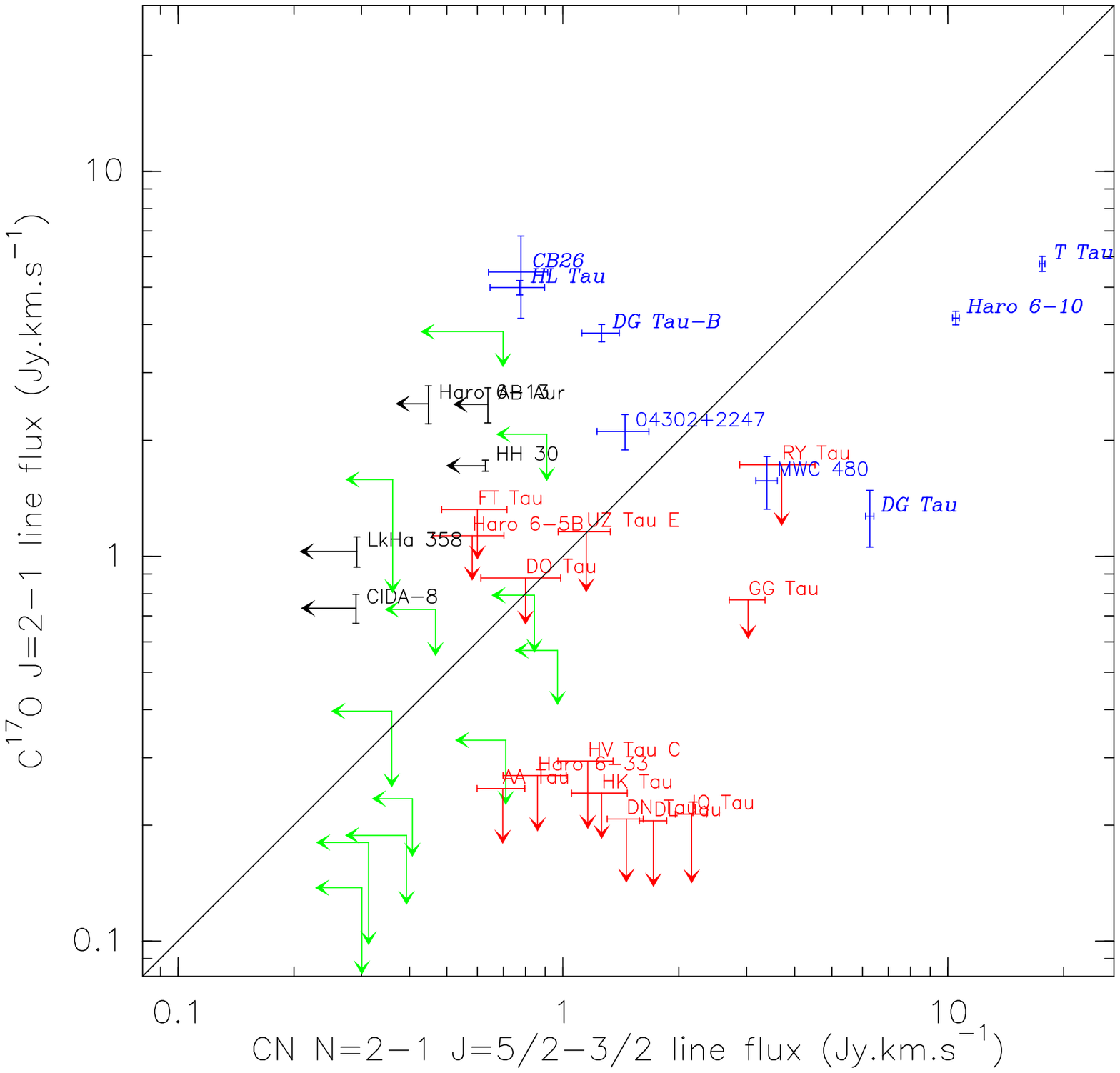}
\caption{As for {Fig.\ref{fig:cn-h2co}} but for C$^{17}$O J=2-1 vs CN N=2-1.}
\label{fig:c17o-cn}
\end{figure*}

\begin{figure*} % FIG.7
\sidecaption
\includegraphics[width=12.0cm]{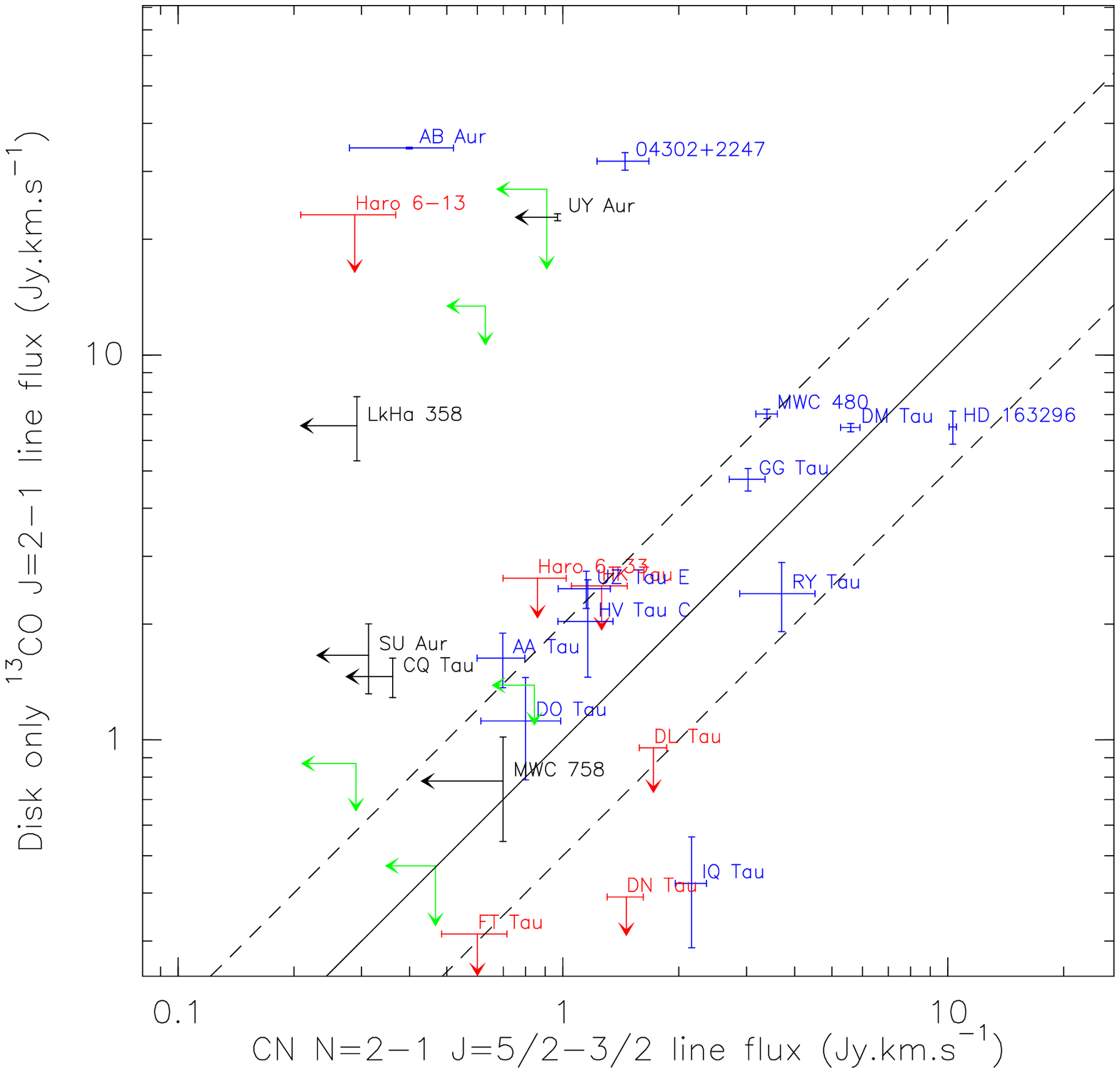}
\caption{As for {Fig.\ref{fig:cn-h2co}} but for $^{13}$CO (disk
component only) vs CN. The dashed lines indicate a factor
2 between the two line fluxes.}
\label{fig:disk-cn}
\end{figure*}

The $^{13}$CO results are seriously affected by confusion. In many
cases, we attempted to  fit only a putative disk component, by masking
the channels in which confusion seemed likely, either because they
appear negative or exceedingly strong, or because of strong C$^{17}$O
emission at the same velocity. A summary of disk detection
is given in Table \ref{tab:radii}. The fits are displayed on top of the $^{13}$CO spectra in
the Figures of Appendix \ref{app:spectra}. The fit results can be found
in Table \ref{tab:cn}. These derived line intensities are presented as a function
of CN line flux in Fig.\ref{fig:disk-cn}. Because of the masking process to ignore confused
parts of the spectrum, the errorbars can be large. However, in sources where both $^{13}$CO and CN are detected, they in general display similar intensities, within a factor 2.

%% HD 163296	&	17:56:21.31	&	-21:57:22.0	&	6.5	\\

\begin{table*}
\caption{Fit Results: CN and $^{13}$CO}
\begin{tabular}{l rrr rrr l}
\hline
Source & \multicolumn{3}{c}{CN} & \multicolumn{3}{c}{$^{13}$CO} \\
       & Area & V$_\mathrm{LSR}$ & $\Delta V$ &  Area & V$_\mathrm{LSR}$ & $\Delta V$ \\
       & Jy.$\kms$ & $\kms$ & $\kms$ & Jy.$\kms$ & $\kms$ & $\kms$ \\
\hline
      FN Tau  & $ <       0.38 $ & & &
\textit{      -1.40 $\pm$      0.09  } & \textit{ 7.44$\pm$ 0.03  } & \textit{ 0.84 $\pm$ 0.05 } & \textit{Cloud}  \\
      CW Tau  &     -0.66  $\pm$      0.07  & 6.80 $\pm$ 0.07  & 0.85  $\pm$ 0.07  &
      2.69  $\pm$      0.45  & 6.20 $\pm$ 0.10  & 3.25  $\pm$ 0.38 \\
      CIDA-1  &     -0.82  $\pm$      0.09  & 6.93 $\pm$ 0.02  & 0.42  $\pm$ 0.03  &
      1.14  $\pm$      0.75  & 8.23 $\pm$ 0.19  & 0.56  $\pm$ 0.27 \\
      BP Tau  & $ <       0.59 $ & & &
\textit{       0.26 $\pm$      0.14  } & \textit{ 3.65$\pm$ 0.39  } & \textit{ 1.06 $\pm$ 0.60 } & \textit{Cloud}  \\
      DE Tau  & $ <       0.39 $ & & &
      0.20  $\pm$      0.15  &11.43 $\pm$ 0.99  & [ 3.00] \\
      RY Tau  &      3.91  $\pm$      0.86  & 5.06 $\pm$ 1.58  &14.22  $\pm$ 2.19  &
      2.54  $\pm$      0.51  & 7.24 $\pm$ 0.29  & [ 3.00] \\
 T Tau  &     18.62  $\pm$      0.31  & 7.96 $\pm$ 0.03  & 3.12  $\pm$ 0.05  &
\textit{     170.37 $\pm$      0.20  } & \textit{ 8.02$\pm$ 0.00  } & \textit{ 2.50 $\pm$ 0.00 } & \textit{Cloud}  \\
   Haro 6-5B  &      0.61  $\pm$      0.13  & 7.67 $\pm$ 0.42  & 3.61  $\pm$ 0.73  &
\textit{      14.94 $\pm$      0.30  } & \textit{ 7.88$\pm$ 0.01  } & \textit{ 0.80 $\pm$ 0.02 } & \textit{Cloud}  \\
      FT Tau  &      0.63  $\pm$      0.12  & 6.95 $\pm$ 0.28  & 2.21  $\pm$ 0.42  &
      0.15  $\pm$      0.07  & 9.38 $\pm$ 0.20  & 0.78  $\pm$ 0.34 \\
DG Tau-B  &      1.33  $\pm$      0.15  & 6.18 $\pm$ 0.18  & 2.34  $\pm$ 0.22  &
\textit{      40.59 $\pm$      0.11  } & \textit{ 6.43$\pm$ 0.00  } & \textit{ 2.03 $\pm$ 0.00 } & \textit{Cloud}  \\
DG Tau  &      6.62  $\pm$      0.16  & 6.52 $\pm$ 0.03  & 2.28  $\pm$ 0.07  &
\textit{      42.08 $\pm$      0.14  } & \textit{ 6.61$\pm$ 0.00  } & \textit{ 2.20 $\pm$ 0.01 } & \textit{Cloud}  \\
Haro 6  &     11.05  $\pm$      0.22  & 6.70 $\pm$ 0.02  & 1.97  $\pm$ 0.05  &
\textit{      41.64 $\pm$      0.47  } & \textit{ 7.15$\pm$ 0.01  } & \textit{ 1.07 $\pm$ 0.02 } & \textit{Cloud}  \\
      IQ Tau  &      2.28  $\pm$      0.21  & 6.22 $\pm$ 0.25  & 5.07  $\pm$ 0.35  &
      0.45  $\pm$      0.14  & 4.70 $\pm$ 0.24  & 1.45  $\pm$ 0.60 \\
    LkHa 358  & $ <       0.27 $ & & &
      6.91  $\pm$      1.31  & 6.82 $\pm$ 0.04  & 1.43  $\pm$ 0.12 \\
       HH 30  & $ <       0.41 $ & & &
      8.21  $\pm$      2.97  & 7.18 $\pm$ 0.07  & 2.34  $\pm$ 0.29 \\
HL Tau  &      0.81  $\pm$      0.13  & 5.00 $\pm$ 0.20  & 1.91  $\pm$ 0.37  &
\textit{      54.88 $\pm$      0.76  } & \textit{ 5.88$\pm$ 0.01  } & \textit{ 1.46 $\pm$ 0.03 } & \textit{Cloud}  \\
      HK Tau  &      1.33  $\pm$      0.22  & 4.49 $\pm$ 0.76  & 9.25  $\pm$ 1.14  &
      1.40  $\pm$      0.62  & 5.25 $\pm$ 0.62  & 3.12  $\pm$ 0.77 \\
   Haro 6-13  &      0.30  $\pm$      0.08  & 5.37 $\pm$ 0.14  & 0.69  $\pm$ 0.16  &
     10.05  $\pm$      7.19  & 5.03 $\pm$ 0.06  & 1.67  $\pm$ 0.12 \\
      GG Tau  &      3.20  $\pm$      0.34  & 6.31 $\pm$ 0.17  & 2.40  $\pm$ 0.22  &
      5.01  $\pm$      0.34  & 6.49 $\pm$ 0.09  & 2.66  $\pm$ 0.17 \\
    UZ Tau E  &      1.21  $\pm$      0.19  & 5.96 $\pm$ 0.47  & 5.75  $\pm$ 0.66  &
      2.61  $\pm$      0.29  & 5.90 $\pm$ 0.18  & 4.47  $\pm$ 0.46 \\
  04302+2247  &      1.53  $\pm$      0.24  & 5.92 $\pm$ 0.48  & 5.37  $\pm$ 0.55  &
     33.71  $\pm$      1.78  & 5.62 $\pm$ 0.02  & 2.92  $\pm$ 0.06 \\
      DL Tau  &      1.82  $\pm$      0.15  & 6.70 $\pm$ 0.10  & 1.41  $\pm$ 0.18  &
      0.53  $\pm$      0.24  & 6.62 $\pm$ 0.29  & 1.29  $\pm$ 0.64 \\
      DM Tau  &      5.90  $\pm$      0.34  & 6.04 $\pm$ 0.07  & 1.86  $\pm$ 0.12  &
      6.84  $\pm$      0.17  & 6.07 $\pm$ 0.02  & 1.58  $\pm$ 0.04 \\
      AA Tau  &      0.74  $\pm$      0.10  & 7.36 $\pm$ 0.11  & 0.96  $\pm$ 0.15  &
      1.72  $\pm$      0.28  & 6.38 $\pm$ 0.53  & 6.16  $\pm$ 1.03 \\
      DN Tau  &      1.54  $\pm$      0.17  & 6.61 $\pm$ 0.20  & 3.17  $\pm$ 0.29  &
      0.24  $\pm$      0.09  & 7.79 $\pm$ 0.22  & 0.98  $\pm$ 0.44 \\
      DO Tau  &      0.84  $\pm$      0.20  & 5.91 $\pm$ 0.22  & 1.44  $\pm$ 0.54  &
      1.18  $\pm$      0.35  & 3.61 $\pm$ 0.85  & 4.96  $\pm$ 1.65 \\
    HV Tau C  &      1.22  $\pm$      0.20  & 6.95 $\pm$ 0.59  & 6.51  $\pm$ 0.67  &
      2.14  $\pm$      0.61  & 6.08 $\pm$ 0.34  & 5.02  $\pm$ 0.87 \\
   Haro 6-33  &      0.91  $\pm$      0.17  & 5.88 $\pm$ 0.33  & 2.98  $\pm$ 0.53  &
      1.59  $\pm$      0.60  & 4.95 $\pm$ 0.55  & 4.52  $\pm$ 0.98 \\
      DQ Tau  & $ <       0.34 $ & & &
\textit{     -13.78 $\pm$      0.14  } & \textit{ 9.62$\pm$ 0.01  } & \textit{ 1.62 $\pm$ 0.01 } & \textit{Cloud}  \\
      DR Tau  & $ <       0.75 $ & & &
      6.73  $\pm$     10.82  & 9.79 $\pm$ 1.95  & 1.80  $\pm$ 1.42 \\
      DS Tau  & $ <       0.28 $ & & &
\textit{       0.13 $\pm$      0.08  } & \textit{ 8.13$\pm$ 0.22  } & \textit{ 0.64 $\pm$ 0.45 } & \textit{Cloud}  \\
      UY Aur  & $ <       0.69 $ & & &
     24.09  $\pm$      0.48  & 6.30 $\pm$ 0.01  & 1.41  $\pm$ 0.02 \\
      AB Aur  &      0.42  $\pm$      0.13  & 6.87 $\pm$ 0.08  & 0.47  $\pm$ 0.09  &
     36.43  $\pm$      0.17  & 5.89 $\pm$ 0.00  & 1.56  $\pm$ 0.01 \\
      SU Aur  & $ <       0.27 $ & & &
      1.75  $\pm$      0.36  & 7.09 $\pm$ 0.33  & 3.05  $\pm$ 0.47 \\
     MWC 480  &      3.58  $\pm$      0.23  & 5.04 $\pm$ 0.16  & 4.44  $\pm$ 0.22  &
      7.42  $\pm$      0.21  & 5.04 $\pm$ 0.06  & 4.18  $\pm$ 0.13 \\
  CB26  &      0.82  $\pm$      0.14  & 6.20 $\pm$ 0.35  & 3.35  $\pm$ 0.45  &
\textit{      48.36 $\pm$      0.15  } & \textit{ 5.68$\pm$ 0.00  } & \textit{ 1.39 $\pm$ 0.00 } & \textit{Cloud}  \\
      CIDA-8  & $ <       0.26 $ & & &
      0.46  $\pm$      0.23  & 7.69 $\pm$ 0.68  & [ 3.00] \\
     CIDA-11  & $ <       0.23 $ & & &
\textit{       0.11 $\pm$      0.05  } & \textit{ 9.04$\pm$ 0.11  } & \textit{ 0.30 $\pm$ 0.33 } & \textit{Cloud}  \\
      RW Aur  & $ <       0.60 $ & & &
      0.84  $\pm$      0.31  & 2.46 $\pm$ 0.59  & 2.86  $\pm$ 1.70 \\
     MWC 758  & $ <       0.86 $ & & &
      0.82  $\pm$      0.25  & 6.27 $\pm$ 0.39  & 2.28  $\pm$ 0.61 \\
      CQ Tau  & $ <       0.28 $ & & &
      1.54  $\pm$      0.18  & 6.01 $\pm$ 0.34  & 5.65  $\pm$ 0.70 \\
   HD 163296  &     10.91  $\pm$      0.23  & 5.85 $\pm$ 0.05  & 4.28  $\pm$ 0.07  &
      6.87  $\pm$      0.68  & 6.05 $\pm$ 0.20  & 4.02  $\pm$ 0.45 \\
\hline
\end{tabular}
\tablefoot{Integrated line flux of CN (main group of hyperfine components) and $^{13}$CO. For
$^{13}$CO, results in \textit{italics} include cloud (or outflow) emission, while
the others were obtained by masking confused channels and trace the disk emission only.}
\label{tab:cn}
\end{table*}
%end input

%\input table-4-result
%
\begin{table*}
\caption{Fit Results: H$_2$CO and C$^{17}$O}
\begin{tabular}{l rrr rrr}
\hline
Source & \multicolumn{3}{c}{H$_2$CO} & \multicolumn{3}{c}{C$^{17}$O} \\
       & Area & V$_\mathrm{LSR}$ & $\Delta V$ &  Area & V$_\mathrm{LSR}$ &  $\Delta V$ \\
       & Jy.$\kms$ & $\kms$ & $\kms$ & Jy.$\kms$ & $\kms$ & $\kms$ \\
\hline
      FN Tau  &      0.24  $\pm$      0.06  &-0.53 $\pm$ 0.14  & 1.10  $\pm$ 0.31  &
    $ <       0.20    $ \\
      CW Tau  &     -0.27  $\pm$      0.03  & 6.80 $\pm$ 0.03  & 0.51  $\pm$ 0.05  &
      0.21  $\pm$      0.05  & 6.28 $\pm$ 0.06  & 0.43  $\pm$ 0.11 \\
      CIDA-1  &     -0.25  $\pm$      0.04  & 6.98 $\pm$ 0.03  & 0.43  $\pm$ 0.09  &
     -0.87  $\pm$      0.07  & 6.91 $\pm$ 0.02  & 0.43  $\pm$ 0.01 \\
      BP Tau  & $ <       0.36 $ & & &
    $ <       0.34    $ \\
      DE Tau  & $ <       0.26 $ & & &
    $ <       0.56    $ \\
      RY Tau  & $ <       0.28 $ & & &
    $ <       1.68    $ \\
 T Tau  &     12.68  $\pm$      0.14  & 7.87 $\pm$ 0.01  & 2.51  $\pm$ 0.03  &
      6.07  $\pm$      0.27  & 7.96 $\pm$ 0.04  & 2.03  $\pm$ 0.11 \\
   Haro 6-5B  &      0.19  $\pm$      0.04  & 8.40 $\pm$ 0.21  & 1.49  $\pm$ 0.29  &
    $ <       0.84    $ \\
      FT Tau  & $ <       0.25 $ & & &
    $ <       1.08    $ \\
DG Tau-B  &      1.54  $\pm$      0.12  & 6.43 $\pm$ 0.05  & 1.61  $\pm$ 0.19  &
      4.01  $\pm$      0.21  & 6.44 $\pm$ 0.02  & 0.98  $\pm$ 0.06 \\
DG Tau  &      2.37  $\pm$      0.08  & 6.43 $\pm$ 0.03  & 1.50  $\pm$ 0.06  &
      1.34  $\pm$      0.22  & 5.87 $\pm$ 0.15  & 2.24  $\pm$ 0.48 \\
Haro 6  &      7.70  $\pm$      0.09  & 6.59 $\pm$ 0.01  & 1.44  $\pm$ 0.02  &
      4.39  $\pm$      0.18  & 6.39 $\pm$ 0.02  & 0.98  $\pm$ 0.05 \\
      IQ Tau  & $ <       0.25 $ & & &
    $ <       0.23    $ \\
    LkHa 358  &      0.12  $\pm$      0.03  & 7.14 $\pm$ 0.03  & 0.28  $\pm$ 0.57  &
      1.09  $\pm$      0.10  & 7.05 $\pm$ 0.02  & 0.45  $\pm$ 0.04 \\
       HH 30  &      0.18  $\pm$      0.04  & 6.06 $\pm$ 0.11  & 0.78  $\pm$ 0.15  &
      1.81  $\pm$      0.06  & 6.52 $\pm$ 0.01  & 0.43  $\pm$ 0.00 \\
HL Tau  &      2.25  $\pm$      0.12  & 6.66 $\pm$ 0.12  & 4.25  $\pm$ 0.28  &
      5.27  $\pm$      0.23  & 6.35 $\pm$ 0.11  & 5.08  $\pm$ 0.28 \\
      HK Tau  & $ <       0.21 $ & & &
    $ <       0.18    $ \\
   Haro 6-13  &      1.13  $\pm$      0.10  & 5.81 $\pm$ 0.17  & 3.85  $\pm$ 0.43  &
      2.63  $\pm$      0.30  & 5.60 $\pm$ 0.01  & 0.43  $\pm$ 0.01 \\
      GG Tau  &      0.91  $\pm$      0.19  & 6.40 $\pm$ 0.24  & 2.02  $\pm$ 0.46  &
    $ <       0.50    $ \\
    UZ Tau E  & $ <       0.25 $ & & &
    $ <       1.11    $ \\
  04302+2247  &      2.82  $\pm$      0.15  & 5.73 $\pm$ 0.13  & 4.88  $\pm$ 0.27  &
      2.23  $\pm$      0.24  & 6.24 $\pm$ 0.20  & 3.77  $\pm$ 0.51 \\
      DL Tau  &      0.65  $\pm$      0.11  & 6.59 $\pm$ 0.26  & 2.98  $\pm$ 0.54  &
    $ <       0.21    $ \\
      DM Tau  &      0.54  $\pm$      0.05  & 6.21 $\pm$ 0.06  & 1.22  $\pm$ 0.13  &
\\
      AA Tau  & $ <       0.46 $ & & &
    $ <       0.22    $ \\
      DN Tau  &      0.26  $\pm$      0.08  & 5.73 $\pm$ 0.28  & 1.62  $\pm$ 0.56  &
    $ <       0.21    $ \\
      DO Tau  &      0.33  $\pm$      0.11  & 5.35 $\pm$ 0.39  & 2.15  $\pm$ 0.54  &
    $ <       0.68    $ \\
    HV Tau C  & $ <       0.18 $ & & &
    $ <       0.31    $ \\
   Haro 6-33  &      0.36  $\pm$      0.05  & 5.82 $\pm$ 0.05  & 0.67  $\pm$ 0.12  &
    $ <       0.24    $ \\
      DQ Tau  & $ <       0.20 $ & & &
    $ <       0.46    $ \\
      DR Tau  & $ <       1.36 $ & & &
    $ <       1.61    $ \\
      DS Tau  &      0.49  $\pm$      0.12  & 6.08 $\pm$ 0.44  & 3.46  $\pm$ 0.80  &
    $ <       0.22    $ \\
      UY Aur  &      0.78  $\pm$      0.17  & 5.92 $\pm$ 0.21  & 1.79  $\pm$ 0.40  &
    $ <       0.50    $ \\
      AB Aur  &      0.90  $\pm$      0.12  & 6.11 $\pm$ 0.19  & 2.56  $\pm$ 0.29  &
      2.61  $\pm$      0.28  & 6.02 $\pm$ 0.10  & 2.03  $\pm$ 0.25 \\
      SU Aur  & $ <       0.16 $ & & &
    $ <       0.26    $ \\
     MWC 480  & $ <       0.14 $ & & &
      1.66  $\pm$      0.26  & 5.51 $\pm$ 0.46  & 5.60  $\pm$ 0.88 \\
  CB26  &      1.56  $\pm$      0.09  & 5.78 $\pm$ 0.06  & 2.45  $\pm$ 0.20  &
      5.77  $\pm$      1.39  & 5.59 $\pm$ 0.04  & 0.82  $\pm$ 0.11 \\
      CIDA-8  &      0.16  $\pm$      0.03  & 6.75 $\pm$ 0.02  & 0.28  $\pm$ 0.58  &
      0.77  $\pm$      0.07  & 6.77 $\pm$ 0.02  & 0.43  $\pm$ 0.00 \\
     CIDA-11  & $ <       0.19 $ & & &
    $ <       0.18    $ \\
      RW Aur  & $ <       0.43 $ & & &
    $ <       0.73    $ \\
     MWC 758  & $ <       0.51 $ & & &
      2.51  $\pm$      0.77  &-0.31 $\pm$ 3.07  &17.47  $\pm$ 4.84 \\
      CQ Tau  &      0.51  $\pm$      0.10  & 6.15 $\pm$ 0.49  & 4.41  $\pm$ 0.77  &
    $ <       2.47    $ \\
    HD163296  &      1.38  $\pm$      0.20  & 5.50 $\pm$ 0.37  & 5.12  $\pm$ 0.79  &
\\
\hline
\end{tabular}
\tablefoot{Integrated line flux of H$_2$CO (average of both transitions) and C$^{17}$O}
\label{tab:h2co}
\end{table*}
%end input
					
\section{Results}
\label{sec:results}

Line emission from these objects can come \textit{a priori} from 4 distinct
regions: molecular cloud(s) along the line of sight, the circumstellar disk, a molecular outflow
and (specially for the youngest objects) a remnant envelope. Molecular cloud will exhibit narrow
lines, and because of our observing technique which differentiate the emission with that 1$'$ away,
may appear in emission or absorption. Disks should lead to symmetric, in general double-peaked, line
profiles, and line widths $\propto \sin{i}$ (with typical values 2-4 $\kms$, face-on objects being rare).
Contribution from outflows or envelopes may vary considerably, and may be difficult to distinguish from disk emission.
Most of the stars in Table \ref{tab:sources}  have a rich observational history
from optical, infrared and radio molecular
line observations which can help interpreting the line profiles.  We have compiled the relevant information and
comment on the observational results for each source in  Appendix \ref{app:sources}.
% so as not to break up the presentation here.  We refer the reader there for information of specific stars.

\subsection{Confusion in $^{13}$CO}
\label{sec:sub:confusion}

The $^{13}$CO spectra are in general complex. Out of the 42 sources, the only two in which $^{13}$CO exhibits
the expected double-peaked profile which is a clear signature of emission from Keplerian disk are
the previously studied objects \object{GG Tau} and \object{MWC 480} (see Fig.\ref{fig:mwc_480}), included
here as reference objects. \object{CQ Tau} and \object{MWC 758} also exhibit $^{13}$CO spectra which can be attributed
solely to disk emission. These 4 sources had been previously studied specifically because they are
relatively far from the molecular clouds.

For the other sources, the spectrum complexity has two main origins. Narrow features appear in emission
or apparent absorption. These narrow features result from small velocity and intensity gradients in the
molecular cloud which are on the same line of sight. Our beam-switching observing procedure reveals
such gradients on the arcminute scale. Note that, as the switching was made at constant elevation, the
gradient direction rotates on the plane of the sky, so that the observed features actually depend on
when the
observations were performed.  A comparison between the spectra of \object{BP Tau} obtained here and
that published by \citet{Dutrey+etal_2003} illustrates  this effect. The BP Tau case also demonstrates
that the lack of apparent confusion does not imply the source is unaffected by the cloud: a sufficiently
optically thick and homogeneous cloud will hide any disk emission situated behind.

Besides contamination by molecular clouds, many sources also exhibit strong, relatively
wide profiles in $^{13}$CO. Based on prior knowledge about these sources, this emission is in general
coming from outflows or envelopes, as in \object{HL Tau}, \object{T Tau},
\object{Haro 6-10}, \object{CB26}, \object{DG Tau} and \object{DG Tau B}, which are all known to drive powerful
jets and CO outflows (see Appendix \ref{app:sources}). This interpretation is corroborated by the
emission lines from other molecules (see below). However, the line widths are not very large, so
this emission can also contain a substantial contribution from the disk, specially for edge-on disks.
An example is \object{IRAS04302+2247} \citep[hereafter 04302+2247, the Butterfly star, see][]{Wolf+etal_2003}, for which IRAM interferometric
observations reveal that the detected lines clearly trace the disk \citep[][in prep]{Dutrey+etal_2012}.

Despite confusion with molecular clouds, many other sources exhibit line wings which
are potential signatures of disk emission. As mentioned in Sec.\ref{sec:sub:analysis},
these sources are identified in Table \ref{tab:radii}, and appear in Fig.\ref{fig:disk-cn}.

\begin{figure}[!ht]
\includegraphics[width=\columnwidth]{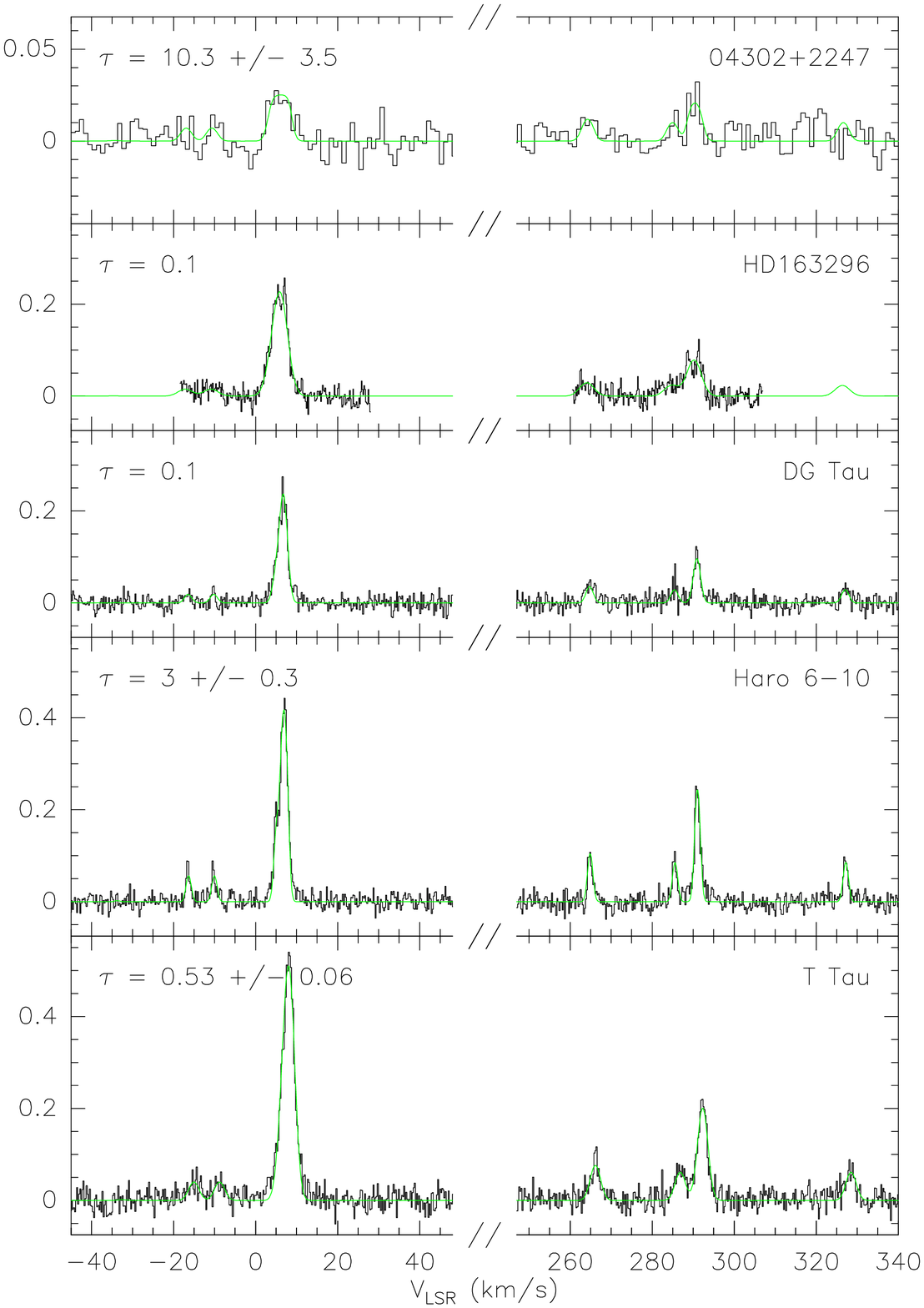}
\caption{Hyperfine components of CN in bright, potentially thick sources.
The best fit with the hyperfine components is overlaid. The intensity scale is $T_A^*$ (K). }
\label{fig:thick-cn}
\end{figure}

\subsection{CN as a confusion-free disk tracer}
\label{sec:sub:cn}

There is essentially no evidence
for confusion with molecular
clouds in CN, except for a few (4 or 5) sources.
This can be seen by the lack of narrow features
corresponding to those detected in $^{13}$CO.

Only two sources exhibit clear evidence of confusion in CN: \object{CW Tau} and \object{CIDA-1}
(see Fig.\ref{fig:CW_TAU} and \ref{fig:CIDA-1}) in which
the CN lines appear in absorption. These two sources are separated by a few arcmin only, and all other lines also
display the same behaviour, clearly demonstrating the confusion along this line of sight.
The lines are very narrow, $0.44 \pm 0.04 \kms$, a further indication that this comes from a cold molecular cloud.
More detailed arguments about the properties of the molecular cloud and its impact as a confusing signal
for disk emission are given in Appendix \ref{app:cloud}.

A closer examination reveals confusion in two other sources, \object{AA Tau} and \object{AB Aur}, and an unclear
situation for \object{Haro 6-13}, which exhibit emission with relatively narrow lines
(see Table \ref{tab:cn} and Appendix \ref{app:sources} for details). Finally,
like for $^{13}$CO, a lack of apparent confusion does not indicate an absence of confusion.

CN emission is compatible with optically thin emission in most sources ($\tau_p < 0.5$,
where $\tau_p$ is the peak line opacity). The HFS method yields $\tau$, the sum of opacities of all hyperfine components.
The main group of components accounts for a little more than half of this sum, but
with local linewidths in the range 0.2-0.5 $\kms$, the peak opacity is $\tau_p \sim 0.27 - 0.4 \tau$,
because the hyperfine components overlap only very moderately, being separated by 0.77 and 1.50 km.s$^{-1}$.
The opacity is constrained in the brightest sources: \object{T Tau} has optically thin
emission ($\tau = 0.53 \pm 0.06$) but \object{Haro 6-10} exhibits substantial optical depth ($\tau = 3.0 \pm 0.3$,
i.e. $\tau_p \approx 1$). The next brightest source, \object{DG Tau}, is marginally thick ($\tau = 0.9 \pm 0.3$).
The spectra also suggest CN could be optically thick in the
Butterfly star 04302+2247, but because of limited signal-to-noise ratio, the significance is low
($\tau = 10 \pm 3$). Spectra of these sources are displayed in Fig.\ref{fig:thick-cn}, together
with the best fit profiles.

Non zero optical depth ($\tau \simeq 1$) is also suggested for some of the other
strong sources, such has \object{DL Tau}, and also \object{DM Tau} and \object{LkCa 15}, but
the error bars remain large. The interferometric study performed by \citet{Chapillon+etal_2012} however confirms
that CN is not completely optically thin in the latter two sources.
Apart from Haro 6-10 and 04302+2247, we found no other source in our sample with optical depth
larger than $\sim 1$;  the weighted mean opacity is $\sim 0.6$ \footnote{The same
value is obtained with or without T Tau.}.

In summary, we detected CN N=2-1 in 24 out of the 42 stars in our sample, and have only 4
cases of confusion, and essentially all spectra are consistent with total optical depth $\sim 0.3 - 0.6$.

\subsection{SO and H$_2$CO as outflow or envelope tracers}
\label{sec:sub:so}

Strong SO emission is found in 7 out of the 41 stars in which it has been observed (see Fig.\ref{fig:so-cn}).
All these sources also display relatively strong H$_2$CO emission (see Fig.\ref{fig:h2co-so}). All of them, except 04302+2247,
are driving well-known jets and outflows. In these sources, CN is either quite strong
(\object{DG Tau}, \object{T Tau}, \object{DG Tau B}, \object{Haro 6-10}) or very weak (\object{CB26}, \object{HL Tau}).
All these sources are also deeply embedded, and surrounded by an extended complex envelope.  We also have a marginal
detection of SO in \object{Haro 6-5 B}, another embedded, jet-driving object. The line profiles of SO and
H$_2$CO are similar, and resemble those of C$^{17}$O when detected.
As the line widths are not very wide, this emission may be attributed to the envelope rather
than the outflow itself.  On the contrary, CN lines often display profiles which are quite
different from those of SO, being either narrower (e.g. \object{HL Tau}, Fig.\ref{fig:HL_TAU})
or with no narrow velocity component (\object{DG Tau B},
Fig.\ref{fig:DG_TAU-B}), or (rarely) broader (\object{DG Tau}, see Fig.\ref{fig:DG_TAU}).

The detection rate of H$_2$CO is rather high, with 20 out of 42 stars observed
(not counting the dubious cases \object{FN Tau} and \object{CIDA-8}).
However, cloud contamination appears more important in H$_2$CO (7 sources with narrow lines out of 42) than in CN.
This can be explained by excitation considerations, because of
the smaller critical densities of the observed transitions of H$_2$CO compared to CN N=2-1. These transitions
have similar Einstein coefficients, but the collisional rates of H$_2$CO with para-H$_2$ \citep{Troscompt+etal_2009} are
typically 5 to 10 times larger than those of  CN with para-H$_2$ at 10 K \citep{Kalugina+etal_2012}.
On average, CN is brighter than H$_2$CO (see Fig.\ref{fig:cn-h2co}), and the high
detection rate of H$_2$CO is obtained because two lines of similar
intensities could be observed simultaneously and added in this study.
Strong H$_2$CO lines clearly correlates with SO detectability (see Fig.\ref{fig:h2co-so}), i.e. in general outflow-driving
sources.

\subsection{C$^{17}$O and other lines}
\label{sec:sub:other}

While $^{13}$CO emission from disk is overwhelmed by confusion, C$^{17}$O is in general too faint
to be a suitable tracer of disks (only 9 detections out of 40 observed sources, see Fig.\ref{fig:c17o-cn}). It is convincingly detected in the disks of \object{AB Aur},
\object{MWC 480} and 04302+2227 only. On another hand, residual confusion from the cloud is still visible in
many sources: \object{CW Tau} and \object{CIDA-1}, \object{CIDA-8}, \object{Haro 6-13}, \object{HK Tau},
\object{Haro 6-33}, \object{Haro 6-5 B}, \object{HH\,30}, and \object{LkHa 358}.  The strong emission from \object{HL Tau} and
\object{CB26} presumably comes from their outflows, as the lineshapes are complex and resemble
those of SO. For other ``outflow'' sources, the linewidths are small enough to leave some doubt
about the origin.

We also have a $5 \sigma$ detection of  $^{13}$C$^{18}$O towards \object{Haro 6-10}, with a relatively narrow
linewidth of $0.7 \pm 0.15 \kms$. A line of similar width is also detected towards \object{AB Aur}, at the $4 \sigma$
level. In the later source, as the disk emission spreads over $2-3 \kms$, this must come from the molecular
cloud, and implies that this cloud also contributes to the signal of the other isotopologues.

Finally, emission at the $3 \sigma$ level appears towards \object{Haro 6-10}, \object{MWC 480} and \object{GG Tau} near the expected
frequency of the HC$_3$N J=23-22 transition, but these features are either too broad, too narrow, or not at the right velocity.
No other significant feature appears anywhere else in the observed bandwidth.

\section{Discussion}
\label{sec:discussion}

\subsection{Using CN as a disk tracer}

Our main approach to test whether CN is a disk tracer is to assume that the
(mean) CN surface density is independent of the disk properties. This assumption is sustained
by several independent arguments. First, the imaging study of  DM Tau, LkCa 15 and MWC 480
by \citet{Chapillon+etal_2012} give similar CN surface densities in the 3 objects. Second,
chemical modeling performed to interpret these CN images predict that the CN surface density
is rather insensitive to the disk mass, varying by only $\pm 30 \%$ for masses differing
a factor 3 in both directions. A similar result was found by \citet{Jonkheid+etal_2007}
in their models for disk around HAe stars.  Last, from an observational point of view,
our analysis shows that the CN line optical depth does not appear to exhibit wide
variations. We thus assume the CN line opacity to be constant (and relatively small), apart from
the 3 sources in which it is clearly determined. Under this assumption, the total flux
is directly related to the total number of CN molecules, i.e. principally by the disk
outer radius ($S_\nu \propto R_\mathrm{out}^2$).

For a quantitative derivation of $R_\mathrm{out}$, we use the approach of
\citet{Guilloteau+Dutrey_1998} which shows that,
at moderate inclinations ($20^\circ < i < 70^\circ$), the line flux from a
Keplerian disk is related to local line width
and disk outer radius:
\begin{equation}
R_\mathrm{out}  =  D \left( \frac{\int S_{\nu} d\mathrm{v} }{B_{\nu}(T_0) (\rho \Delta V) \pi {\rm cos}(i)} \right) ^{1/2}
\label{eq:main-rout}
\end{equation}
where $D$ is the distance (140 pc), $T_0$ the excitation temperature, $\Delta V$ the local line width,
$i$ the inclination, and $\rho$ a factor depending on line opacity along the line of sight. The rationale
for and derivation of Eq.\ref{eq:main-rout} are given in Appendix \ref{app:disk}.
In the optically thin regime, $\rho$ is proportional to the opacity, which is itself inversely proportional
to $\cos(i)$, so the dependency on inclination disappears, and the flux just scales with the total number of molecules.
In this optically thin case, Eq.\ref{eq:main-rout} can be extrapolated to low inclinations $i < 20^\circ$.
For nearly edge-on objects, where the above formula breaks, \citet{Beckwith+Sargent_1993} have shown that
the expected flux is not a strong function of inclination. Hence the apparent dependence on $i$ essentially vanishes.
We thus use $\rho \cos(i) = 0.3 \cos(45^\circ)$ for all sources in which the opacity cannot be determined from our spectra,
and, for the few sources in which $\tau$ is determined by the HFS fit, we use $\rho \cos(i) = \tau \cos(45^\circ)$.
However, larger disks with smaller
average line opacity can equally fit the data: in this respect, the derived radii are thus likely
to be \textit{lower limits} to the true CN disk size.

For the other parameters, we take $\Delta V = 0.2 \kms$, a value consistent with the local line width
determined in the disks of \object{DM Tau}, \object{LkCa 15} and \object{MWC 480} by \citet{Pietu+etal_2007} and
\citet{Chapillon+etal_2012}. For the temperature, \citet{Chapillon+etal_2012} derived
30 K for \object{MWC 480}, but values (10 - 15 K) for \object{DM Tau} and \object{LkCa 15}. We used 15 or 30 K according
to our best knowledge of source properties. Note that the derived disk radius will scale as  $1/\sqrt{T_0}$ (Eq.\ref{eq:main-rout}).
The assumed temperatures and derived disk radii are given in Table \ref{tab:radii} for a uniform disk distance of 140 pc.
Given the uncertainty on the CN opacity, these radii should only be interpreted as estimates of the disk size.
They range from $< 200$ AU for non detection, to $> 1000$ AU for the brightest sources.
For \object{MWC 480} and \object{DM Tau}, the derived radius is in good agreement with the more accurate
determination from interferometric data of \citet{Chapillon+etal_2012}, justifying a posteriori our
choice of parameter values.

For \object{T Tau} and \object{Haro 6-10}, the CN line opacity is reasonably well constrained.
The large disk radii which are derived make the disk interpretation very unlikely,
as both objects contain multiple systems with strong evidence for tidal truncation
of the embedded circumstellar disks to outer radii $< 60$ AU \citep{Akeson+etal_1998,Guilloteau+etal_2011}.
Both systems also display clear evidence for high velocity wings in CO, indicating powerful molecular
jets. Hence, for these objects, CN may come from the outflows rather than from a circum-stellar/binary disk.
As the line width are however limited, it is also possible that a substantial fraction of
CN comes from the extended envelope in which these objects are embedded.

For \object{DG Tau}, although modeling performed by \citet{Salter+etal_2011} shows the CN line intensity
can be produced by a large Keplerian disk, we suspect that a similar situation occurs because of
the strong outflow and complex, clearly non Keplerian, circumstellar environment (see Appendix
\ref{app:sources} for more details on this source).

\begin{figure}[!h]
\includegraphics[width=\columnwidth]{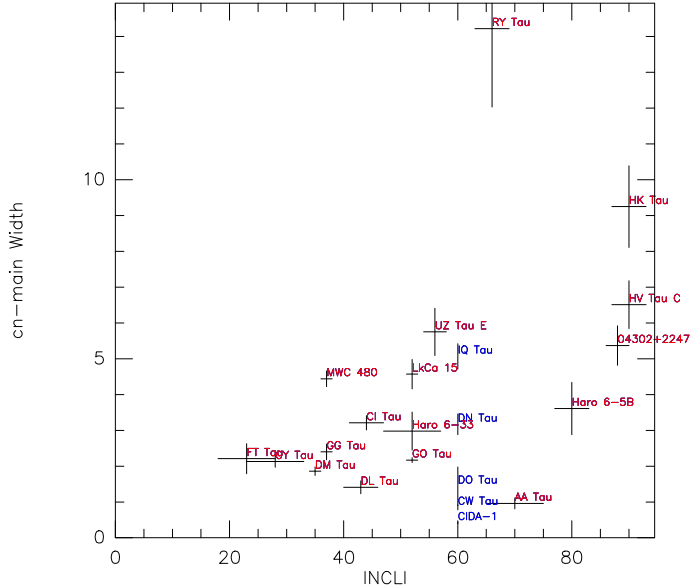} % {dv-CN-MAIN-INCLI-a.eps}
\caption{CN line width (km.s$^{-1}$) as a function of disk inclination.
Sources with unknown inclinations are in blue: we used $i=60^\circ$ as the
most likely value. Sources with outflows were excluded.}
\label{fig:width}
\end{figure}

For other sources, a disk origin for the CN emission is inferred from the following qualitative arguments:
\begin{itemize}
\item The line-width $\Delta v$ derived from the Gaussian fit is of the order of $2-4 \kms$, as expected for
a Keplerian disk orbiting a star of 0.5-2 $\msun$.
\item Larger line widths are found for the more inclined objects (see Fig.\ref{fig:width}). This figure
does not correct for stellar mass: doing so would improve the correlation between line width and inclination.
\item The CN outer radius is within 300-600 AU, in good agreement with the range of interferometrically
derived radii in previously imaged Keplerian disks.
\item In objects which have already been imaged in CN, \object{MWC 480}, \object{DM Tau}, \object{LkCa 15} \citep{Chapillon+etal_2012},
\object{GM Aur} and \object{AA Tau}
\citep{Oberg+etal_2010}, CN unambiguously comes from the disk,
even in the confused case of \object{AA Tau}.
\end{itemize}

Note that none of the main CN line profiles appears double peaked (e.g. compare
CN and $^{13}$CO in MWC\,480 Fig.\ref{fig:mwc_480}). This is a result
of the overlap between the three strongest J=5/2-3/2 hyperfine components which are separated
by 0.77 and 1.50 km.s$^{-1}$ respectively. These separations are smaller than the overall
  width due to Keplerian shear, significantly blurring the expected double peaked profiles emerging
  from such disks in (spatially) unresolved observations. However, the velocity separations are larger than the
  local line width (0.3 - 0.5 km.s$^{-1}$), so that in resolved (interferometric) images, the patterns from each of
  the three components are spatially separated. Furthermore, the strongest J=3/2-1/2 component, which
  is separated by 17 km.s$^{-1}$ from the nearest one, is only 1.6 times weaker than the
  J=5/2-3/2 F=5/2-3/2 component. In practice, the hyperfine structure of CN is only a minor complication in
  the derivation of the kinematics from interferometric measurements, especially at the high
  densities encountered in disks, which ensures that all hyperfine components will have the
  same excitation temperature.

To the 24 sources detected in CN in our sample, we should add 5 more disks: \object{GM Aur}
\citep{Oberg+etal_2010}, \object{LkCa 15} \citep{Chapillon+etal_2012}, \object{CI Tau}, \object{CY Tau}, and \object{GO Tau}
\citep[][in prep]{Guilloteau+etal_2012}, leading to a total of 29 detections
out of 47 objects. 3 (\object{Haro 6-10}, \object{T Tau} and \object{DG Tau}) are very unlikely to be disks, and, on the basis
of the existence of outflows and SO detection, 2 more are perhaps outflows (\object{HL Tau} and \object{CB26}). This
still leads to a disk detection rate of at least 24 / 47, i.e. $\gtrsim 50$\% at our sensitivity level of $\sim 0.1$ Jy.km.s$^{-1}$.
Furthermore, in most disks, the CN N=2-1 intensity is within a
factor 2 of that of $^{13}$CO J=2-1 (see Fig.\ref{fig:disk-cn}).

This demonstrates the ability of the CN 2-1 line to be a good tracer
of disk kinematics for stars still embedded in molecular clouds whose CO and $^{13}$CO line emission
could confuse the disk emission profile.

\begin{table*}
\caption{CN outer radii and line detections}
\label{tab:radii}
\begin{tabular}{l rll lll}
\hline
Source & Radius & $T_0$ & Line width & \multicolumn{3}{c}{Detection ?} \\ % Flux
  & (AU) &  (K) & $\kms$ & $^{13}$CO & H$_2$CO & SO  \\ % Jy\,$\kms$
\hline
      FN Tau &
 $<       170 $ & 15 &
 &
 & Cloud % Yes
 &
\\ % FN_TAU
      CW Tau &
 $<        170 $ & 15 &
 & Disk
 &
 &
\\ % CW_TAU
      CIDA-1 &
 $<        170 $ & 15 &
 &
 &
 &
\\ % CIDA-1
      BP Tau &
 $<        250 $ & 30 &
 &
 &
 &
\\ % BP_TAU
      DE Tau &
 $<        250 $ & 15 &
 &
 &
 &
\\ % DE_TAU
      RY Tau &
   310 &      30  &       0.82 $\pm$       0.27
 & Disk
 &
 &
\\ % & 0.62 % RY_TAU
       T Tau &
  1230 &    30 &       2.96 $\pm$       0.02
 &
 & Yes
 & Yes
\\ % & 9.55 % T_TAU
   Haro 6-5B &
   310 &       15  &       3.12 $\pm$       1.17
 &
 & Yes
 &
\\ % & 0.31 % HARO6-5B
      FT Tau &
   310 &      15  &       1.69 $\pm$       0.34
 &
 &
 &
\\ % & 0.32 % FT_TAU
    DG Tau-B &
   450 &      15 &       2.12 $\pm$       0.30
 &
 & Yes
 & Yes
\\ %  & 0.66 % DG_TAU-B
      DG Tau &
   700 &     30 &       2.11 $\pm$       0.09
 &
 & Yes
 & Yes
\\ %  & 3.15 % DG_TAU
   Haro 6-10 &
   660 &  30  &       1.50 $\pm$       0.04
 &
 & Yes
 & Yes
\\ % &      7.15 % HARO6-10N
      IQ Tau &
   560 &  15  &       4.92 $\pm$       0.53
 & Disk
 &
 &
\\ % &  1.02 % IQ_TAU
    LkHa 358 &
 $<        240 $ & 15  &
 & Disk
 & Yes
 &
\\ % LKHA_358
       HH 30 &
 $<        270 $ & 15 &
 & (Disk)
 & Yes
 &
\\ % HH30
      HL Tau &
   250 &  30    &       1.79 $\pm$       0.44
 &
 & Yes
 & Yes
\\ % &  0.39 % HL_TAU
      HK Tau &
   320 &  30    &       8.56 $\pm$       1.96
 &
 &
 &
\\ % &  0.64 % HK_TAU
   Haro 6-13 &
 $\lessapprox       280 $ & 15 &
 &
 & Yes
 &
\\ % HARO6-13
      GG Tau &
   490 &       30  &       2.28 $\pm$       0.32
 & Disk
 & Yes
 &
\\ % & 1.53 % GG_TAU
    UZ Tau E &
   310 & 30 &        6.58 $\pm$       1.23
 & Disk
 &
 &
\\ % &   0.63 % UZTAU_E
  04302+2247 &
   190 & 30  &       2.99 $\pm$       0.47
 & Disk
 & Yes
 & Yes
\\ % & 1.99 % 04302+2247
      DL Tau &
   560 & 15  &       1.96 $\pm$       0.16
 &
 & Yes
 &
\\ % &       1.01 % DL_TAU
      DM Tau &
   610 & 15  &       1.44 $\pm$       0.11
 & Disk
 & Yes
\\ % & 4.38 % DM_TAU
      AA Tau &
   300 &  15  &       0.95 $\pm$       0.26
 & Disk
 &
 &
\\ % &  0.28 % AA_TAU
      DN Tau &
   490 & 15  &       3.21 $\pm$       0.42
 &
 & Yes
 &
\\ %  0.77 & DN_TAU
      DO Tau &
   310 & 15  &       0.50 $\pm$       0.17
 & Disk
 & Yes
 &
\\ % &    0.31 % DO_TAU
    HV Tau C &
   310 & 30  &     5.97 $\pm$       0.87
 & Disk
 &
 &
\\ %       0.61 % HV_TAU-C
   Haro 6-33 &
   260 & 30  &       3.05 $\pm$       0.68
 &
 & Yes
 &
\\ %  & 0.44 % HARO6-33
      DQ Tau &
 $<        210 $ & 15 &
 &
 &
 &
\\ % DQ_TAU
      DR Tau &
 $<        420 $ & 15 &
 &
 &
 &
\\ % DR_TAU
      DS Tau &
 $<        310 $ & 30 &
 &
 & Yes
 &
\\ % DS_TAU
      UY Aur &
 $<        300 $ & 30 &
 & Disk
 & Yes
 &
\\ % UY_AUR
      AB Aur &
 $<        250 $ & 30  &
 & Disk
 & Yes
 &
\\ % AB_AUR
      SU Aur &
 $<        140 $ & 30 &
 & Disk
 &
 &
\\ % SU_AUR
     MWC 480 &
   520 &  30     &       4.54 $\pm$       0.34
 & Disk
 &
 &
\\ % &  1.72 % MWC_480
       CB 26 &
   350 &    15  &       2.92 $\pm$       1.00
 &
 & Yes
 & Yes
\\ %  & 0.41 % CB_26
      CIDA-8 &
 $<        290 $ & 15 &
 &
 & Cloud
 &
\\ % CIDA-8
     CIDA-11 &
 $<        540 $ & 15 &
 &
 &
 &
\\ % CIDA-11
      RW Aur &
 $<        280 $ & 30 &
 &
 &
 &
\\ % RW_AUR
     MWC 758 &
 $<        450 $ & 30 &
 & Disk
 &
 &
\\ % MWC_758
      CQ Tau &
 $<        190 $ & 30 &
 & Disk
 & Yes
 &
\\ % CQ_TAU
   HD 163296($^a$) &
   760 &    30  &       4.16 $\pm$       0.11
 & Disk
 & Yes
 &
\\ %  &  5.05 % HD163296
\end{tabular}
\tablefoot{The assumed distance is D=140 pc, except for HD\,163296 ($^a$, D=120 pc).
For $^{13}$CO, sources in which the emission partially originates from a disk are
indicated. For H$_2$CO, no distinction is made upon the origin of the emission, apart
from the two likely cases of cloud confusion FN Tau and CIDA-8.}
\end{table*}
%end input

\subsection{Molecular disk properties}

Our study is the first large, sensitive survey for $^{13}$CO, CN, ortho-H$_2$CO, and SO towards disk
around Class II PMS stars.  Detailed sensitive studies of several molecules concerned only a few objects:
in Taurus only \object{GG Tau}, \object{DM Tau}, \object{LkCa 15} and \object{MWC 480} \citep{Dutrey+etal_1997,Guilloteau+etal_1999,Thi+etal_2004,Pietu+etal_2007,Chapillon+etal_2012,Dutrey+etal_2011}.
Previous multi-molecule surveys were limited to a dozen objects, and in general biased towards
the brightest disks, either in CO \citep[e.g. the DISCS survey of][]{Oberg+etal_2011}, or in continuum \citep[][used a minimum
1.3\,mm continuum flux of 75 mJy]{Salter+etal_2011}. We reach here a sensitivity level ($ \sim 0.1$ Jy.km.s$^{-1}$) comparable
to or better than that of \citet{Oberg+etal_2011},
and 5 to 10 times better than that of \citet{Salter+etal_2011}. Furthermore, the smaller beam size of the 30-m allows to better
isolate confusion cases compared to the JCMT study of \citet{Salter+etal_2011}. The improved sensitivity allows
us to detect many more sources, such as \object{FT Tau}, \object{IQ Tau}, V806 Tau (\object{Haro 6-13}), \object{UZ Tau},
\object{DN Tau}, \object{DO Tau} in the Salter et al.~sample.
Excluding sources detected in SO which are presumably contaminated by outflows, we have a total of 32 sources\footnote{This
includes the 16 sources detected in CN that have no SO emission, plus the Butterfly star, plus 8 sources in $^{13}$CO (including HH\,30)
and 2 sources in H$_2$CO (Haro 6-13 and DS Tau), to which we add the 5 sources detected in CN by other studies}  (out of 47) which exhibit
CN, $^{13}$CO or H$_2$CO emission from a disk, i.e. 68 \% of the sample.

\begin{figure}
\includegraphics[width=\columnwidth]{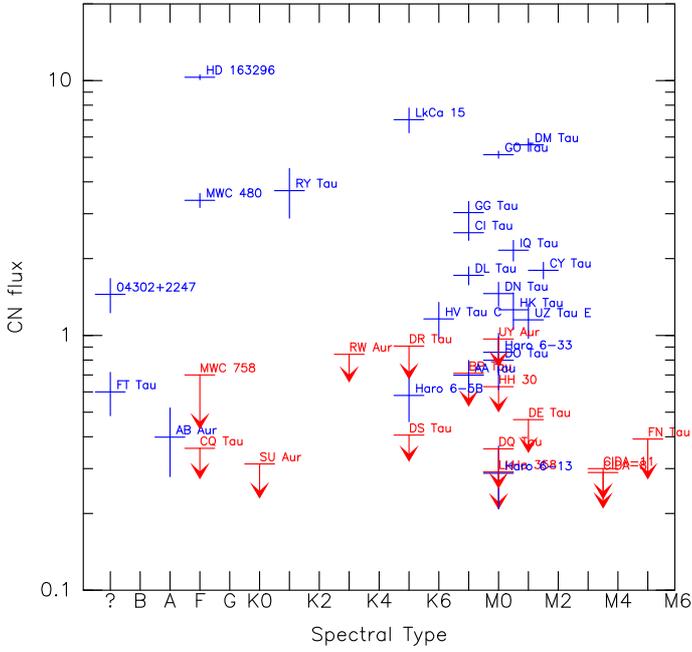}
\caption{Correlation between CN line flux (main fine structure group, in Jy.km.s$^{-1}$)
and stellar spectral type.
Blue is for detected sources, red for upper limits. Sources with outflows were excluded.
Emission in AB Aur is from the envelope.}
\label{fig:correl1}
\end{figure}
\begin{figure}
\includegraphics[width=\columnwidth]{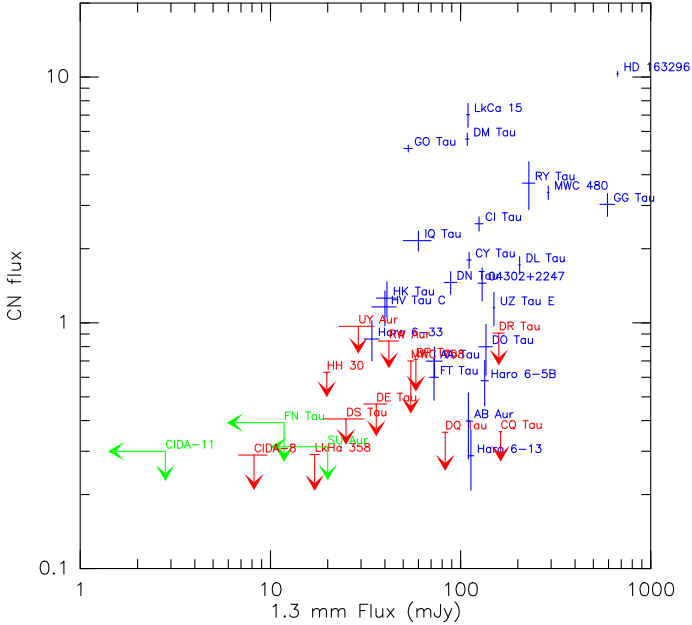}
\caption{As Fig.\ref{fig:correl1}, but for CN line flux (main fine structure group) vs 1.3 mm continuum flux (in mJy).}
\label{fig:correl2}
\end{figure}
\begin{figure}
\includegraphics[width=\columnwidth]{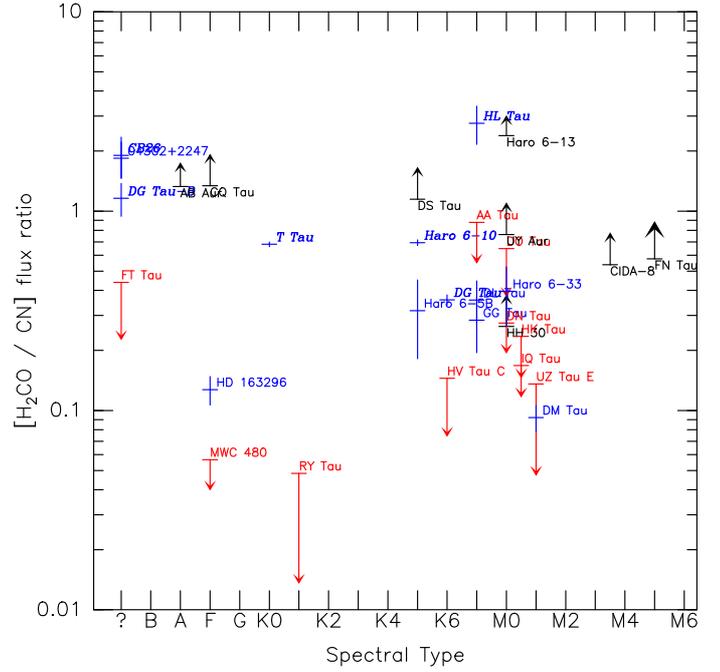}
\caption{Correlation between the H$_2$CO to CN flux ratio and stellar spectral type.
Red indicates upper limits (no H$_2$CO), black lower limits (no CN). H$_2$CO in FN Tau and CIDA 8 is likely to be cloud contamination. Outflow sources are in italics. }
\label{fig:correl3}
\end{figure}

Figures \ref{fig:correl1} and \ref{fig:correl2} explore possible correlation of
the CN intensity with spectral type of the star and 1.3\,mm continuum flux of the disk
respectively. Figure \ref{fig:correl3} considers the H$_2$CO/CN intensity ratio vs the stellar
spectral type. Despite the large sample size, no correlations are apparent. This lack
of correlation is also observed for other stellar parameters (mass loss, X-ray luminosity).
However, a number of properties emerge from these correlation plots.

First, CN line is relatively weak (in fact undetected so far) in the hottest stars in our sample
(\object{SU Aur}, \object{CQ Tau}, \object{MWC 758}, \object{AB Aur}), although \object{MWC 480} and \object{HD 163296}
exhibit strong CN emission (Fig.\ref{fig:correl1}). The rather poor molecular content of \object{AB Aur} has been shown to be
a consequence of its strong UV flux by \citet{Schreyer+etal_2008}.  The UV flux penetration
may furthermore be enhanced by the large inner cavity (radius $\sim 100$ AU) of \object{AB Aur} \citep{Pietu+etal_2005}. A similar situation
may occur for \object{MWC 758} due to its $\sim 40$ AU cavity \citep{Isella+etal_2010a}. \citet{Chapillon+etal_2008,Chapillon+etal_2010}
argue that \object{CQ Tau} is gas deficient.
Little is known about the gas disk of \object{SU Aur}, but the strong X-ray luminosity of \object{SU Aur} may play
a role in lowering the molecular abundances. \object{MWC 480} and \object{HD 163296} have strong CN. Contrary
to \object{AB Aur}, \object{CQ Tau} and \object{MWC 758} which are Type I sources, these two sources are Type II according to
the classification of \citet{Meeus+etal_2001}, which can indicate self-shadowed disks. Such a
configuration would seriously limit the UV penetration, leading to a chemical composition much
more similar to those of T Tauri disks.

Second, at the level of our sensitivity, we do not detect the CN line for spectral types cooler than M1.5 (Fig.\ref{fig:correl1}).
Assuming similar CN column densities than in the warmer stars, our sensitivity translates into
upper limits of order 150 - 250 AU for the disk outer radius. Thus, our negative result may only
be the consequence of smaller disk sizes for lower mass stars, rather than implying a change
in chemistry at this spectral type threshold.

Third, although CN intensity does not correlate with the continuum flux at 1.3\,mm (Fig.\ref{fig:correl2}), which is a tracer
of the dust mass, there is a clear threshold. Only 4 sources with $S_\nu$(1.3\,mm) $ > 60 $mJy are undetected
in CN (\object{CW Tau}, \object{CQ Tau}, \object{DQ Tau}, and \object{DR Tau} which has been observed with relatively limited sensitivity)
while only 4 sources are detected among the 17 sources with $S_\nu$(1.3\,mm) $ \le 60 $mJy.

Comparing the emissions from CN and  H$_2$CO, we find that to first order, the H$_2$CO / CN ratio divides
the sources in two categories (see Fig.\ref{fig:correl3}).
[H$_2$CO/CN]$ > 0.5$ is found in warm sources (Herbig Ae stars) or outflow/envelope dominated objects,
while it is below 0.3 for the disk sources.
This trend is consistent with our current understanding of chemistry.
In disks, H$_2$CO is formed in the gas phase and on grains \citep{Semenov+Wiebe_2011}.
Therefore, H$_2$CO can be released in the gas phase
at high enough temperatures, either because of large stellar flux (HAe stars) or because of shock heating (outflow sources).
Enhanced abundance of H$_2$CO has been observed in many outflows
\citep[e.g.][]{Bachiller+etal_2001,Tafalla+etal_2010}.
As a consequence, sources with weak or no H$_2$CO emission mostly correspond to cold objects without outflows.
While CN appears as a clear disk tracer in the M1-K5 spectral type range, the situation is more complex
for the HAe stars, for which H$_2$CO is sometimes a better probe than CN, but remains weak.
Using the \citet{Baraffe+etal_1998} tracks, the M1 limit corresponds
to about $0.5 \msun$ for stars in the 1 to 3 Myr age.

Sulfur-bearing molecules  (CS, SO, H$_2$S and CCS) have been recently searched by \citet{Dutrey+etal_2011}
and \citet{Chapillon+etal_2012} in the four disks surrounding \object{MWC 480}, \object{LkCa 15}, \object{GO Tau} and \object{DM Tau}. So far,
only CS has been detected, neither SO nor H$_2$S. Our survey provides a much larger sample for SO, and
confirms its general weakness in disks (only one source detected, 04302+2242, our results contradicting
the detection in \object{AB Aur} reported by \citet{Fuente+etal_2010}).  On the other hand, we find SO to be ubiquitous
in outflow sources. The SO chemistry in disks is not fully understood: current models predict too much SO and H$_2$S molecules \citet{Dutrey+etal_2011}, most likely as a result of incomplete treatment of Sulfur chemistry on grain surfaces. However, SO is a clear tracer of shocks, with high abundances in extremely high velocity outflows \citep{Tafalla+etal_2010}.

\begin{figure}
\begin{center}
\includegraphics[width=\columnwidth]{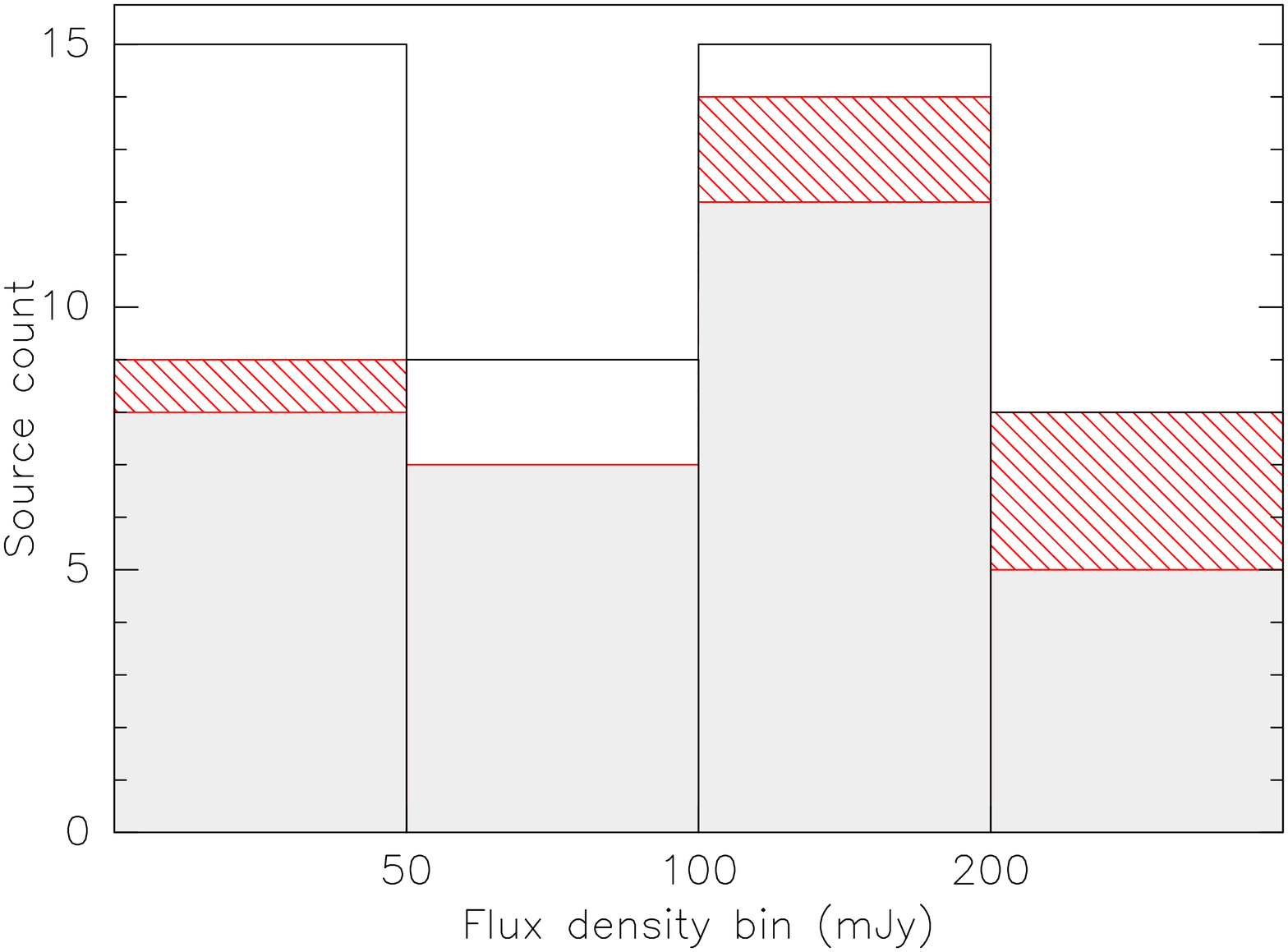}
\end{center}
\caption{Histogram of detections as a function of 1.3\,mm continuum flux range. Filled: disks, hatched: outflow contaminated sources,
empty: non detections.}
\label{fig:histo}
\end{figure}

The absence of correlations in Fig.\ref{fig:correl1}-\ref{fig:correl2}
justifies \textit{a posteriori} our simple source-independent hypothesis (essentially constant CN surface density)
to interpret the line emission. In this framework, the CN flux is a simple tracer of disk size, and our
study shows that \textit{at least 50\%} of the young stars are surrounded by \textit{large} ($\gtrsim 300$ AU) disks, in our
specific case of isolated star formation. The overall \textit{disk} detection rate is displayed in Fig.\ref{fig:histo}.
Hatched areas are for sources with molecular outflows (traced by SO), filled areas are disks detected in CN, $^{13}$CO or H$_2$CO,
and empty histograms non detections. The distribution of disk sizes exhibits a strong peak near 300 AU; however, because
of sensitivity limitations, our survey becomes incomplete for smaller disk sizes.

It is beyond the scope of our angularly unresolved data to provide detailed modeling
of the disk emission for individual objects. The above considerations
on the detected lines are broadly consistent with prior knowledge of disk chemistry. The large CN detection
rate supports the simple idea that CN is present only in a photodissociation layer above the disk plane,
with a relatively uniform surface density over a large range of source properties \citep{Jonkheid+etal_2007,Chapillon+etal_2012}.
The lower detection rate for sources with low 1.3 mm flux may be the result of a change of chemical
regime for disks with low mass. However, it is also consistent with a simpler idea, namely that these
faint disks are just much smaller than the others. Higher sensitivity, spatially resolved observations can solve this issue.

\section{Summary}
\label{sec:summary}

We have performed a sensitive survey of 42 young stars probing isolated star formation regions
in CN, ortho-H$_2$CO, SO, $^{13}$CO and C$^{17}$O rotational lines near 206 - 228 GHz with
the IRAM 30-m telescope. It reveals that :
\begin{itemize}
\item $^{13}$CO is strongly affected by confusion, either from molecular clouds, or from outflows
and envelopes.
\item CN is a good, nearly confusion free, tracer of disks for stars in the M1-K5 spectral type range (i.e.
stars more massive than about $0.5 \msun$ according to the \citet{Baraffe+etal_1998} tracks).
\item The warmest stars exhibit two very different behaviours: some have strong CN, but most have undetectable CN and weak H$_2$CO emission.
\item SO is ubiquitously found in outflow-driving, embedded sources, but exceptional in disks
(only 1 source). When SO is present, H$_2$CO is stronger than CN.
\item CN N=2-1 is strong enough to be a good tracer of disk kinematics in at least half of the objects.
\item The disk detection rate in CN lines exceeds 50 \%. The detected disks are large, $\gtrsim 300$ AU in radius.
\item The overall gas disk detection rate (by any molecular tracer) is $\approx 68$ \%. However, gas disk detection
is much more difficult for sources with small ($\leq 60$ mJy) 1.3\,mm continuum flux.
We cannot distinguish at present whether this is due to smaller disk masses
or smaller disk sizes ($< 200$ AU radius).
\end{itemize}
Our results only apply to regions of isolated star formation. In more crowded regions
(e.g. $\rho$ Oph), confusion will unavoidably be larger. The disk size distribution may also be
different. It is however likely that CN will remain a better tracer of disk kinematics than
the much more confused $^{12}$CO or $^{13}$CO, or the significantly weaker H$_2$CO emission.
%
%While the current result is using the IRAM 30-m at its best capabilities,
% the situation will change with the advent of ALMA.  ALMA has 8 times more collecting
Looking to the advent of ALMA, with 8 times greater collecting area
than the IRAM 30-m, better antenna efficiency, and expected system temperatures
around 100 K at this frequency, will provide a sensitivity gain of order 16. Thus ALMA
should be able to obtain, in the same integration time, a 4 times better limit on the
outer radius. This may open for the first time the possibility to probe beyond the M1 limit, unless
the chemistry changes to reduce drastically the CN abundances around the lower mass stars.

\begin{acknowledgements}
This work was supported by ``Programme National de Physique Stellaire'' (PNPS) and ``Programme
National de Physique Chimie du Milieu Interstellaire'' (PCMI) from INSU/CNRS. The work of MS
was supported in part by NSF grant AST 09-07745.
This research has made use of the SIMBAD database,
operated at CDS, Strasbourg, France
\end{acknowledgements}

\clearpage

\bibliography{mays-cn}

\begin{thebibliography}{89}
\expandafter\ifx\csname natexlab\endcsname\relax\def\natexlab#1{#1}\fi

\bibitem[{{Akeson} {et~al.}(2002){Akeson}, {Ciardi}, {van Belle}, \&
  {Creech-Eakman}}]{Akeson+etal_2002}
{Akeson}, R.~L., {Ciardi}, D.~R., {van Belle}, G.~T., \& {Creech-Eakman}, M.~J.
  2002, \apj, 566, 1124

\bibitem[{{Akeson} {et~al.}(1998){Akeson}, {Koerner}, \&
  {Jensen}}]{Akeson+etal_1998}
{Akeson}, R.~L., {Koerner}, D.~W., \& {Jensen}, E.~L.~N. 1998, \apj, 505, 358

\bibitem[{{Andrews} \& {Williams}(2005)}]{Andrews+etal_2005}
{Andrews}, S.~M. \& {Williams}, J.~P. 2005, \apj, 631, 1134

\bibitem[{{Bachiller} {et~al.}(2001){Bachiller}, {P{\'e}rez Guti{\'e}rrez},
  {Kumar}, \& {Tafalla}}]{Bachiller+etal_2001}
{Bachiller}, R., {P{\'e}rez Guti{\'e}rrez}, M., {Kumar}, M.~S.~N., \&
  {Tafalla}, M. 2001, \aap, 372, 899

\bibitem[{{Banzatti} {et~al.}(2011){Banzatti}, {Testi}, {Isella}, {Natta},
  {Neri}, \& {Wilner}}]{Banzatti+etal_2011}
{Banzatti}, A., {Testi}, L., {Isella}, A., {et~al.} 2011, \aap, 525, A12

\bibitem[{{Baraffe} {et~al.}(1998){Baraffe}, {Chabrier}, {Allard}, \&
  {Hauschildt}}]{Baraffe+etal_1998}
{Baraffe}, I., {Chabrier}, G., {Allard}, F., \& {Hauschildt}, P.~H. 1998, \aap,
  337, 403

\bibitem[{{Beckwith} \& {Sargent}(1993)}]{Beckwith+Sargent_1993}
{Beckwith}, S.~V.~W. \& {Sargent}, A.~I. 1993, \apj, 402, 280

\bibitem[{{Bertout} {et~al.}(1999){Bertout}, {Robichon}, \&
  {Arenou}}]{Bertout+etal_1999}
{Bertout}, C., {Robichon}, N., \& {Arenou}, F. 1999, \aap, 352, 574

\bibitem[{{Bertout} {et~al.}(2007){Bertout}, {Siess}, \&
  {Cabrit}}]{Bertout+etal_2007}
{Bertout}, C., {Siess}, L., \& {Cabrit}, S. 2007, \aap, 473, L21

\bibitem[{{Bouvier} {et~al.}(1999){Bouvier}, {Chelli}, {Allain}, {Carrasco},
  {Costero}, {Cruz-Gonzalez}, {Dougados}, {Fern{\'a}ndez}, {Mart{\'{\i}}n},
  {M{\'e}nard}, {Mennessier}, {Mujica}, {Recillas}, {Salas}, {Schmidt}, \&
  {Wichmann}}]{Bouvier+etal_1999}
{Bouvier}, J., {Chelli}, A., {Allain}, S., {et~al.} 1999, \aap, 349, 619

\bibitem[{{Brice{\~n}o} {et~al.}(2002){Brice{\~n}o}, {Luhman}, {Hartmann},
  {Stauffer}, \& {Kirkpatrick}}]{Briceno+etal_2002}
{Brice{\~n}o}, C., {Luhman}, K.~L., {Hartmann}, L., {Stauffer}, J.~R., \&
  {Kirkpatrick}, J.~D. 2002, \apj, 580, 317

\bibitem[{{Briceno} {et~al.}(1993){Briceno}, {Calvet}, {Gomez}, {Hartmann},
  {Kenyon}, \& {Whitney}}]{Briceno+etal_1993}
{Briceno}, C., {Calvet}, N., {Gomez}, M., {et~al.} 1993, \pasp, 105, 686

\bibitem[{{Cabrit} {et~al.}(1996){Cabrit}, {Guilloteau}, {Andre}, {Bertout},
  {Montmerle}, \& {Schuster}}]{Cabrit+etal_1996}
{Cabrit}, S., {Guilloteau}, S., {Andre}, P., {et~al.} 1996, \aap, 305, 527

\bibitem[{{Cabrit} {et~al.}(2006){Cabrit}, {Pety}, {Pesenti}, \&
  {Dougados}}]{Cabrit+etal_2006}
{Cabrit}, S., {Pety}, J., {Pesenti}, N., \& {Dougados}, C. 2006, \aap, 452, 897

\bibitem[{{Chakraborty} \& {Ge}(2004)}]{Chakraborty+Ge_2007}
{Chakraborty}, A. \& {Ge}, J. 2004, \aj, 127, 2898

\bibitem[{{Chapillon} {et~al.}(2008){Chapillon}, {Guilloteau}, {Dutrey}, \&
  {Pi{\'e}tu}}]{Chapillon+etal_2008}
{Chapillon}, E., {Guilloteau}, S., {Dutrey}, A., \& {Pi{\'e}tu}, V. 2008, \aap,
  488, 565

\bibitem[{{Chapillon} {et~al.}(2012){Chapillon}, {Guilloteau}, {Dutrey},
  {Pi{\'e}tu}, \& {Gu{\'e}lin}}]{Chapillon+etal_2012}
{Chapillon}, E., {Guilloteau}, S., {Dutrey}, A., {Pi{\'e}tu}, V., \&
  {Gu{\'e}lin}, M. 2012, \aap, 537, A60

\bibitem[{{Chapillon} {et~al.}(2010){Chapillon}, {Parise}, {Guilloteau},
  {Dutrey}, \& {Wakelam}}]{Chapillon+etal_2010}
{Chapillon}, E., {Parise}, B., {Guilloteau}, S., {Dutrey}, A., \& {Wakelam}, V.
  2010, \aap, 520, A61

\bibitem[{{Crutcher} {et~al.}(1984){Crutcher}, {Churchwell}, \&
  {Ziurys}}]{Crutcher+etal_1984}
{Crutcher}, R.~M., {Churchwell}, E., \& {Ziurys}, L.~M. 1984, \apj, 283, 668

\bibitem[{{Duch{\^e}ne} {et~al.}(2010){Duch{\^e}ne}, {McCabe}, {Pinte},
  {Stapelfeldt}, {M{\'e}nard}, {Duvert}, {Ghez}, {Maness}, {Bouy}, {Barrado y
  Navascu{\'e}s}, {Morales-Calder{\'o}n}, {Wolf}, {Padgett}, {Brooke}, \&
  {Noriega-Crespo}}]{Duchene+etal_2010}
{Duch{\^e}ne}, G., {McCabe}, C., {Pinte}, C., {et~al.} 2010, \apj, 712, 112

\bibitem[{{Dutrey} {et~al.}(1996){Dutrey}, {Guilloteau}, {Duvert}, {Prato},
  {Simon}, {Schuster}, \& {Menard}}]{Dutrey+etal_1996}
{Dutrey}, A., {Guilloteau}, S., {Duvert}, G., {et~al.} 1996, \aap, 309, 493

\bibitem[{{Dutrey} {et~al.}(2013){Dutrey}, {Guilloteau}, {Graefe}, \&
  {Wolf}}]{Dutrey+etal_2012}
{Dutrey}, A., {Guilloteau}, S., {Graefe}, C., \& {Wolf}, S. 2013, {in prep.}

\bibitem[{{Dutrey} {et~al.}(1997){Dutrey}, {Guilloteau}, \&
  {Guelin}}]{Dutrey+etal_1997}
{Dutrey}, A., {Guilloteau}, S., \& {Guelin}, M. 1997, \aap, 317, L55

\bibitem[{{Dutrey} {et~al.}(2003){Dutrey}, {Guilloteau}, \&
  {Simon}}]{Dutrey+etal_2003}
{Dutrey}, A., {Guilloteau}, S., \& {Simon}, M. 2003, \aap, 402, 1003

\bibitem[{{Dutrey} {et~al.}(2011){Dutrey}, {Wakelam}, {Boehler}, {Guilloteau},
  {Hersant}, {Semenov}, {Chapillon}, {Henning}, {Pi{\'e}tu}, {Launhardt},
  {Gueth}, \& {Schreyer}}]{Dutrey+etal_2011}
{Dutrey}, A., {Wakelam}, V., {Boehler}, Y., {et~al.} 2011, \aap, 535, A104

\bibitem[{{Duvert} {et~al.}(1998){Duvert}, {Dutrey}, {Guilloteau}, {Menard},
  {Schuster}, {Prato}, \& {Simon}}]{Duvert+etal_1998}
{Duvert}, G., {Dutrey}, A., {Guilloteau}, S., {et~al.} 1998, \aap, 332, 867

\bibitem[{{Duvert} {et~al.}(2000){Duvert}, {Guilloteau}, {M{\'e}nard}, {Simon},
  \& {Dutrey}}]{Duvert+etal_2000}
{Duvert}, G., {Guilloteau}, S., {M{\'e}nard}, F., {Simon}, M., \& {Dutrey}, A.
  2000, \aap, 355, 165

\bibitem[{{Eisl{\"o}ffel} \& {Mundt}(1998)}]{Eisloffel+Mundt_1998}
{Eisl{\"o}ffel}, J. \& {Mundt}, R. 1998, \aj, 115, 1554

\bibitem[{{Franciosini} {et~al.}(2007){Franciosini}, {Scelsi}, {Pallavicini},
  \& {Audard}}]{Franciosini+etal_2007}
{Franciosini}, E., {Scelsi}, L., {Pallavicini}, R., \& {Audard}, M. 2007, \aap,
  471, 951

\bibitem[{{Fuente} {et~al.}(2010){Fuente}, {Cernicharo}, {Ag{\'u}ndez},
  {Bern{\'e}}, {Goicoechea}, {Alonso-Albi}, \& {Marcelino}}]{Fuente+etal_2010}
{Fuente}, A., {Cernicharo}, J., {Ag{\'u}ndez}, M., {et~al.} 2010, \aap, 524,
  A19

\bibitem[{{Greaves}(2005)}]{Greaves_2005}
{Greaves}, J.~S. 2005, \mnras, 364, L47

\bibitem[{{Guilloteau} \& {Dutrey}(1998)}]{Guilloteau+Dutrey_1998}
{Guilloteau}, S. \& {Dutrey}, A. 1998, \aap, 339, 467

\bibitem[{{Guilloteau} {et~al.}(2011){Guilloteau}, {Dutrey}, {Pi{\'e}tu}, \&
  {Boehler}}]{Guilloteau+etal_2011}
{Guilloteau}, S., {Dutrey}, A., {Pi{\'e}tu}, V., \& {Boehler}, Y. 2011, \aap,
  529, A105

\bibitem[{{Guilloteau} {et~al.}(1999){Guilloteau}, {Dutrey}, \&
  {Simon}}]{Guilloteau+etal_1999}
{Guilloteau}, S., {Dutrey}, A., \& {Simon}, M. 1999, \aap, 348, 570

\bibitem[{{Guilloteau} {et~al.}(2013){Guilloteau}, {Dutrey}, {Wakelam},
  {Boehler}, {Guilloteau}, {Hersant}, {Semenov}, {Chapillon}, {Henning},
  {Pi{\'e}tu}, \& {Gueth}}]{Guilloteau+etal_2012}
{Guilloteau}, S., {Dutrey}, A., {Wakelam}, V., {et~al.} 2013, {in prep.}

\bibitem[{{Guilloteau} {et~al.}(2006){Guilloteau}, {Pi{\'e}tu}, {Dutrey}, \&
  {Gu{\'e}lin}}]{Guilloteau+etal_2006}
{Guilloteau}, S., {Pi{\'e}tu}, V., {Dutrey}, A., \& {Gu{\'e}lin}, M. 2006,
  \aap, 448, L5

\bibitem[{{Hern{\' a}ndez} {et~al.}(2004){Hern{\' a}ndez}, {Calvet}, {Brice{\~
  n}o}, {Hartmann}, \& {Berlind}}]{Hernandez+etal_2004}
{Hern{\' a}ndez}, J., {Calvet}, N., {Brice{\~ n}o}, C., {Hartmann}, L., \&
  {Berlind}, P. 2004, \aj, 127, 1682

\bibitem[{{Hirth} {et~al.}(1994){Hirth}, {Mundt}, \& {Solf}}]{Hirth+etal_1994}
{Hirth}, G.~A., {Mundt}, R., \& {Solf}, J. 1994, \aap, 285, 929

\bibitem[{{Horne} \& {Marsh}(1986)}]{Horne+Marsh_1986}
{Horne}, K. \& {Marsh}, T.~R. 1986, \mnras, 218, 761

\bibitem[{{Hughes} {et~al.}(2011){Hughes}, {Wilner}, {Andrews}, {Qi}, \&
  {Hogerheijde}}]{Hughes+etal_2011}
{Hughes}, A.~M., {Wilner}, D.~J., {Andrews}, S.~M., {Qi}, C., \& {Hogerheijde},
  M.~R. 2011, \apj, 727, 85

\bibitem[{{Isella} {et~al.}(2010{\natexlab{a}}){Isella}, {Carpenter}, \&
  {Sargent}}]{Isella+etal_2010b}
{Isella}, A., {Carpenter}, J.~M., \& {Sargent}, A.~I. 2010{\natexlab{a}}, \apj,
  714, 1746

\bibitem[{{Isella} {et~al.}(2010{\natexlab{b}}){Isella}, {Natta}, {Wilner},
  {Carpenter}, \& {Testi}}]{Isella+etal_2010a}
{Isella}, A., {Natta}, A., {Wilner}, D., {Carpenter}, J.~M., \& {Testi}, L.
  2010{\natexlab{b}}, \apj, 725, 1735

\bibitem[{{Isella} {et~al.}(2007){Isella}, {Testi}, {Natta}, {Neri}, {Wilner},
  \& {Qi}}]{Isella+etal_2007}
{Isella}, A., {Testi}, L., {Natta}, A., {et~al.} 2007, \aap, 469, 213

\bibitem[{{Jonkheid} {et~al.}(2007){Jonkheid}, {Dullemond}, {Hogerheijde}, \&
  {van Dishoeck}}]{Jonkheid+etal_2007}
{Jonkheid}, B., {Dullemond}, C.~P., {Hogerheijde}, M.~R., \& {van Dishoeck},
  E.~F. 2007, \aap, 463, 203

\bibitem[{{Kalugina} {et~al.}(2012){Kalugina}, {Lique}, \&
  {K{\l}os}}]{Kalugina+etal_2012}
{Kalugina}, Y., {Lique}, F., \& {K{\l}os}, J. 2012, \mnras, 422, 812

\bibitem[{{Kenyon} \& {Hartmann}(1995)}]{Kenyon+Hartmann_1995}
{Kenyon}, S.~J. \& {Hartmann}, L. 1995, \apjs, 101, 117

\bibitem[{{Kitamura} {et~al.}(1996){Kitamura}, {Kawabe}, \&
  {Saito}}]{Kitamura+etal_1996}
{Kitamura}, Y., {Kawabe}, R., \& {Saito}, M. 1996, \apj, 457, 277

\bibitem[{{Krist} {et~al.}(1998){Krist}, {Stapelfeldt}, {Burrows}, {Ballester},
  {Clarke}, {Crisp}, {Evans}, {Gallagher}, {Griffiths}, {Hester}, {Hoessel},
  {Holtzman}, {Mould}, {Scowen}, {Trauger}, {Watson}, \&
  {Westphal}}]{Krist+etal_1998}
{Krist}, J.~E., {Stapelfeldt}, K.~R., {Burrows}, C.~J., {et~al.} 1998, \apj,
  501, 841

\bibitem[{{Launhardt} {et~al.}(2009){Launhardt}, {Pavlyuchenkov}, {Gueth},
  {Chen}, {Dutrey}, {Guilloteau}, {Henning}, {Pi{\'e}tu}, {Schreyer}, \&
  {Semenov}}]{Launhardt+etal_2009}
{Launhardt}, R., {Pavlyuchenkov}, Y., {Gueth}, F., {et~al.} 2009, \aap, 494,
  147

\bibitem[{{Liu} {et~al.}(2012){Liu}, {Shang}, {Pyo}, {Takami}, {Walter}, {Yan},
  {Wang}, {Ohashi}, \& {Hayashi}}]{Liu+etal_2012}
{Liu}, C.-F., {Shang}, H., {Pyo}, T.-S., {et~al.} 2012, \apj, 749, 62

\bibitem[{{Luhman} {et~al.}(2010){Luhman}, {Allen}, {Espaillat}, {Hartmann}, \&
  {Calvet}}]{Luhman+etal_2010}
{Luhman}, K.~L., {Allen}, P.~R., {Espaillat}, C., {Hartmann}, L., \& {Calvet},
  N. 2010, \apjs, 186, 111

\bibitem[{{McCabe} {et~al.}(2011){McCabe}, {Duch{\^e}ne}, {Pinte},
  {Stapelfeldt}, {Ghez}, \& {M{\'e}nard}}]{McCabe+etal_2011}
{McCabe}, C., {Duch{\^e}ne}, G., {Pinte}, C., {et~al.} 2011, \apj, 727, 90

\bibitem[{{Meeus} {et~al.}(2001){Meeus}, {Waters}, {Bouwman}, {van den Ancker},
  {Waelkens}, \& {Malfait}}]{Meeus+etal_2001}
{Meeus}, G., {Waters}, L.~B.~F.~M., {Bouwman}, J., {et~al.} 2001, \aap, 365,
  476

\bibitem[{{Mitchell} {et~al.}(1997){Mitchell}, {Sargent}, \&
  {Mannings}}]{Mitchell+etal_1997}
{Mitchell}, G.~F., {Sargent}, A.~I., \& {Mannings}, V. 1997, \apjl, 483, L127+

\bibitem[{{Momose} {et~al.}(1996){Momose}, {Ohashi}, {Kawabe}, {Hayashi}, \&
  {Nakano}}]{Momose+etal_1996}
{Momose}, M., {Ohashi}, N., {Kawabe}, R., {Hayashi}, M., \& {Nakano}, T. 1996,
  \apj, 470, 1001

\bibitem[{{Monin} {et~al.}(1996){Monin}, {Pudritz}, \&
  {Lazareff}}]{Monin+etal_1996}
{Monin}, J.-L., {Pudritz}, R.~E., \& {Lazareff}, B. 1996, \aap, 305, 572

\bibitem[{{M{\"u}ller} {et~al.}(2001){M{\"u}ller}, {Thorwirth}, {Roth}, \&
  {Winnewisser}}]{CDMS_2001}
{M{\"u}ller}, H.~S.~P., {Thorwirth}, S., {Roth}, D.~A., \& {Winnewisser}, G.
  2001, \aap, 370, L49

\bibitem[{{Nguyen} {et~al.}(2012){Nguyen}, {Brandeker}, {van Kerkwijk}, \&
  {Jayawardhana}}]{Nguyen+etal_2012}
{Nguyen}, D.~C., {Brandeker}, A., {van Kerkwijk}, M.~H., \& {Jayawardhana}, R.
  2012, \apj, 745, 119

\bibitem[{{{\"O}berg} {et~al.}(2010){{\"O}berg}, {Qi}, {Fogel}, {Bergin},
  {Andrews}, {Espaillat}, {van Kempen}, {Wilner}, \&
  {Pascucci}}]{Oberg+etal_2010}
{{\"O}berg}, K.~I., {Qi}, C., {Fogel}, J.~K.~J., {et~al.} 2010, \apj, 720, 480

\bibitem[{{{\"O}berg} {et~al.}(2011){{\"O}berg}, {Qi}, {Fogel}, {Bergin},
  {Andrews}, {Espaillat}, {Wilner}, {Pascucci}, \& {Kastner}}]{Oberg+etal_2011}
{{\"O}berg}, K.~I., {Qi}, C., {Fogel}, J.~K.~J., {et~al.} 2011, \apj, 734, 98

\bibitem[{{Osterloh} \& {Beckwith}(1995)}]{Osterloh+Beckwith_1995}
{Osterloh}, M. \& {Beckwith}, S.~V.~W. 1995, \apj, 439, 288

\bibitem[{{Padgett} {et~al.}(1999){Padgett}, {Brandner}, {Stapelfeldt},
  {Strom}, {Terebey}, \& {Koerner}}]{Padgett+etal_1999}
{Padgett}, D.~L., {Brandner}, W., {Stapelfeldt}, K.~R., {et~al.} 1999, \aj,
  117, 1490

\bibitem[{{Pety} {et~al.}(2006){Pety}, {Gueth}, {Guilloteau}, \&
  {Dutrey}}]{Pety+etal_2006}
{Pety}, J., {Gueth}, F., {Guilloteau}, S., \& {Dutrey}, A. 2006, \aap, 458, 841

\bibitem[{{Pi{\'e}tu} {et~al.}(2007){Pi{\'e}tu}, {Dutrey}, \&
  {Guilloteau}}]{Pietu+etal_2007}
{Pi{\'e}tu}, V., {Dutrey}, A., \& {Guilloteau}, S. 2007, \aap, 467, 163

\bibitem[{{Pi{\'e}tu} {et~al.}(2005){Pi{\'e}tu}, {Guilloteau}, \&
  {Dutrey}}]{Pietu+etal_2005}
{Pi{\'e}tu}, V., {Guilloteau}, S., \& {Dutrey}, A. 2005, \aap, 443, 945

\bibitem[{{Prato} {et~al.}(2009){Prato}, {Lockhart}, {Johns-Krull}, \&
  {Rayner}}]{Prato+etal_2009}
{Prato}, L., {Lockhart}, K.~E., {Johns-Krull}, C.~M., \& {Rayner}, J.~T. 2009,
  \aj, 137, 3931

\bibitem[{{Pyo} {et~al.}(2003){Pyo}, {Kobayashi}, {Hayashi}, {Terada}, {Goto},
  {Takami}, {Takato}, {Gaessler}, {Usuda}, {Yamashita}, {Tokunaga}, {Hayano},
  {Kamata}, {Iye}, \& {Minowa}}]{Pyo+etal_2003}
{Pyo}, T., {Kobayashi}, N., {Hayashi}, M., {et~al.} 2003, \apj, 590, 340

\bibitem[{{Qi} {et~al.}(2011){Qi}, {D'Alessio}, {{\"O}berg}, {Wilner},
  {Hughes}, {Andrews}, \& {Ayala}}]{Qi+etal_2011}
{Qi}, C., {D'Alessio}, P., {{\"O}berg}, K.~I., {et~al.} 2011, \apj, 740, 84

\bibitem[{{Ratzka} {et~al.}(2009){Ratzka}, {Schegerer}, {Leinert},
  {{\'A}brah{\'a}m}, {Henning}, {Herbst}, {K{\"o}hler}, {Wolf}, \&
  {Zinnecker}}]{Ratzka+etal_2009}
{Ratzka}, T., {Schegerer}, A.~A., {Leinert}, C., {et~al.} 2009, \aap, 502, 623

\bibitem[{{Roccatagliata} {et~al.}(2011){Roccatagliata}, {Ratzka}, {Henning},
  {Wolf}, {Leinert}, \& {Bouwman}}]{Roccatagliata+etal_2011}
{Roccatagliata}, V., {Ratzka}, T., {Henning}, T., {et~al.} 2011, \aap, 534, A33

\bibitem[{{Salter} {et~al.}(2011){Salter}, {Hogerheijde}, {van der Burg},
  {Kristensen}, \& {Brinch}}]{Salter+etal_2011}
{Salter}, D.~M., {Hogerheijde}, M.~R., {van der Burg}, R.~F.~J., {Kristensen},
  L.~E., \& {Brinch}, C. 2011, \aap, 536, A80

\bibitem[{{Schaefer} {et~al.}(2009){Schaefer}, {Dutrey}, {Guilloteau}, {Simon},
  \& {White}}]{Schaefer+etal_2009}
{Schaefer}, G.~H., {Dutrey}, A., {Guilloteau}, S., {Simon}, M., \& {White},
  R.~J. 2009, \apj, 701, 698

\bibitem[{{Schreyer} {et~al.}(2008){Schreyer}, {Guilloteau}, {Semenov},
  {Bacmann}, {Chapillon}, {Dutrey}, {Gueth}, {Henning}, {Hersant}, {Launhardt},
  {Pety}, \& {Pi{\'e}tu}}]{Schreyer+etal_2008}
{Schreyer}, K., {Guilloteau}, S., {Semenov}, D., {et~al.} 2008, \aap, 491, 821

\bibitem[{{Semenov} \& {Wiebe}(2011)}]{Semenov+Wiebe_2011}
{Semenov}, D. \& {Wiebe}, D. 2011, \apjs, 196, 25

\bibitem[{{Simon} {et~al.}(2000){Simon}, {Dutrey}, \&
  {Guilloteau}}]{Simon+etal_2000}
{Simon}, M., {Dutrey}, A., \& {Guilloteau}, S. 2000, \apj, 545, 1034

\bibitem[{{Skatrud} {et~al.}(1983){Skatrud}, {de Lucia}, {Blake}, \&
  {Sastry}}]{Skatrud+etal_1983}
{Skatrud}, D.~D., {de Lucia}, F.~C., {Blake}, G.~A., \& {Sastry}, K.~V.~L.~N.
  1983, Journal of Molecular Spectroscopy, 99, 35

\bibitem[{{Tafalla} {et~al.}(2010){Tafalla}, {Santiago-Garc{\'{\i}}a}, {Hacar},
  \& {Bachiller}}]{Tafalla+etal_2010}
{Tafalla}, M., {Santiago-Garc{\'{\i}}a}, J., {Hacar}, A., \& {Bachiller}, R.
  2010, \aap, 522, A91

\bibitem[{{Testi} {et~al.}(2002){Testi}, {Bacciotti}, {Sargent}, {Ray}, \&
  {Eisl{\"o}ffel}}]{Testi+etal_2002}
{Testi}, L., {Bacciotti}, F., {Sargent}, A.~I., {Ray}, T.~P., \&
  {Eisl{\"o}ffel}, J. 2002, \aap, 394, L31

\bibitem[{{Thi} {et~al.}(2004){Thi}, {van Zadelhoff}, \& {van
  Dishoeck}}]{Thi+etal_2004}
{Thi}, W.-F., {van Zadelhoff}, G.-J., \& {van Dishoeck}, E.~F. 2004, \aap, 425,
  955

\bibitem[{{Tilling} {et~al.}(2012){Tilling}, {Woitke}, {Meeus}, {Mora},
  {Montesinos}, {Riviere-Marichalar}, {Eiroa}, {Thi}, {Isella}, {Roberge},
  {Martin-Zaidi}, {Kamp}, {Pinte}, {Sandell}, {Vacca}, {M{\'e}nard},
  {Mendigut{\'{\i}}a}, {Duch{\^e}ne}, {Dent}, {Aresu}, {Meijerink}, \&
  {Spaans}}]{Tilling+etal_2012}
{Tilling}, I., {Woitke}, P., {Meeus}, G., {et~al.} 2012, \aap, 538, A20

\bibitem[{{Troscompt} {et~al.}(2009){Troscompt}, {Faure}, {Wiesenfeld},
  {Ceccarelli}, \& {Valiron}}]{Troscompt+etal_2009}
{Troscompt}, N., {Faure}, A., {Wiesenfeld}, L., {Ceccarelli}, C., \& {Valiron},
  P. 2009, \aap, 493, 687

\bibitem[{{van den Ancker} {et~al.}(1998){van den Ancker}, {de Winter}, \&
  {Tjin A Djie}}]{vandenAncker+etal_1998}
{van den Ancker}, M.~E., {de Winter}, D., \& {Tjin A Djie}, H.~R.~E. 1998,
  \aap, 330, 145

\bibitem[{{van Leeuwen}(2007)}]{vanLeeuwen_2007}
{van Leeuwen}, F. 2007, \aap, 474, 653

\bibitem[{{White} \& {Basri}(2003)}]{White+Basri_2003}
{White}, R.~J. \& {Basri}, G. 2003, \apj, 582, 1109

\bibitem[{{White} \& {Ghez}(2001)}]{White+Ghez_2001}
{White}, R.~J. \& {Ghez}, A.~M. 2001, \apj, 556, 265

\bibitem[{{White} \& {Hillenbrand}(2004)}]{White+Hillenbrand_2004}
{White}, R.~J. \& {Hillenbrand}, L.~A. 2004, \apj, 616, 998

\bibitem[{{Wilking} {et~al.}(2012){Wilking}, {Marvel}, {Claussen}, {Gerling},
  {Wootten}, \& {Gibb}}]{Wilking+etal_2012}
{Wilking}, B.~A., {Marvel}, K.~B., {Claussen}, M.~J., {et~al.} 2012, ArXiv
  e-prints

\bibitem[{{Wolf} {et~al.}(2003){Wolf}, {Padgett}, \&
  {Stapelfeldt}}]{Wolf+etal_2003}
{Wolf}, S., {Padgett}, D.~L., \& {Stapelfeldt}, K.~R. 2003, \apj, 588, 373

\bibitem[{{Yokogawa} {et~al.}(2002){Yokogawa}, {Kitamura}, {Momose}, \&
  {Kawabe}}]{Yokogawa+etal_2002}
{Yokogawa}, S., {Kitamura}, Y., {Momose}, M., \& {Kawabe}, R. 2002, in 8th
  Asian-Pacific Regional Meeting, Volume II, ed. S.~{Ikeuchi}, J.~{Hearnshaw},
  \& T.~{Hanawa}, 239--240

\end{thebibliography}
\bibliographystyle{aa}
%\end{document}

\clearpage
\appendix

\section{Comments on individual sources}
\paragraph{\object{FN Tau}:}
the tentative detection of H$_2$CO is at an unusual velocity of -0.5 $\kms$.

\paragraph{\object{CW Tau}} is dominated by confusion (see discussion in Appendix \ref{app:cloud}).
Despite strong confusion, line wings of $^{13}$CO could indicate a
disk at velocity $6.2 \kms$ and with a line width of $3.3 \pm 0.4 \kms$.
CW Tau drives a bipolar outflow that can be traced in optical forbidden lines
over $4" - 6"$  along both outflow directions \citep{Hirth+etal_1994}.

\paragraph{\object{CIDA 1}} \citet{Briceno+etal_1993} is a late spectral type object \citep[M5.5][]{White+Basri_2003}, with
weak 1.3 mm continuum emission \citep{Schaefer+etal_2009}. The line of sight towards CIDA 1 intercepts the
same molecular cloud than that of CW Tau and is highly confused. The tentative fit of the
$^{13}$CO line wing is not conclusive on the presence of a disk.

\paragraph{\object{BP Tau}} harbors a small (120 AU radius) disk \citep{Dutrey+etal_2003} and strong
confusion in $^{12}$CO. No signal at all is detected in these new observations, but the sensitivity
towards this source is lower than average.

\paragraph{\object{DE Tau}} has relatively weak 1.3 \,mm continuum (36 mJy). There is some confusion in $^{13}$CO,
but no evidence for disk emission.

\paragraph{\object{RY Tau}:}
the CN detection is marginal (3.5 $\sigma$). A fit of a $3 \kms$ wide
gaussian to the $^{13}$CO spectrum also indicate a potential disk,
at the $5 \sigma$ level, but with a different velocity.
RY Tau is a suspected close binary, with separation $> 3$\,AU \citep{Bertout+etal_1999}.
Velocity measurements by \citet{Nguyen+etal_2012} indicate a possible SB1 with
radial heliocentric velocity in the range $16.7-19.4$\,km/s,
which is equivalent to $7.7-10.4$\,km/s in the LSR,
and more consistent with the $^{13}$CO velocity.
High resolution image at 1.3\,mm by \citet{Isella+etal_2010b} indicate an inclination of
$66^\circ$ and a morphology consistent with a deficit of emission in the inner 15 AU.

\paragraph{\object{T Tau}} is a triple system, with a compact ($\sim 60$ AU) circumstellar
disk around the single T Tau N \citep{Akeson+etal_1998,Guilloteau+etal_2011}.
Its molecular environment is complex, and all
the detected lines are most likely coming from an outflow or heated envelope,
perhaps the outflowing cavity walls imaged in $^{13}$CO by \citet{Momose+etal_1996}.

\paragraph{\object{Haro 6-5 B},} also known as \object{FS Tau B}, is a nearly edge-on, deeply embedded object.
Confusion in $^{13}$CO is very strong. Previous interferometric measurements
of $^{13}$CO J=1-0 revealed a complex region \citep{Dutrey+etal_1996}. HST
images indicate a 300 AU radius disk at high inclination, 70-80$^\circ$ \citep{Krist+etal_1998,Padgett+etal_1999}.
FS Tau B possess a jet and a counterjet, as well as an optically visible cavity wall \citep{Liu+etal_2012}.
Our combined detection of CN and H$_2$CO may indicate a compact, warm disk or some emission from an outflow or envelope.
5$''$ resolution images in $^{13}$CO J=1-0 with the NMA \citep{Yokogawa+etal_2002} revealed a disk at $V_\mathrm{LSR} \approx
7.0 \kms$, with a kinematics dominated by rotation. They indicate motions compatible with Keplerian motions
around a central star of 0.25 $\msun$ only. Given the high confusion level, this value may be unreliable.
Our CN line width, though unprecise, suggests a higher stellar mass.  The asymmetric H$_2$CO spectrum may be affected by
confusion in the blueshifted part.

\paragraph{\object{FT Tau}} is a deeply embedded, low luminosity star with unknown spectral type.
A compact ($60$ AU) dust disk was resolved by \citet{Guilloteau+etal_2011},
who also detected $^{12}$CO emission, suggesting a small disk at low inclination
around a $0.7 - 1.0 \msun$ star. The clear detection of CN confirms the
disk interpretation.  $^{13}$CO  emission from the disk may be present at redshifted
velocities compatible with CN, but confusion makes any quantitative estimate
impossible.

\paragraph{\object{DG Tau}} is a well studied object, driving an optical microjet.
The system inclination remain unclear: between 38 and 45$^\circ$ from
the jet \citep{Pyo+etal_2003,Eisloffel+Mundt_1998}, and 20 to 32$^\circ$
from the dust disk \citep{Isella+etal_2010b,Guilloteau+etal_2011}.
Images in $^{13}$CO (2-1) \citep{Testi+etal_2002} indicate a complex
morphology, a large fraction of the emission coming from a remnant
envelope \citep[see also][]{Kitamura+etal_1996}. There is no clear evidence for
Keplerian rotation. The relatively narrow
lines of H$_2$CO and SO could originate from this remnant envelope.
CN(2-1) was previously detected by \citet{Salter+etal_2011}\footnote{Although their Figs. 10 and 12 mark it as the 3-2 transition}, who modeled this as coming from a Keplerian disk.
Our CN line width is consistent with the measurement of \citet{Salter+etal_2011},
but the large derived outer radius, combined with the complex morphology,
suggests CN is most likely coming from the outflow or the remnant envelope.

\paragraph{\object{DG Tau B}} is a totally obscured object, driving a powerful one-sided molecular
outflow \citet{Mitchell+etal_1997}. High inclination \citep[$64 - 75^\circ$, see][]{Guilloteau+etal_2011,
Eisloffel+Mundt_1998} however yield moderate line-of-sight velocities for the outflow.
The rich spectrum may originate from the outflow, but CN lacks the narrow component
detected in H$_2$CO and C$^{17}$O, which is presumably due to the
molecular cloud.

\paragraph{\object{Haro 6-10}} (also known as GV Tau) is a close binary, with two compact
($< 30$ AU), optically thick, circumstellar disks
\citep{Guilloteau+etal_2011}, but no known circumbinary disk. The circumstellar disks are misaligned
 and the system is embedded in a common envelope \citep{Roccatagliata+etal_2011}.
 H$_2$O maser emission has been recently reported, and suggest that the Southern component is itself
 a close binary \citep{Wilking+etal_2012}. As for DG Tau B, the CN line
 is much wider than those of H$_2$CO and C$^{17}$O. The proeminent red wing in $^{13}$CO is
 likely to trace the outflow.

\paragraph{\object{IQ Tau}} was observed in $^{12}$CO by \citet{Schaefer+etal_2009}, who did
not detect any significant emission with the IRAM Plateau de Bure
interferometer at 5$''$ resolution.
The CN emission exhibits the clear signature of Keplerian rotation (although
the expected double peaked aspect is reduced because of the hyperfine component
blending). CN is apparently stronger than $^{13}$CO in this source. A tentative
fit of a disk component to the $^{13}$CO spectrum suggest the redshifted emission
may still be masked by the cloud.

\paragraph{\object{LkHa 358}:}
$^{13}$CO is heavily confused.  A small molecular disk was reported
from interferometric measurements by \citet{Schaefer+etal_2009} from $^{12}$CO emission.
An attempt to fit a disk component to the $^{13}$CO line wings
is consistent with this interpretation, provided $^{13}$CO has significant
optical depth ($> 2$ on average).
The narrow emission from C$^{17}$O most likely comes from the surrounding cloud,
as the $\sim 3.5 \sigma$ signal from H$_2$CO.

\paragraph{\object{HH 30}:}
\citet{Pety+etal_2006} had previously imaged the $^{13}$CO emission from this
emblematic edge-on object. The narrow line from C$^{17}$O likely comes
from the cloud, as the $\sim 3 \sigma$ signal from H$_2$CO.

\paragraph{\object{HL Tau}} is the archetype of the deeply embedded, outflow driving sources. The redshifted
side of the outflow is known to be brighter \citep{Cabrit+etal_1996,Monin+etal_1996}. From the line profiles,
C$^{17}$O,  H$_2$CO and especially SO (which is significantly redshifted),
must come from the outflow. However, the CN line is much narrower and exhibit
a systemic velocity of $4.7 \kms$, and thus could be a pure disk tracer.

\paragraph{\object{HK Tau}} is a binary star, with one component exhibiting an edge-on disk \citep{McCabe+etal_2011}.
The relatively broad CN line is consistent with emission from this edge-on disk around \object{HK Tau B}. The blue
shifted side of the disk may also have been detected in $^{13}$CO.

\paragraph{\object{Haro 6-13},} also known as \object{V 806 Tau}, was imaged in $^{12}$CO by \citet{Schaefer+etal_2009}, who
based on agreement in position angle between CO and dust emission, attributed the emission
to a disk of at least 180 AU radius. HCO$^+$3-2 emission has been detected
by \citet{Salter+etal_2011}. The H$_2$CO spectra are complex, as the C$^{17}$O one, and perhaps
result of superposition of relatively broad emission with a narrow ($0.5 \kms$) absorption feature.

\paragraph{\object{GG Tau}} was included as a reference source in our sample. Our results agree with the
initial detections of \citet{Dutrey+etal_1997}. H$_2$CO is well detected in the combined spectrum.

\paragraph{\object{UZ Tau} E} is a spectroscopic binary part of the UZ Tau hierarchical quadruple system. The CO disk
was detected by \citet{Simon+etal_2000}, who found an inclination of $55^\circ$ and an outer
radius of 300 AU. Our derived radius from CN is in excellent agreement with
this value. $^{13}$CO emission from the disk is also detected, in the line wings, but still
exhibits confusion.

\paragraph{\object{IRAS04302+2247}}  \citep[the Butterfly star][]{Wolf+etal_2003}, is an edge-on system.
Recent IRAM interferometric measurements indicate a stellar mass of order $1.6 \msun$
and a disk radius of 260 AU \citep[][in prep]{Dutrey+etal_2012}.
Although this source exhibits a
large H$_2$CO / CN ratio, which we in general attribute to
outflows, the H$_2$CO and SO emissions come from the disk.
From the hyperfine ratios, CN appears potentially optically thick
in this source, in which case the emission must be dominated by smaller radii.
The rather large linewidth concurs with this interpretation.

\paragraph{\object{DL Tau}} was imaged in $^{12}$CO by \citet{Simon+etal_2000} who pointed out
the strong confusion making the inclination derivation unreliable. Confusion is also
large in $^{13}$CO, masking all the emission except for the red-shifted wing.
Like IQ Tau, DL Tau shows strong, double peaked line profile in CN. The outer
radius derived from CN is consistent with the value quoted by \citet{Simon+etal_2000} for CO. H$_2$CO is
also detected.

\paragraph{\object{AA Tau}:} CN is clearly detected, and weak emission is also visible
in $^{13}$CO, despite confusion.  However, the disk contribution
in $^{13}$CO is much smaller than the intensity reported by \citet{Greaves_2005}
(0.6 K.$\kms$ at JCMT, i.e. 12 Jy$\kms$), which must have included cloud
emission. CN in AA Tau was discovered by \citet{Oberg+etal_2010}.
We recover only about half of their quoted flux (1.7 Jy$\kms$), and our derived
velocity suggests that the blue-shifted part of the CN spectrum is masked by confusion,
like the $^{13}$CO one.

\paragraph{\object{DN Tau}} is very similar to \object{DL Tau} and \object{IQ Tau}, with CN stronger than
$^{13}$CO. \citet{Schaefer+etal_2009} failed to detect $^{12}$CO
from this object with the IRAM interferometer, again presumably because of strong confusion
with the cloud.

\paragraph{\object{DO Tau}} shows clear CN detection ($4 \sigma$), but the line width
is poorly constrained. The line can be narrow, so (positive)
contamination by the cloud is not fully excluded.
Identification of a disk component in $^{13}$CO is difficult,
as the redshifted side is dominated by the cloud.

\paragraph{\object{HV Tau C},} like \object{HK Tau B}, is a compact (80 AU radius) edge on disk in a multiple system,
here a hierarchical triple.  $^{12}$CO emission from
the disk was resolved by \citet{Duchene+etal_2010} with the
IRAM array. The broad CN line ($\Delta V = 6 \kms$) is consistent
with an origin in a compact disk. The CN flux
suggests a somewhat larger radius ($ 300$ AU in Table \ref{tab:radii}),
or equivalently, non negligible optical depth. The observed
hyperfine ratios are indeed compatible with substantial optical
depth, $\tau = 7.0 \pm 2.7$, which would also yield a lower line width,
$4.2 \pm 0.6 \kms$.
$^{13}$CO emission from the disk produces the line wings visible
in Fig.\ref{fig:HV_TAU-C}

\paragraph{\object{Haro 6-33}} (also known as
\object{IRAS04385+2550}) was detected in $^{12}$CO by \citet{Schaefer+etal_2009}, who
interpreted the emission as coming from a Keplerian disk, despite strong
confusion. There is also substantial cloud contamination in $^{13}$CO, but CN and H$_2$CO are detected.
The derived disk radius in CN is in good agreement with the $^{12}$CO result.

\paragraph{\object{DQ Tau}} exhibits huge confusion, and no detection at all.

\paragraph{\object{DR Tau}} was reported as tentatively detected in CN(2-1)
by \citet{Salter+etal_2011} at a velocity of 11 $\kms$
with an integrated flux of $3.5 \pm 1.3$ Jy\,$\kms$. Our spectrum shows
no line at all, with a $3 \sigma$ upper limit on the flux of $1.4 $ Jy\,$\kms$.
Confusion with the cloud is huge in $^{13}$CO: only the red wing between 10 and 11 $\kms$ may
be attributable to a disk component.

\paragraph{\object{DS Tau}} shows no apparent confusion. A weak ($4  \sigma$) line
is detected when combining both o-H$_2$CO transitions, with a velocity (6.0 $\kms$)
and linewidth (3 $\kms$) consistent with disk emission, but no CN or $^{13}$CO.

\paragraph{\object{UY Aur}} is a 0.8$''$ separation binary system.
The $^{13}$CO  line is strong, and consistent with an origin in the complex
circumbinary structure imaged by \citet{Duvert+etal_2000}. There is a $4 \sigma$ detection
of H$_2$CO with a compatible kinematics.

\paragraph{\object{AB Aur}:} the strong $^{13}$CO line emanates from a combination of the disk/ring structure
imaged by \citet{Pietu+etal_2005} and the surrounding envelope. The clear
detection of C$^{17}$O confirms the large CO content of this source.
H$_2$CO is convincingly detected, confirming the result of \citet{Fuente+etal_2010}
using the para-H$_2$CO $3_{0,3}-2_{0,2}$ line at 218.2 GHz. However, we find no evidence at all
for CN or SO from the disk, which were claimed to be detected by these authors. We only have
a narrow (0.4 $\kms$), 3.5$\sigma$ feature in CN(2-1) at a velocity consistent
with cloud emission.  The CN (1-0) line quoted by \citet{Fuente+etal_2010}
must have been residual emission from the cloud. For SO, they report a line
intensity of 26 mK\,$\kms$ for the $3_4-2_3$ line, and 60 mK\,$\kms$ for the $5_6-4_5$ transitions,
while we obtain a $3 \sigma$ upper limit of 21 mK$\kms$ for the $5_4-4_3$ using the AB Aur velocity
and line widths. Note that  \citet{Fuente+etal_2010} used a larger throw (120$''$) than us
in the wobbler switching, so that their results are more susceptible to residual emission
from the envelope.

\paragraph{\object{SU Aur}:} despite confusion with the cloud, the $^{13}$CO spectrum exhibits a broad component
compatible with emission from a disk.  No other molecule is detected. \object{SU Aur} is a strong X ray source
\citet{Franciosini+etal_2007}. Near IR imaging indicates a highly inclined disk ($\simeq 65^\circ$) and
outflow cavity walls \citep{Akeson+etal_2002,Chakraborty+Ge_2007}.

\paragraph{\object{MWC 480}} was included here as a reference source in our sample. It has been
extensively studied with the IRAM mm array \citet{Pietu+etal_2007,Chapillon+etal_2012}.
We find the double peaked
$^{13}$CO profile characteristic of an isolated Keplerian disk.  CN emission is
in excellent agreement with the interferometric results of \citet{Chapillon+etal_2012}. The
detection of C$^{17}$O for the first time in this source confirms the significant
CO abundance in this object. Note that, as expected, the C$^{17}$O profile is
apparently larger than the $^{13}$CO one.

\paragraph{\object{CB26}} is a deeply embedded, edge-on object driving an unusual outflow \citet{Launhardt+etal_2009}
in an isolated Bok globule. Like in \object{HL Tau}, we detect here strong H$_2$CO and SO emission, but only a weak CN line.
If interpreted by disk emission, the CN line is consistent with a disk size about 250 AU.

\paragraph{\object{CIDA 8}} has weak continuum emission and no CO disk from interferometric measurements of
\citet{Schaefer+etal_2009}. Cloud contamination exists, but appears limited
to a very narrow velocity range. There is a surprising detection of H$_2$CO just
outside the range visibly contaminated in $^{13}$CO, but with a narrow line width which suggests
it originates from a cloud.

\paragraph{\object{CIDA 11}} shows no detection at all. It was also not detected in continuum and CO
by \citet{Schaefer+etal_2009}.

\paragraph{\object{RW Aur}} is a binary system that was imaged in $^{12}$CO by \citet{Cabrit+etal_2006}, who
find evidence for a very small (50 AU radius) disk, but did not
detect $^{13}$CO emission. Our negative result is consistent with their findings.

\paragraph{\object{MWC 758}} is a 1.8 $\msun$ star surrounded by a disk where CO is apparently depleted
and mostly optically thin \citep{Chapillon+etal_2008}, despite high temperatures (30 K).
We find a relatively weak line of $^{13}$CO, consistent
with this interpretation.  No other molecule
is detected: the marginal signal in C$^{17}$O is not at the expected disk velocity.
The limited CO opacity could be related to the
large cavity detected by \citet{Isella+etal_2010a} in continuum.

\paragraph{\object{CQ Tau}} is a ``twin'' of MWC 758, also exhibiting a warm ($> 50$ K) disk with optically thin CO
emission \citep{Chapillon+etal_2008}, but its continuum emission is strongly centrally peaked
\citep{Banzatti+etal_2011,Guilloteau+etal_2011}. Although different distances are often quoted based on Hipparcos parallax, these
differences are not significant and \citet{Chapillon+etal_2008} argue that both \object{CQ Tau}
and \object{MWC 758} are also located at 140 pc.
% Testi+etal_2003
We also detect $^{13}$CO with a similar intensity than in \object{MWC 758}. H$_2$CO is
detected at the $5 \sigma$ level, with kinematic parameters
quite consistent with those derived from $^{13}$CO.
However, we find no evidence for CN. Our 3 $\sigma$ limit of 0.28 Jy$\kms$ is marginally consistent
with the flux of $0.22 \pm 0.11$ Jy$\kms$ reported by \citet{Oberg+etal_2010}.

\paragraph{\object{HD 163296}} is an isolated Herbig Ae star located at 120 pc \citep{vanLeeuwen_2007}.
It harbours a large ($\sim 550$ AU) CO disk inclined
at about $45^\circ$ \citep{Isella+etal_2007},
and has been extensively studied in CO isotopologues, including C$^{17}$O  \citep{Qi+etal_2011}. CN N=3-2 was
detected by \citet{Thi+etal_2004}. The strong CN N=2-1 emission is compatible with an optically thin line.
The outer radius in Table \ref{tab:radii} is slightly larger than that measured from CO. The discrepancy may
have several origins. First, as the star is very luminous, CN may be in a warmer region. Second, \citet{Hughes+etal_2011}
report a rather large turbulence level ($\simeq 0.3 \kms$) from CO measurements, larger than assumed.
We also report the first detection of H$_2$CO, through the ortho-H$_2$CO $3_{13}-2_{12}$ transition.
Comparison of our $^{13}$CO spectrum to the interferometric result of \citet{Qi+etal_2011} suggest
some contamination near $2 - 4 \kms$ (and also $13 \kms$ as in the $^{12}$CO spectra of
\citet{Qi+etal_2011}).

%end input
\label{app:sources}

\section{Spectra for individual sources}

This appendix displays the spectra towards the various sources. For each
source, the top
panel shows the $^{13}$CO J=2-1 spectrum. The continuous red line,
if present, is a Gaussian fit obtained after masking the confused velocity
range. The bottom panel displays on a common  scale from
top to bottom the spectra of: $^{13}$CO J=2-1 (with fit as in top panel),
C$^{17}$O J=2-1, CN N=2-1, ortho H$_2$CO $3_{13}-2_{12}$, ortho H$_2$CO $3_{12}-2_{11}$,
the average of both o-H$_2$CO transitions, and SO $5_4-4_3$, arbitrarily shifted in intensity
to avoid overlap. The intensity scale is antenna temperature (T$_A^*$ in K):
conversion to flux density can be obtained
using a factor of 9 Jy/K.

\label{app:spectra}
\clearpage

\clearpage
\begin{figure}
\includegraphics[height=11.5cm]{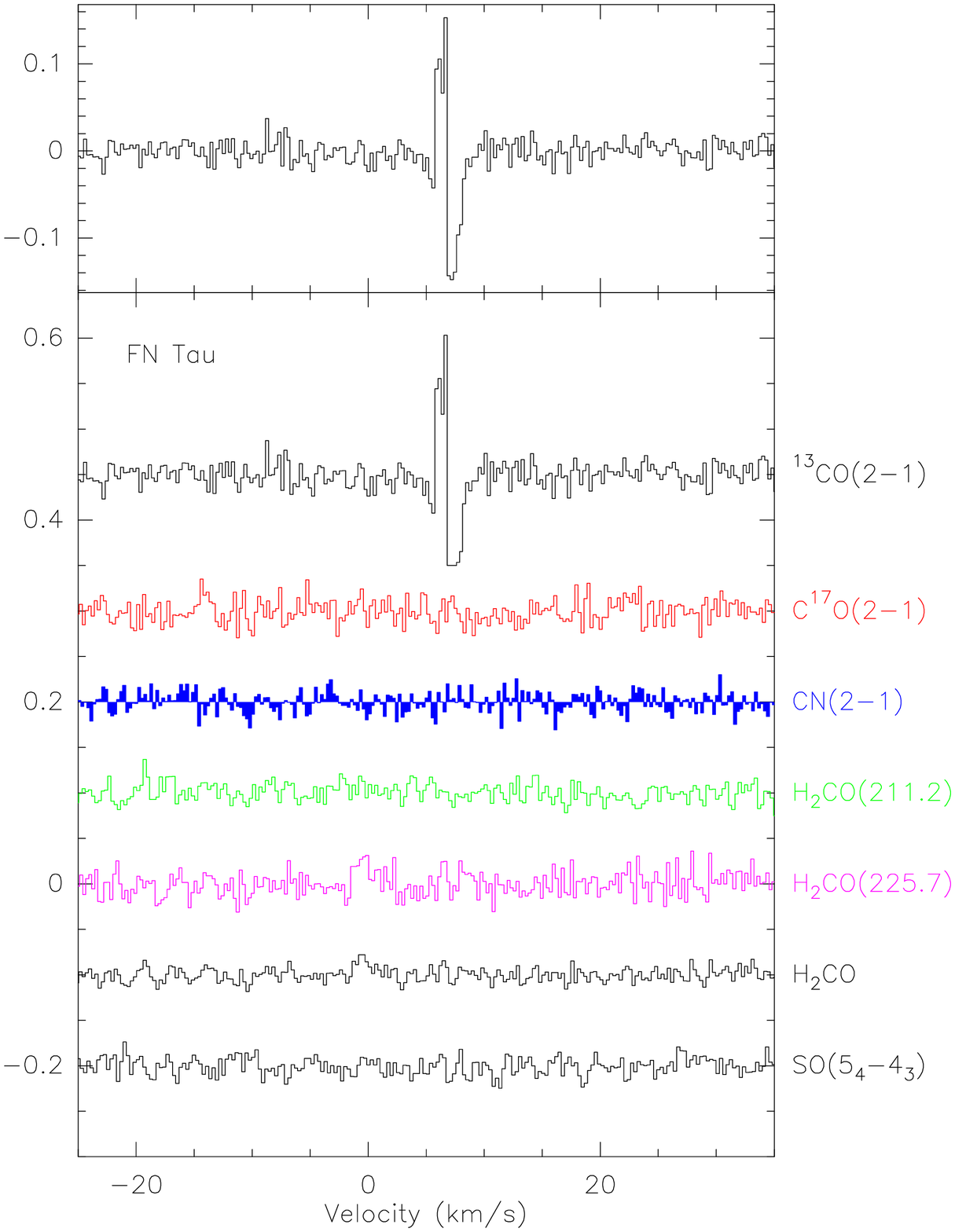}
\caption{Spectra of the observed transitions towards FN Tau}
\label{fig:FN_TAU}
\end{figure}
\begin{figure}
\includegraphics[height=11.5cm]{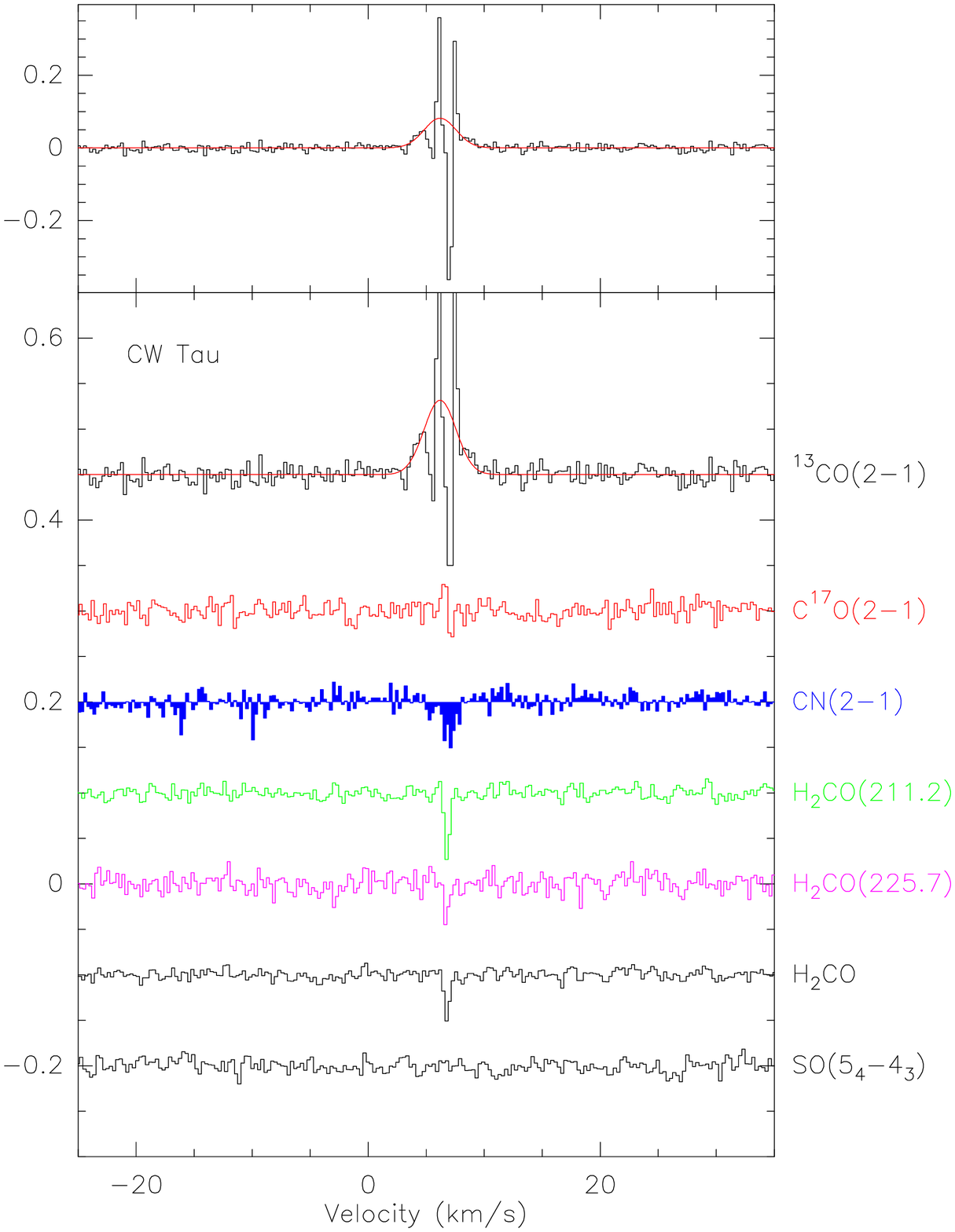}
\caption{Spectra of the observed transitions towards CW Tau}
\label{fig:CW_TAU}
\end{figure}
\begin{figure}
\includegraphics[height=11.5cm]{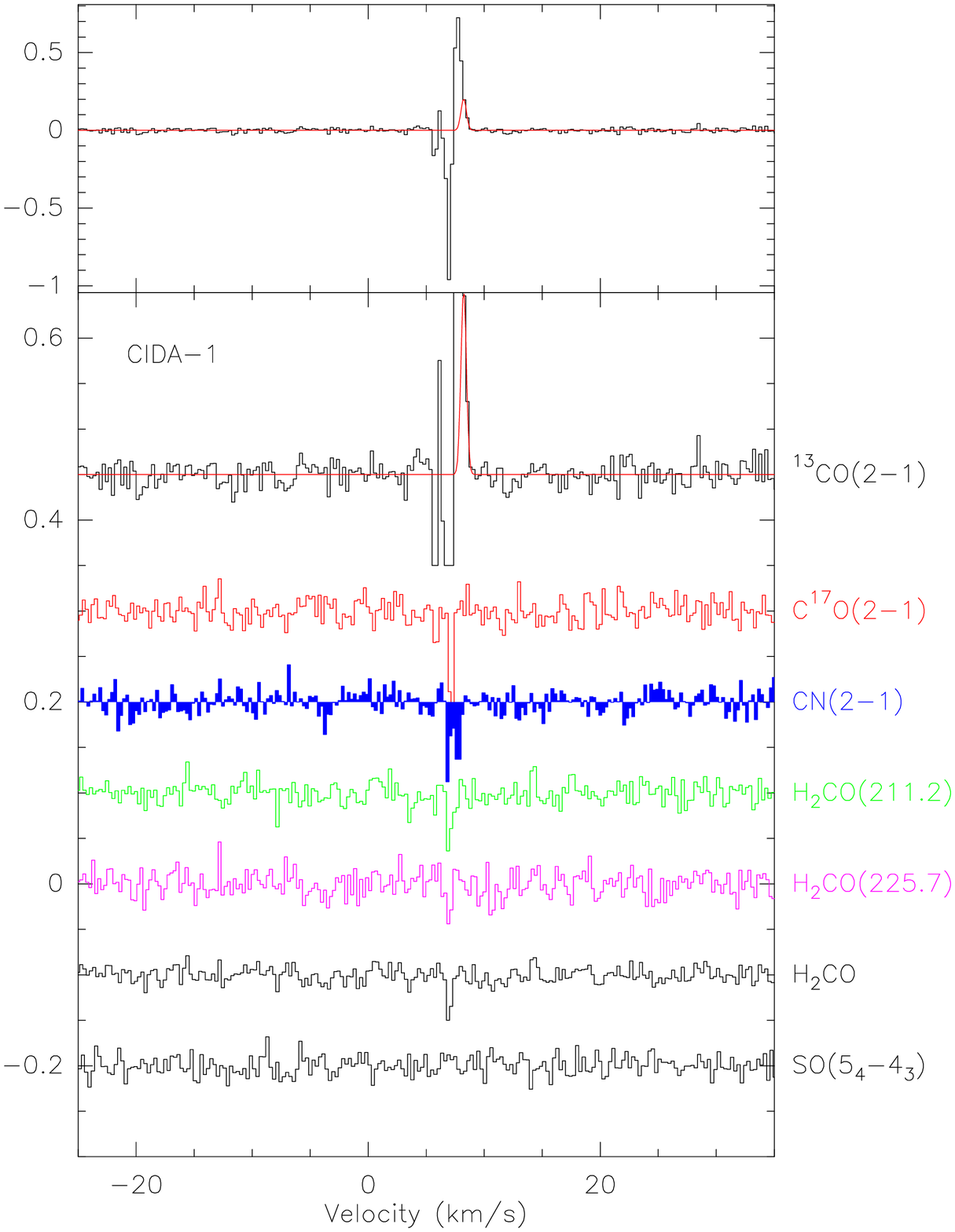}
\caption{Spectra of the observed transitions towards CIDA-1}
\label{fig:CIDA-1}
\end{figure}
\begin{figure}
\includegraphics[height=11.5cm]{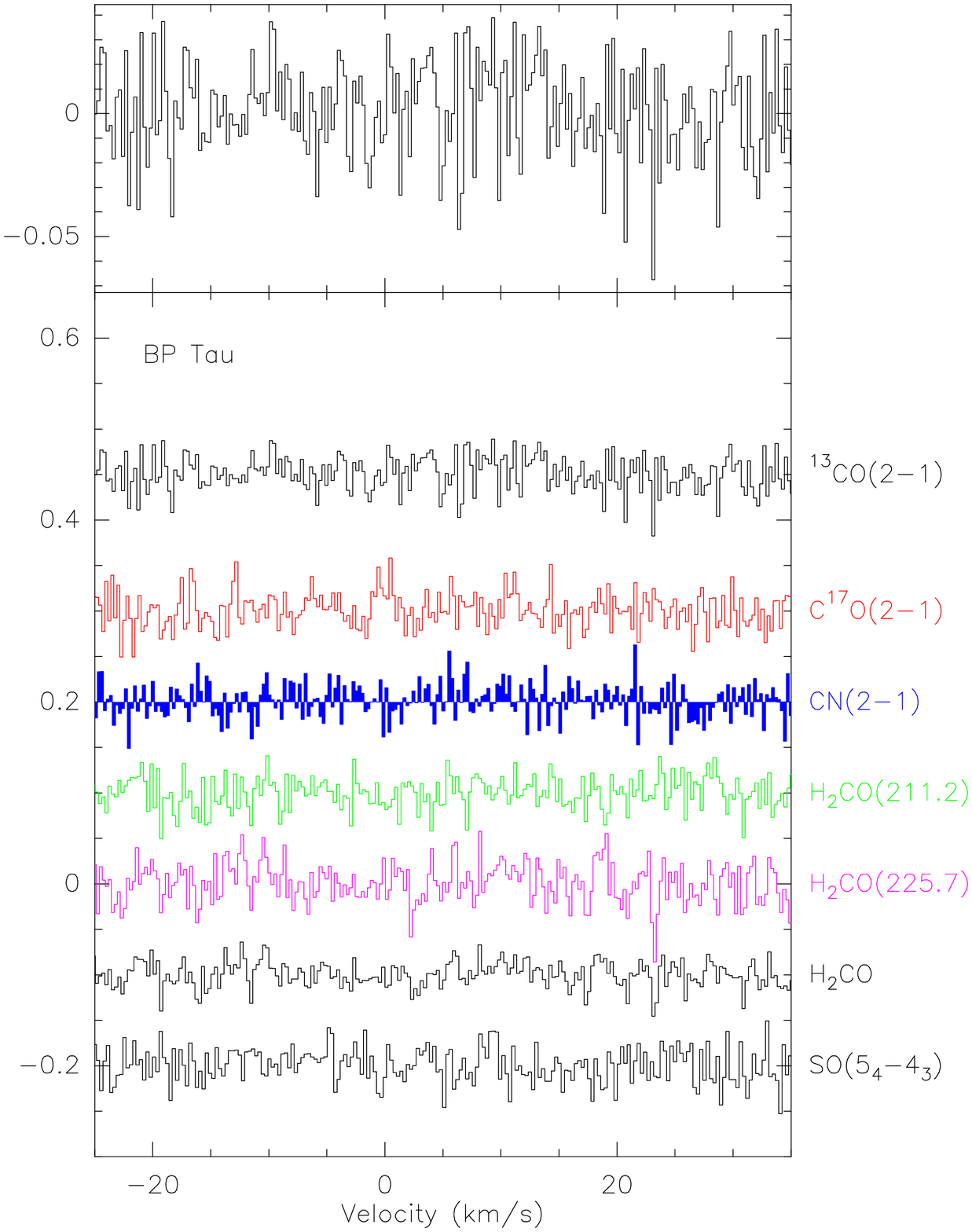}
\caption{Spectra of the observed transitions towards BP Tau}
\label{fig:BP_TAU}
\end{figure}
\clearpage
\begin{figure}
\includegraphics[height=11.5cm]{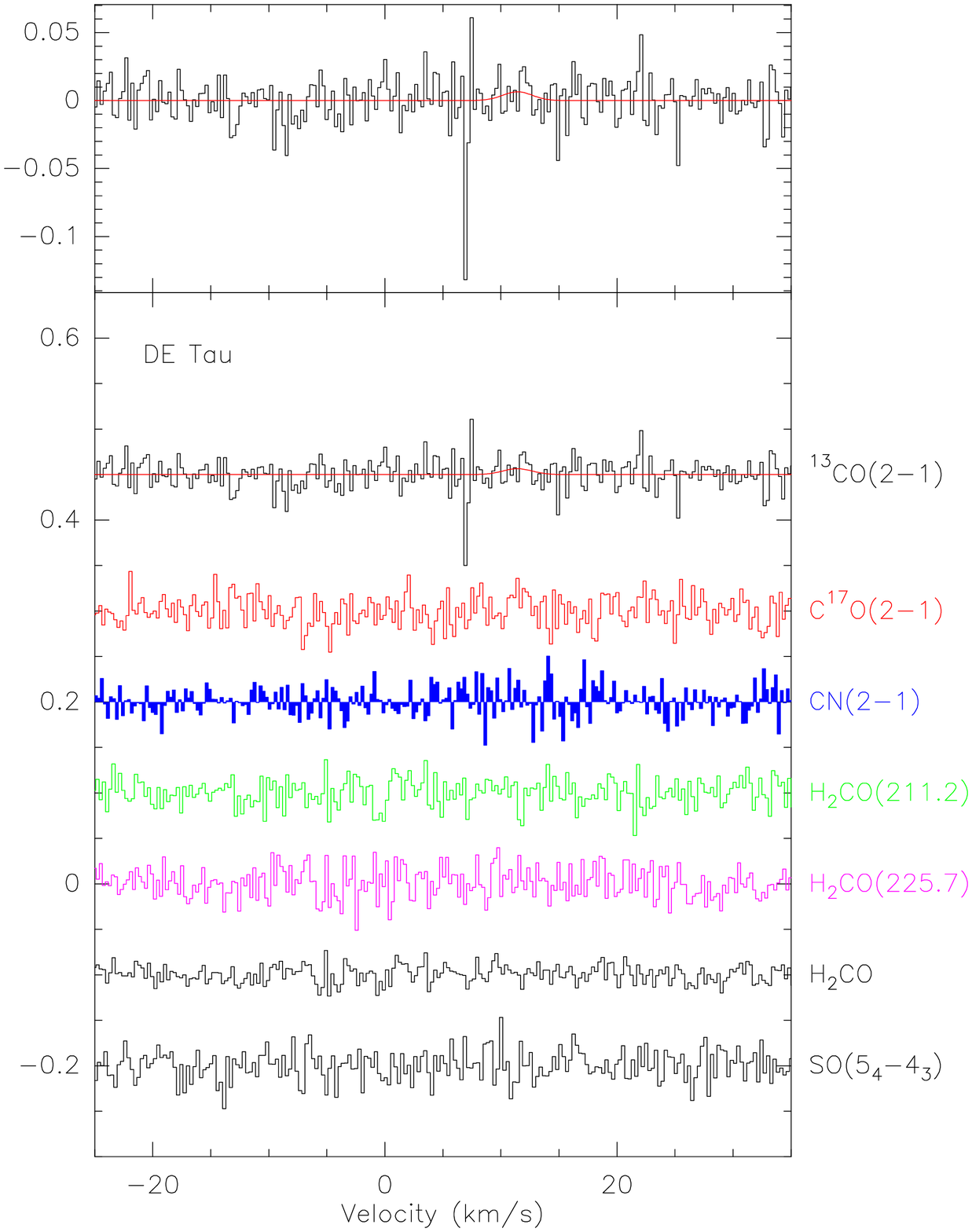}
\caption{Spectra of the observed transitions towards DE Tau}
\label{fig:DE_TAU}
\end{figure}
\begin{figure}
\includegraphics[height=11.5cm]{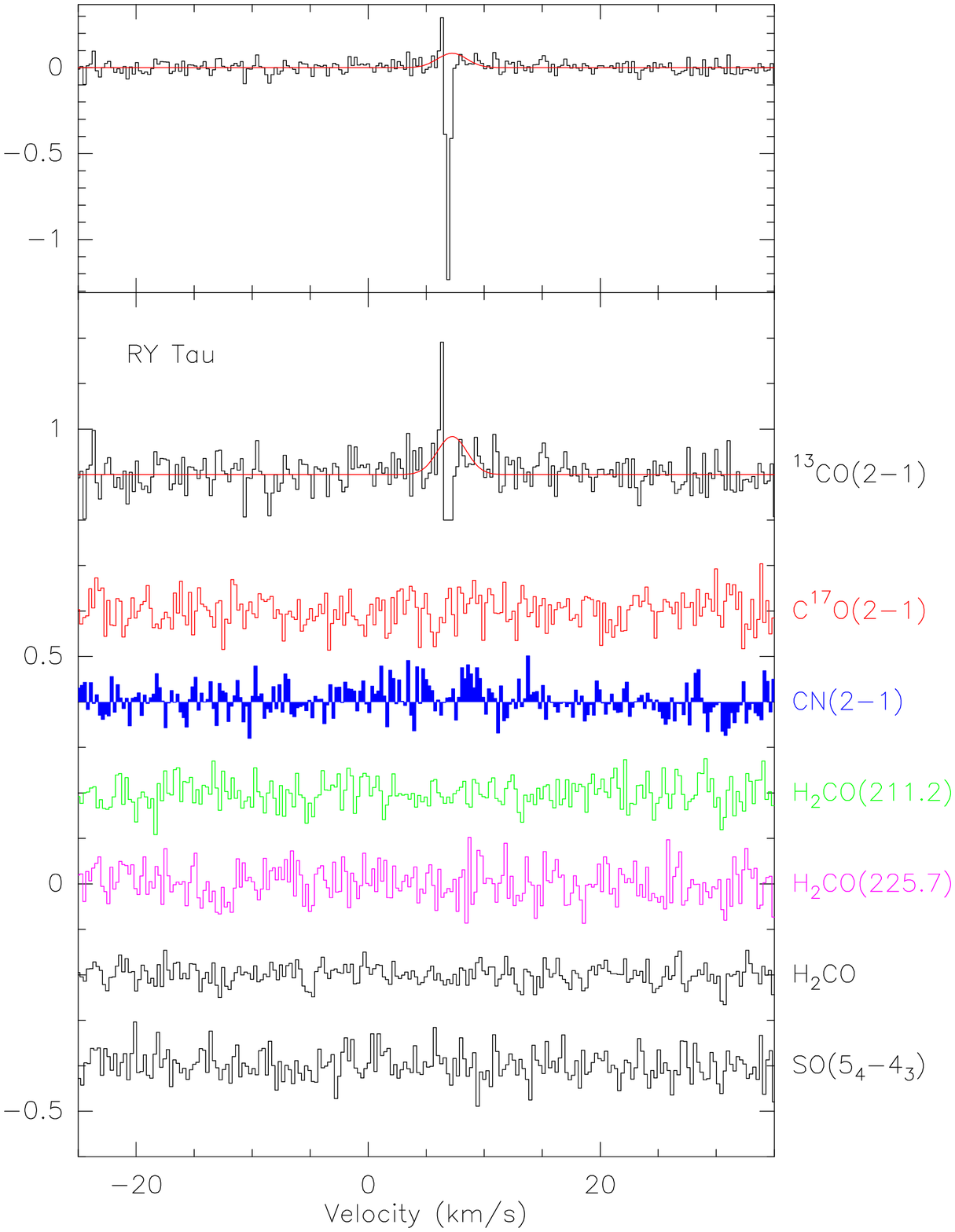}
\caption{Spectra of the observed transitions towards RY Tau}
\label{fig:RY_TAU}
\end{figure}
\begin{figure}
\includegraphics[height=11.5cm]{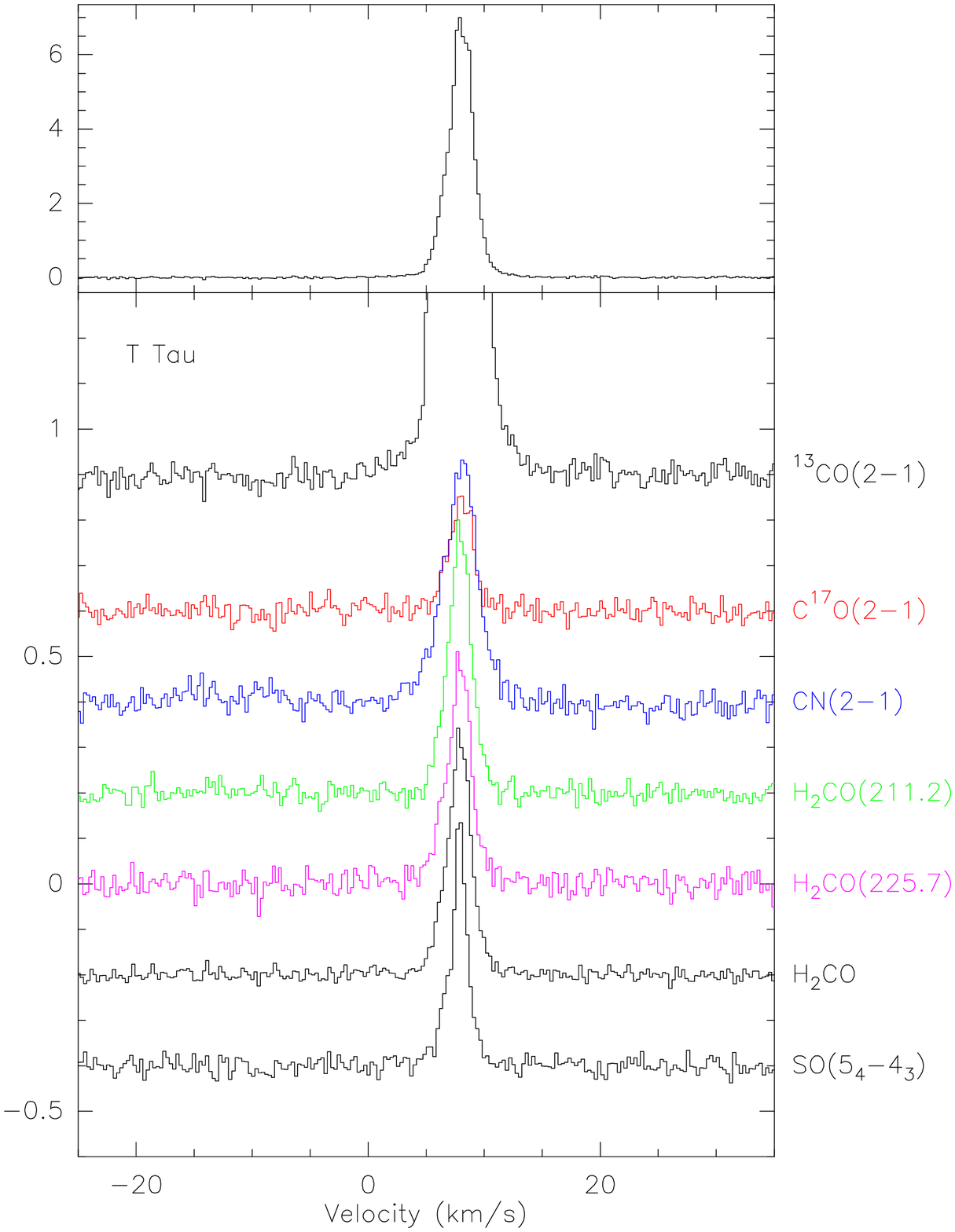}
\caption{Spectra of the observed transitions towards T Tau}
\label{fig:T_TAU}
\end{figure}
\begin{figure}
\includegraphics[height=11.5cm]{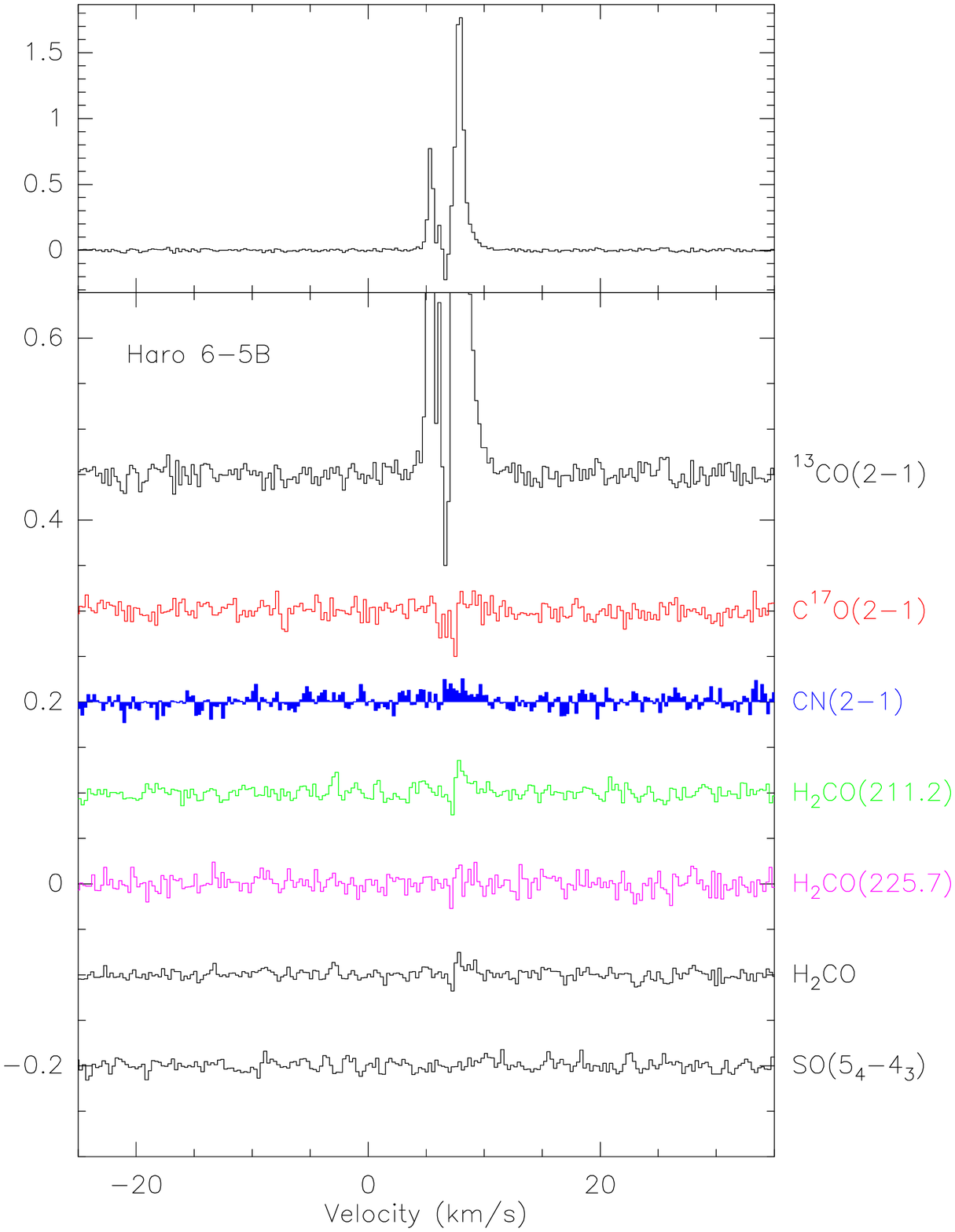}
\caption{Spectra of the observed transitions towards Haro 6-5 B}
\label{fig:HARO6-5B}
\end{figure}
\clearpage
\begin{figure}
\includegraphics[height=11.5cm]{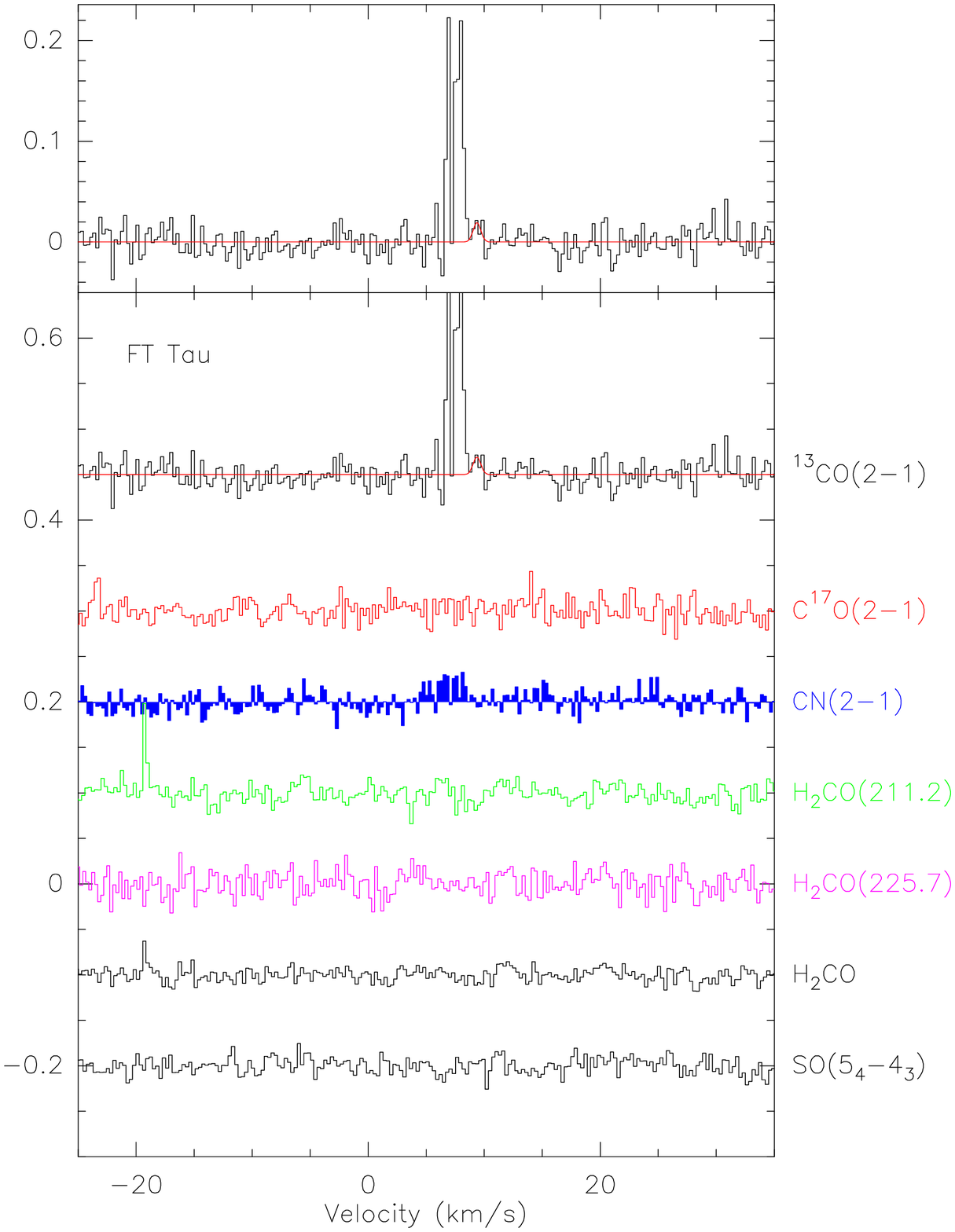}
\caption{Spectra of the observed transitions towards FT Tau}
\label{fig:FT_TAU}
\end{figure}
\begin{figure}
\includegraphics[height=11.5cm]{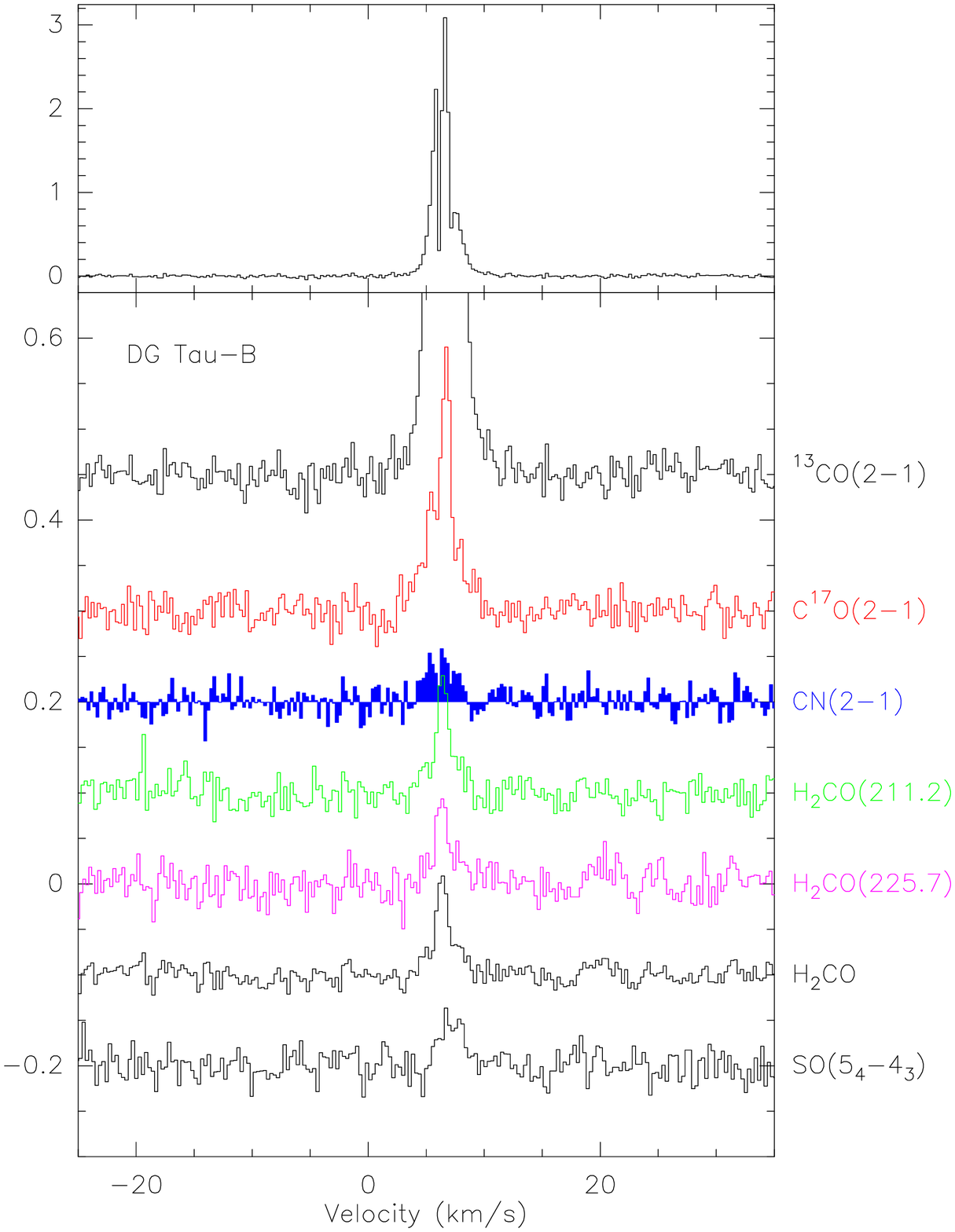}
\caption{Spectra of the observed transitions towards DG Tau B}
\label{fig:DG_TAU-B}
\end{figure}
\begin{figure}
\includegraphics[height=11.5cm]{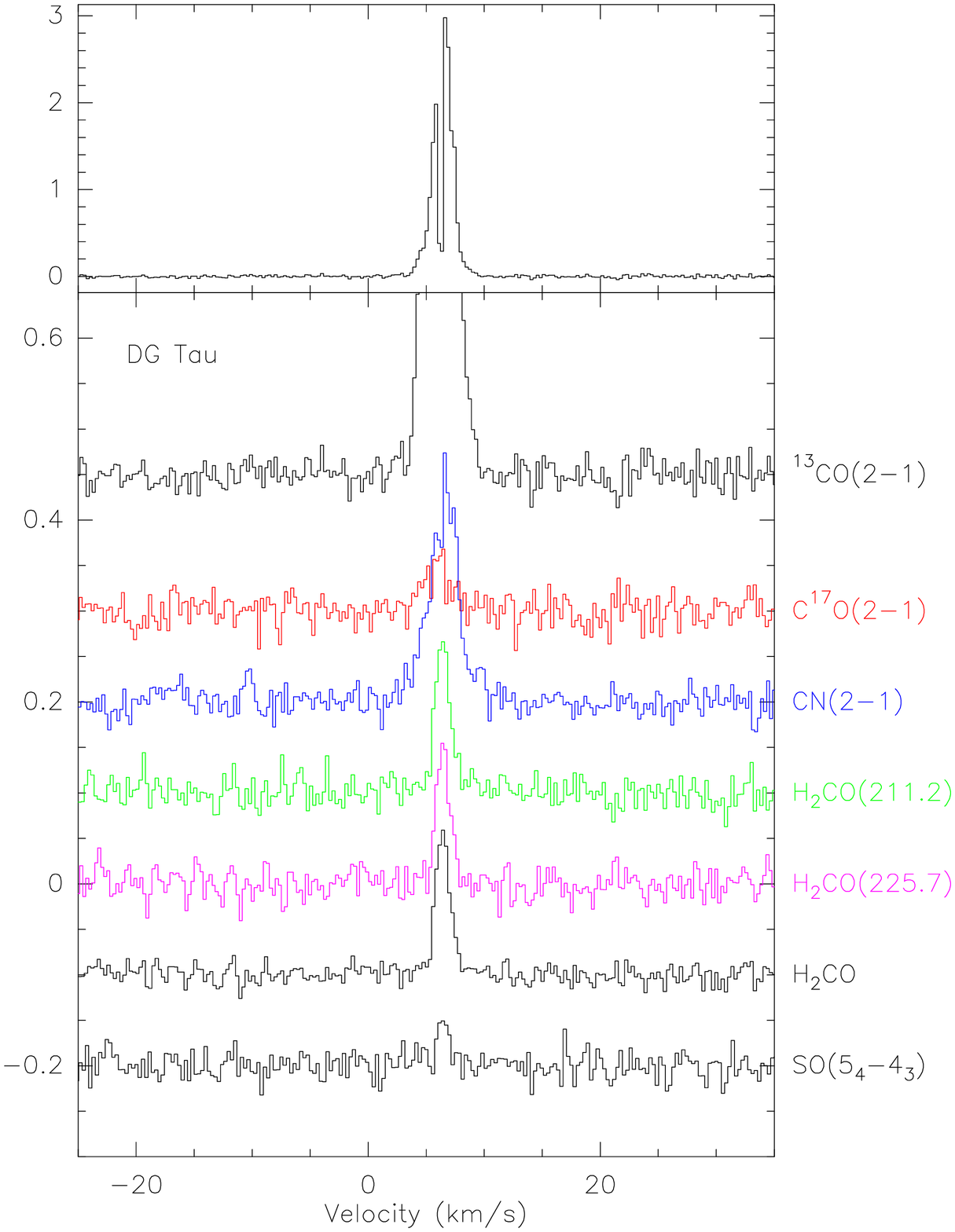}
\caption{Spectra of the observed transitions towards DG Tau}
\label{fig:DG_TAU}
\end{figure}
\begin{figure}
\includegraphics[height=11.5cm]{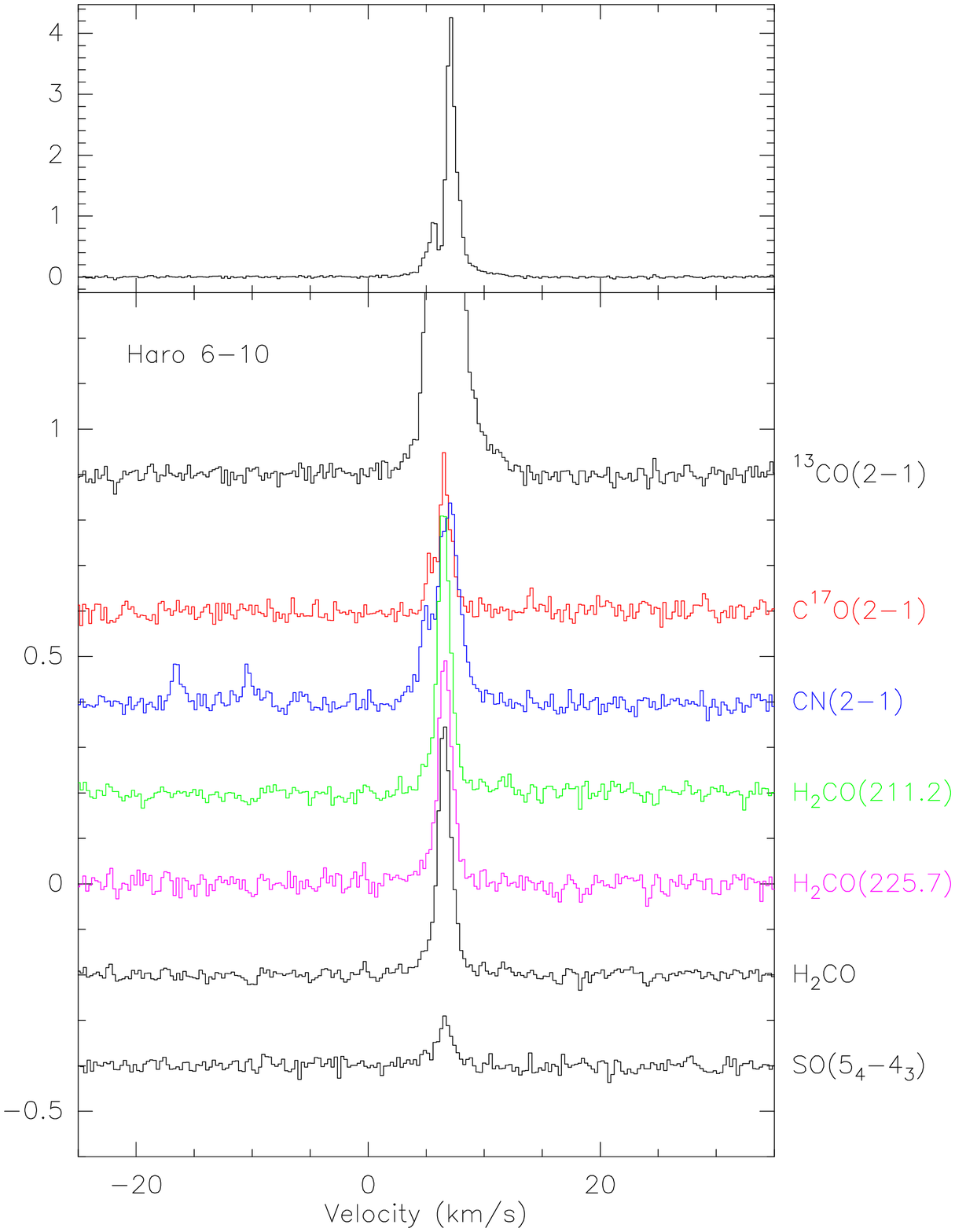}
\caption{Spectra of the observed transitions towards Haro 6-10}
\label{fig:HARO6-10N}
\end{figure}
\clearpage
\begin{figure}
\includegraphics[height=11.5cm]{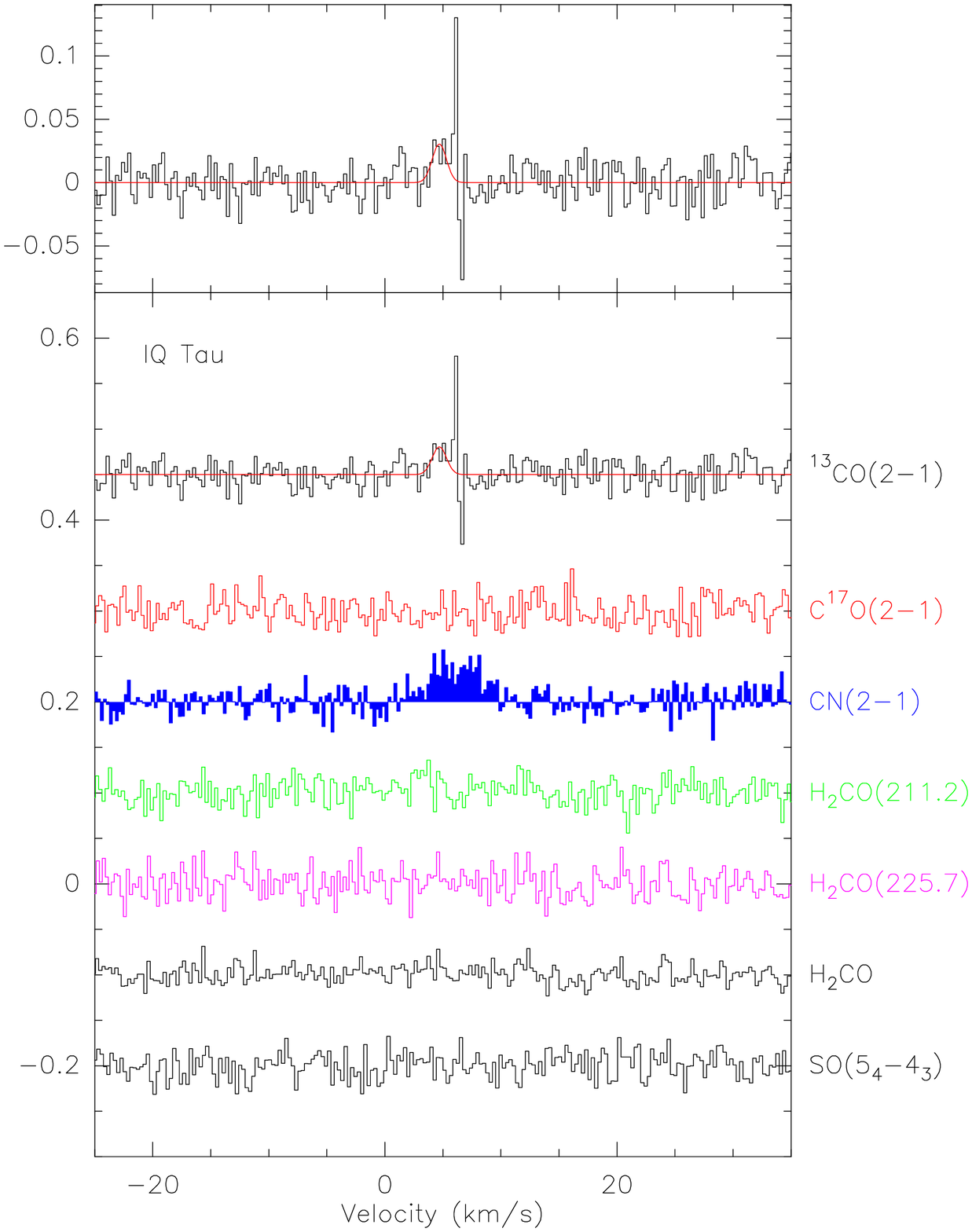}
\caption{Spectra of the observed transitions towards IQ Tau}
\label{fig:IQ_TAU}
\end{figure}
\begin{figure}
\includegraphics[height=11.5cm]{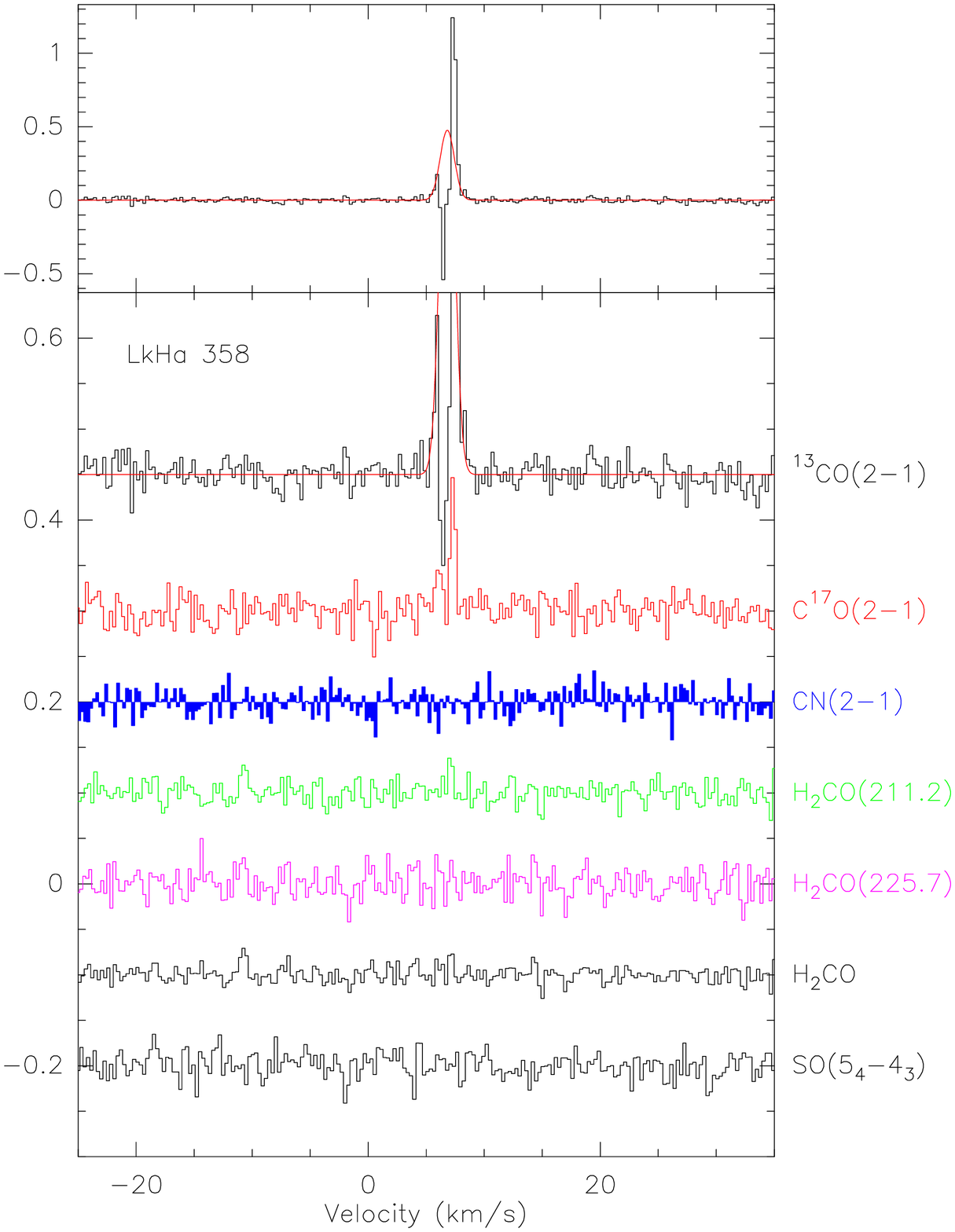}
\caption{Spectra of the observed transitions towards LkHa 358}
\label{fig:LKHA_358}
\end{figure}
\begin{figure}
\includegraphics[height=11.5cm]{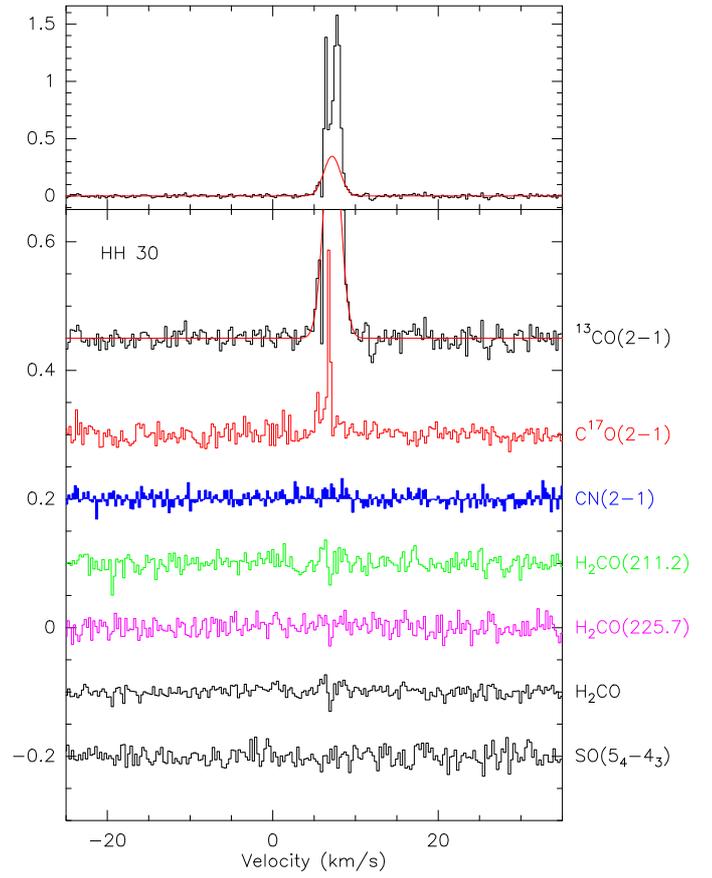}
\caption{Spectra of the observed transitions towards HH 30}
\label{fig:HH30}
\end{figure}
\begin{figure}
\includegraphics[height=11.5cm]{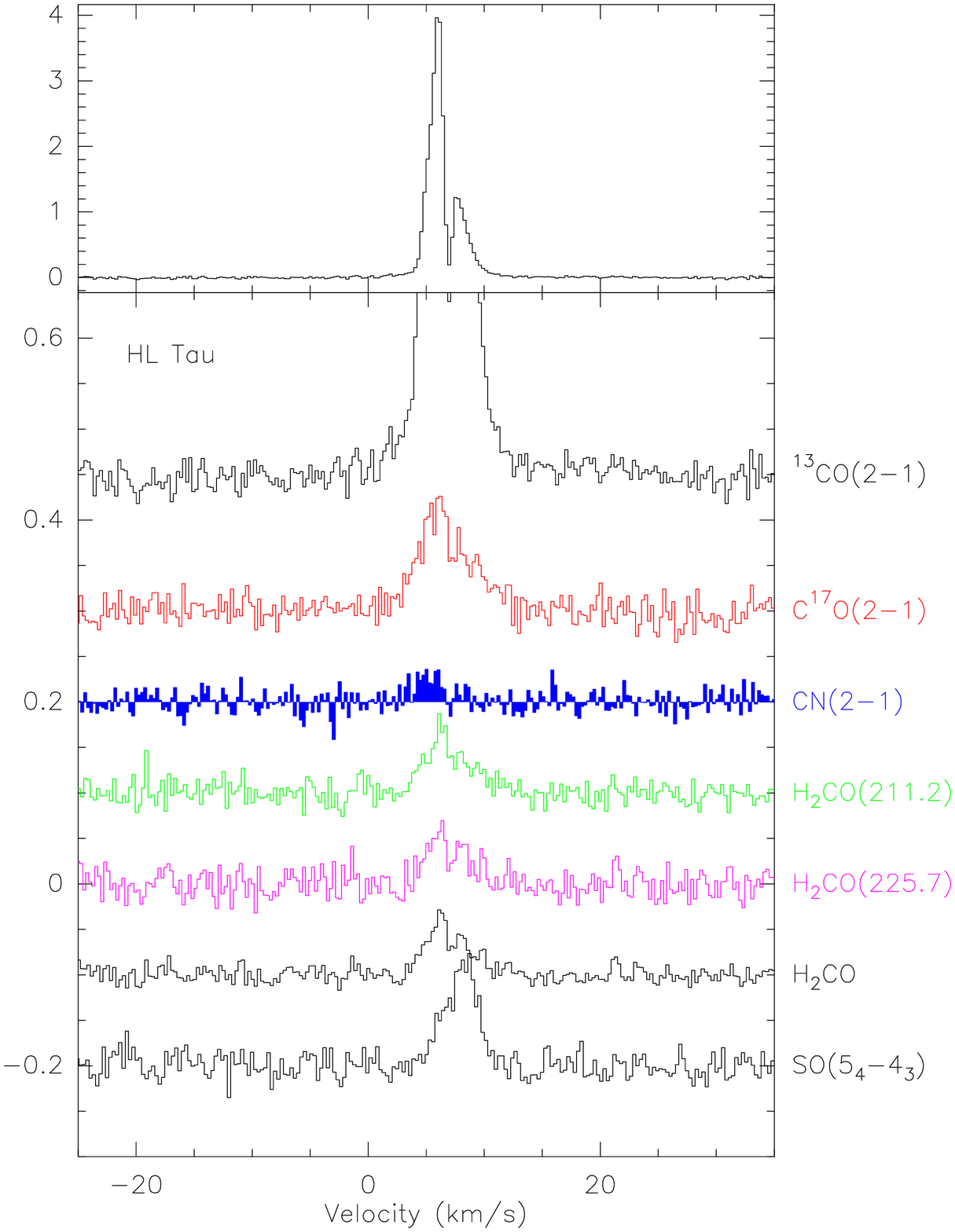}
\caption{Spectra of the observed transitions towards HL Tau}
\label{fig:HL_TAU}
\end{figure}
\clearpage
\begin{figure}
\includegraphics[height=11.5cm]{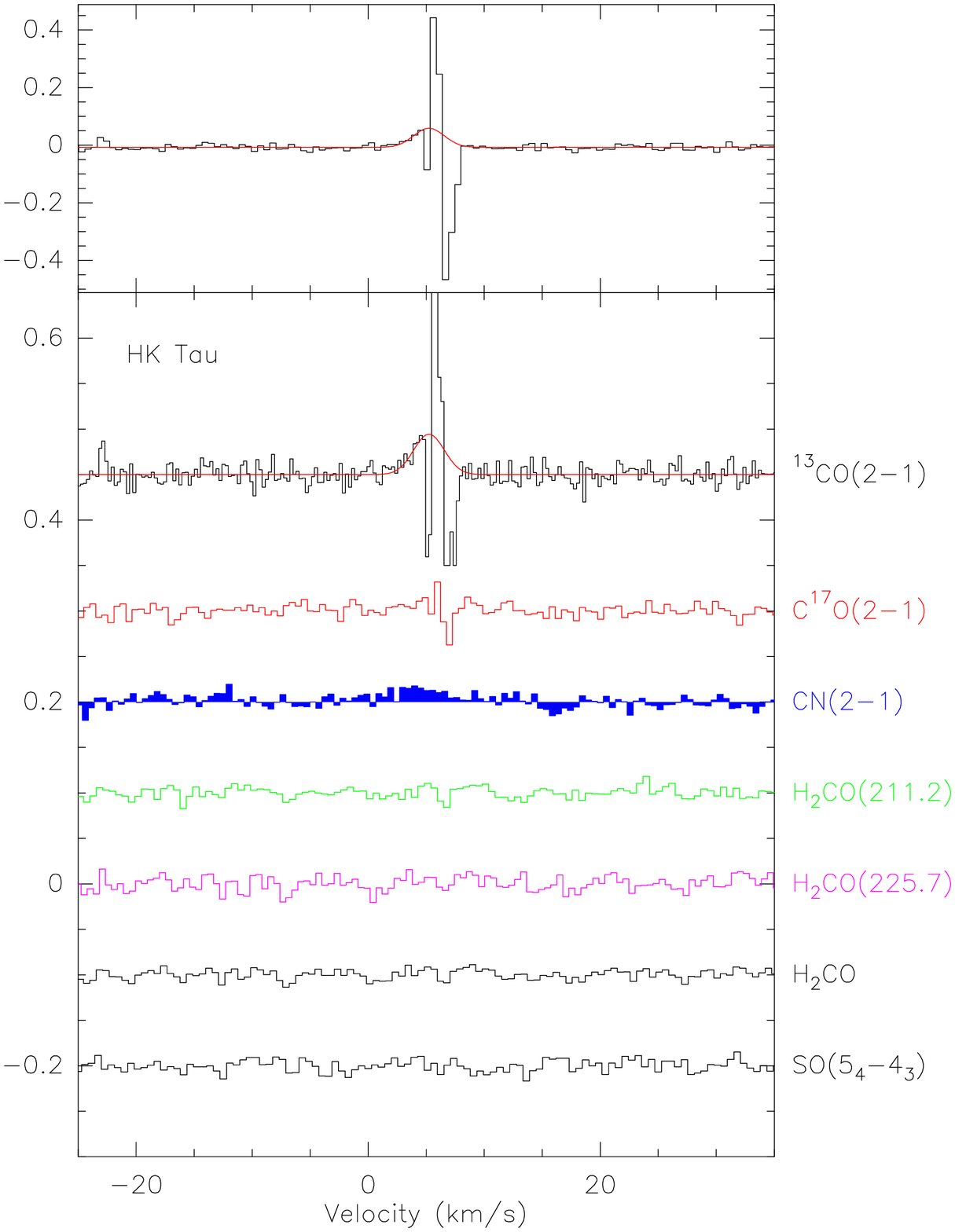}
\caption{Spectra of the observed transitions towards HK Tau}
\label{fig:HK_TAU}
\end{figure}
\begin{figure}
\includegraphics[height=11.5cm]{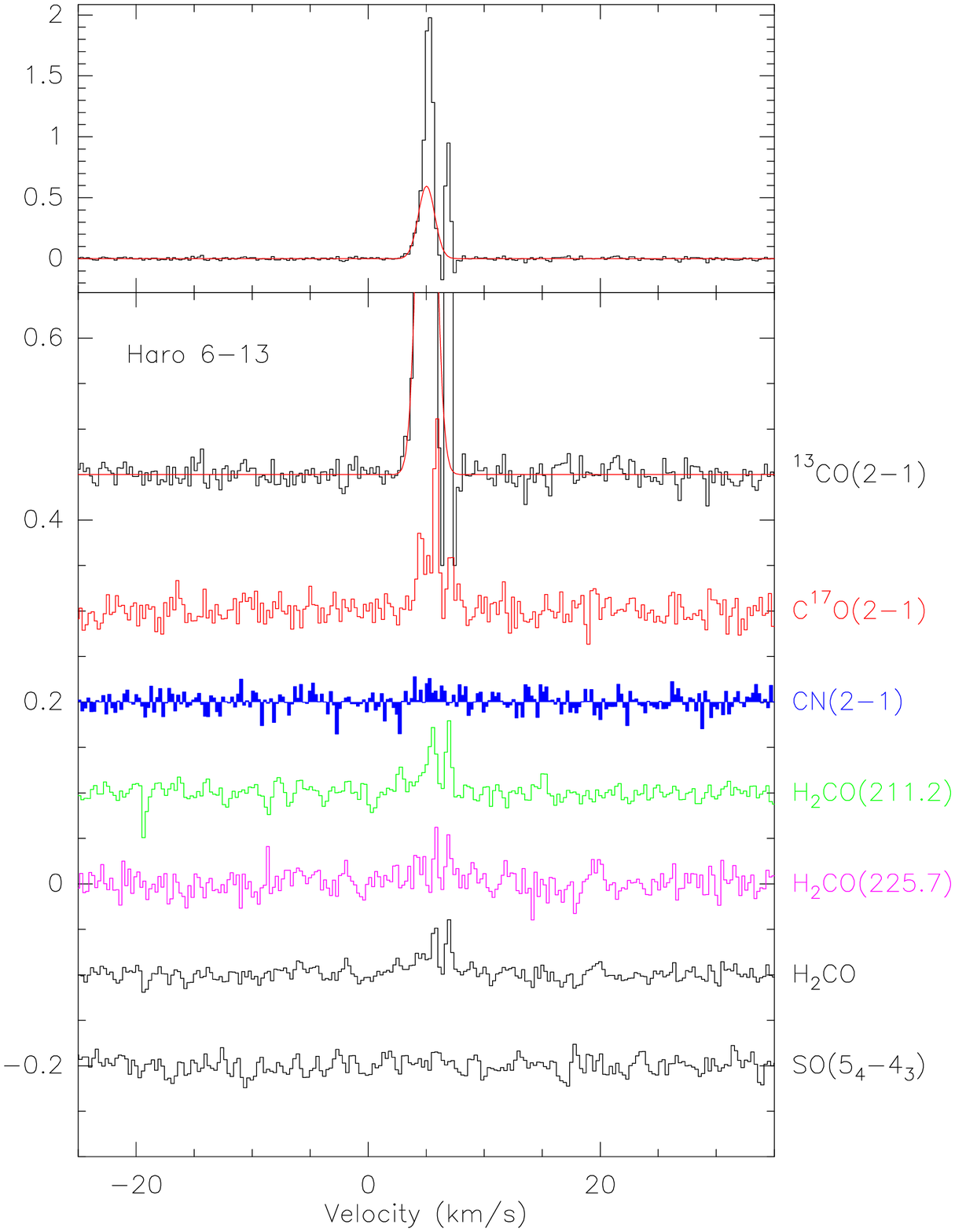}
\caption{Spectra of the observed transitions towards Haro 6-13}
\label{fig:HARO6-13}
\end{figure}
\begin{figure}
\includegraphics[height=11.5cm]{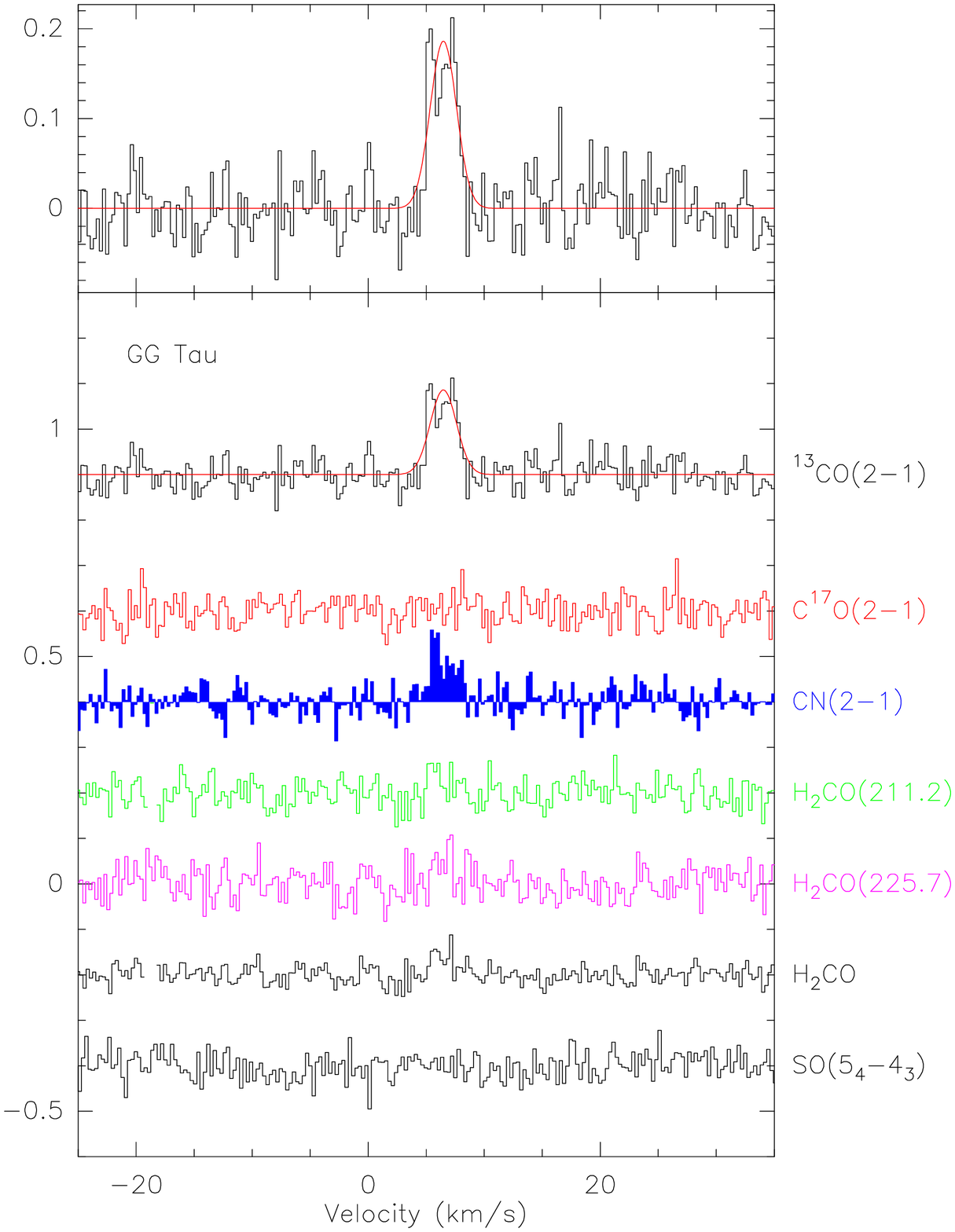}
\caption{Spectra of the observed transitions towards GG Tau}
\label{fig:GG_TAU}
\end{figure}
\begin{figure}
\includegraphics[height=11.5cm]{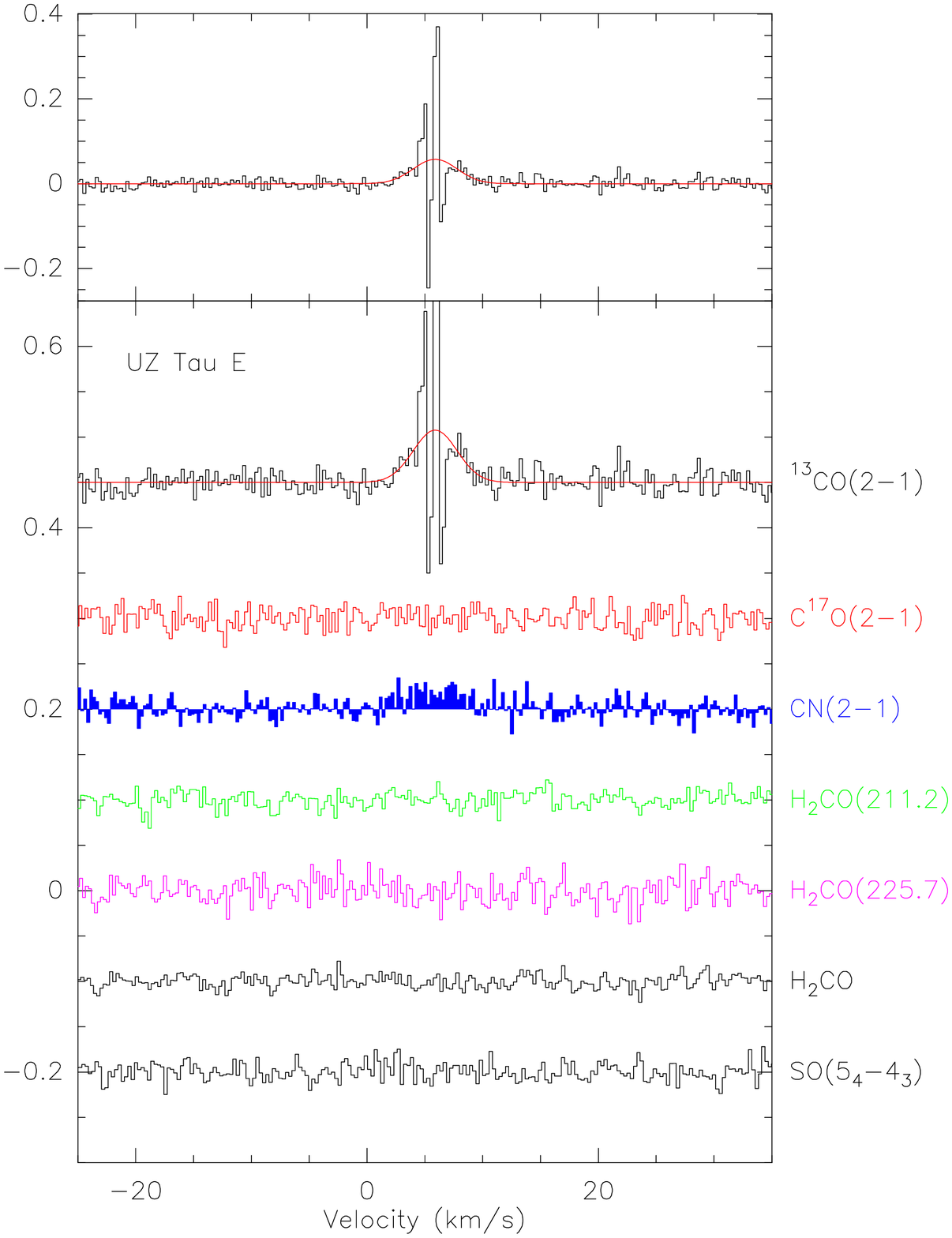}
\caption{Spectra of the observed transitions towards UZ Tau E}
\label{fig:UZTAU_E}
\end{figure}
\clearpage
\begin{figure}
\includegraphics[height=11.5cm]{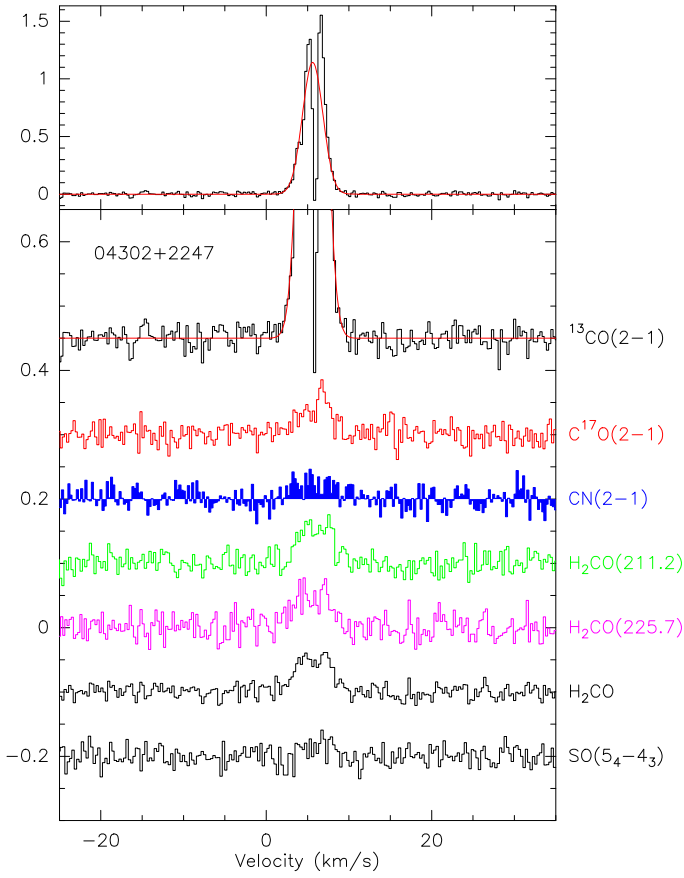}
\caption{Spectra towards 04302+2247}
\label{fig:04302+2247}
\end{figure}
\begin{figure}
\includegraphics[height=11.5cm]{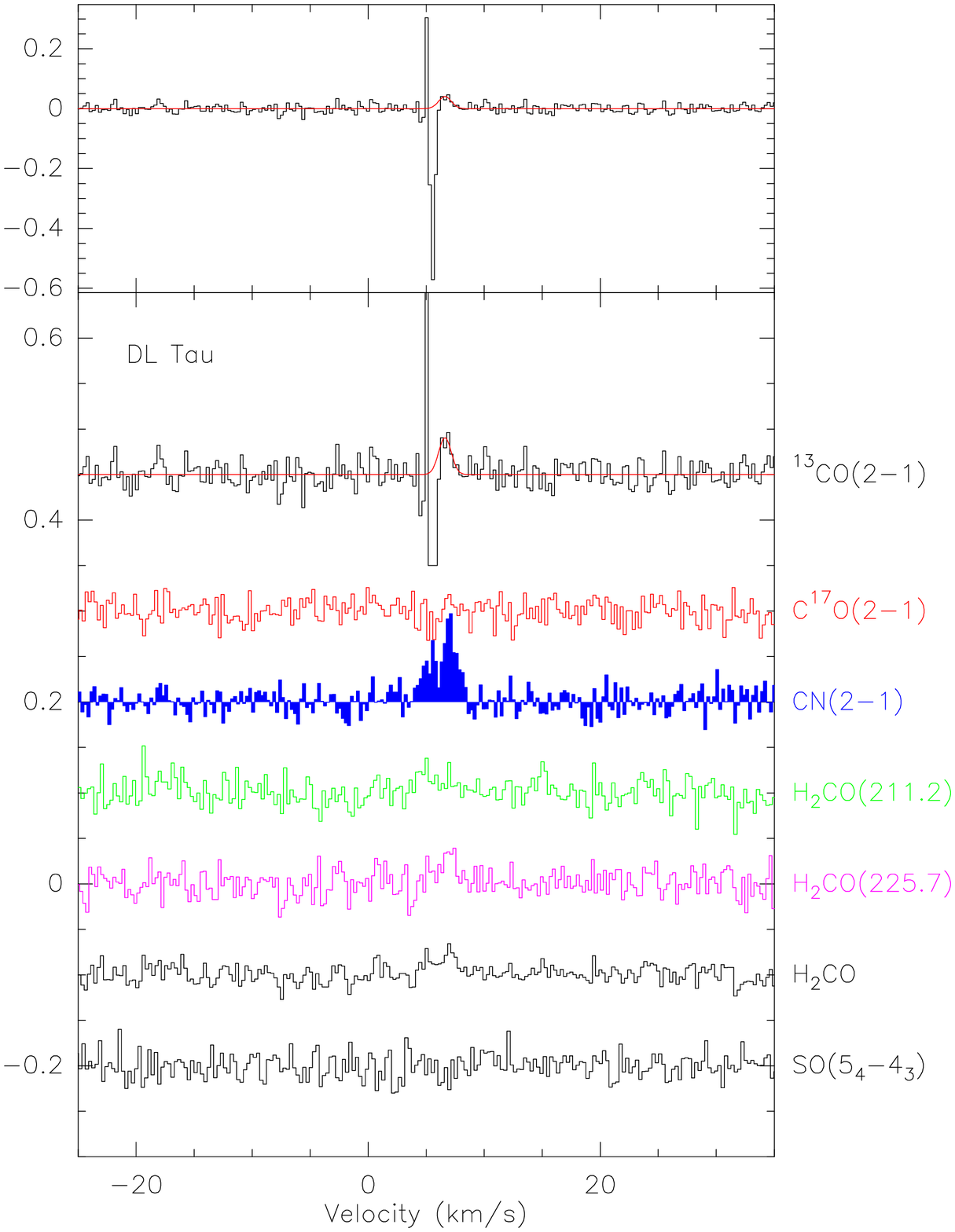}
\caption{Spectra of the observed transitions towards DL Tau}
\label{fig:DL_TAU}
\end{figure}
\begin{figure}
\includegraphics[height=11.5cm]{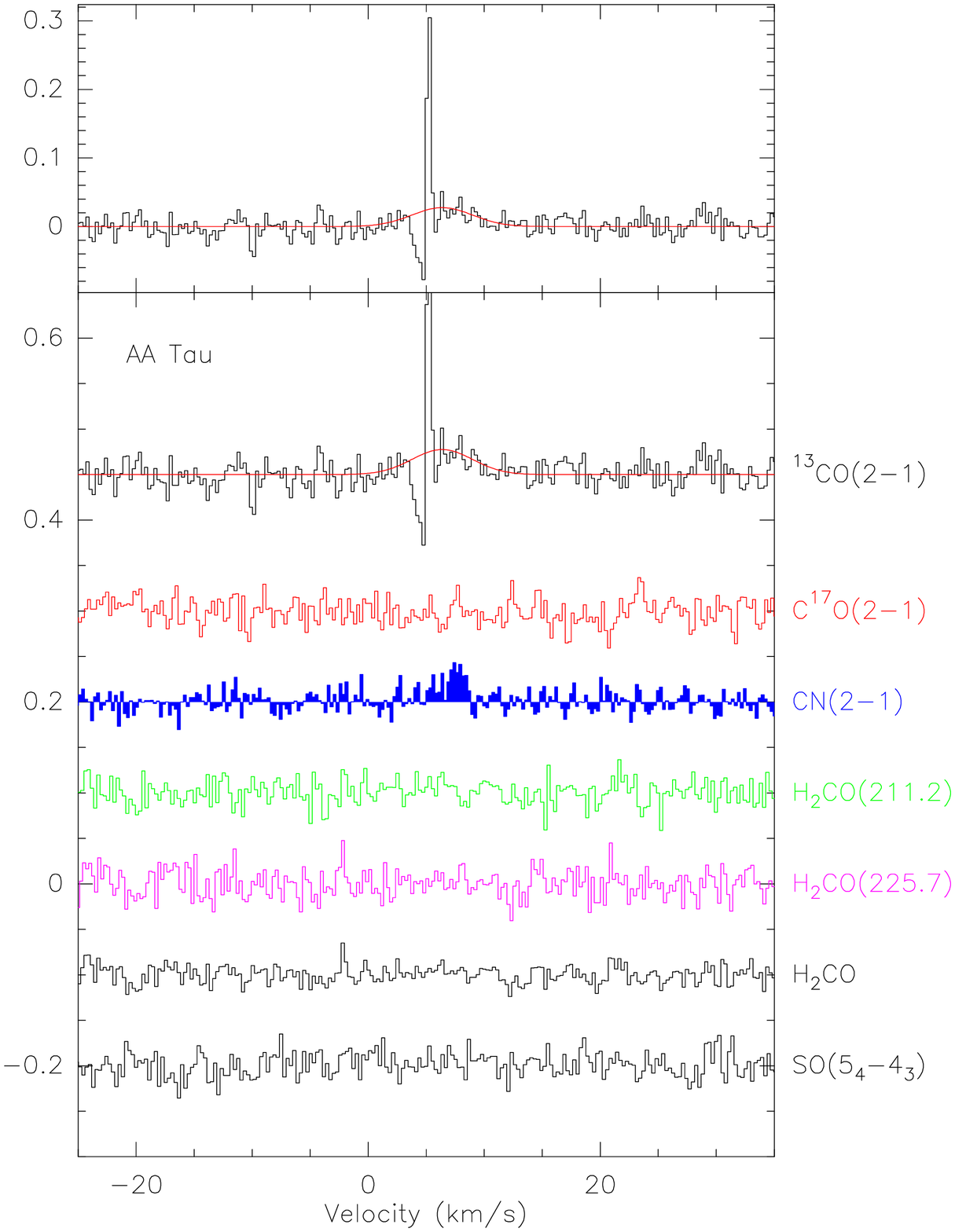}
\caption{Spectra of the observed transitions towards AA Tau}
\label{fig:AA_TAU}
\end{figure}
\begin{figure}
\includegraphics[height=11.5cm]{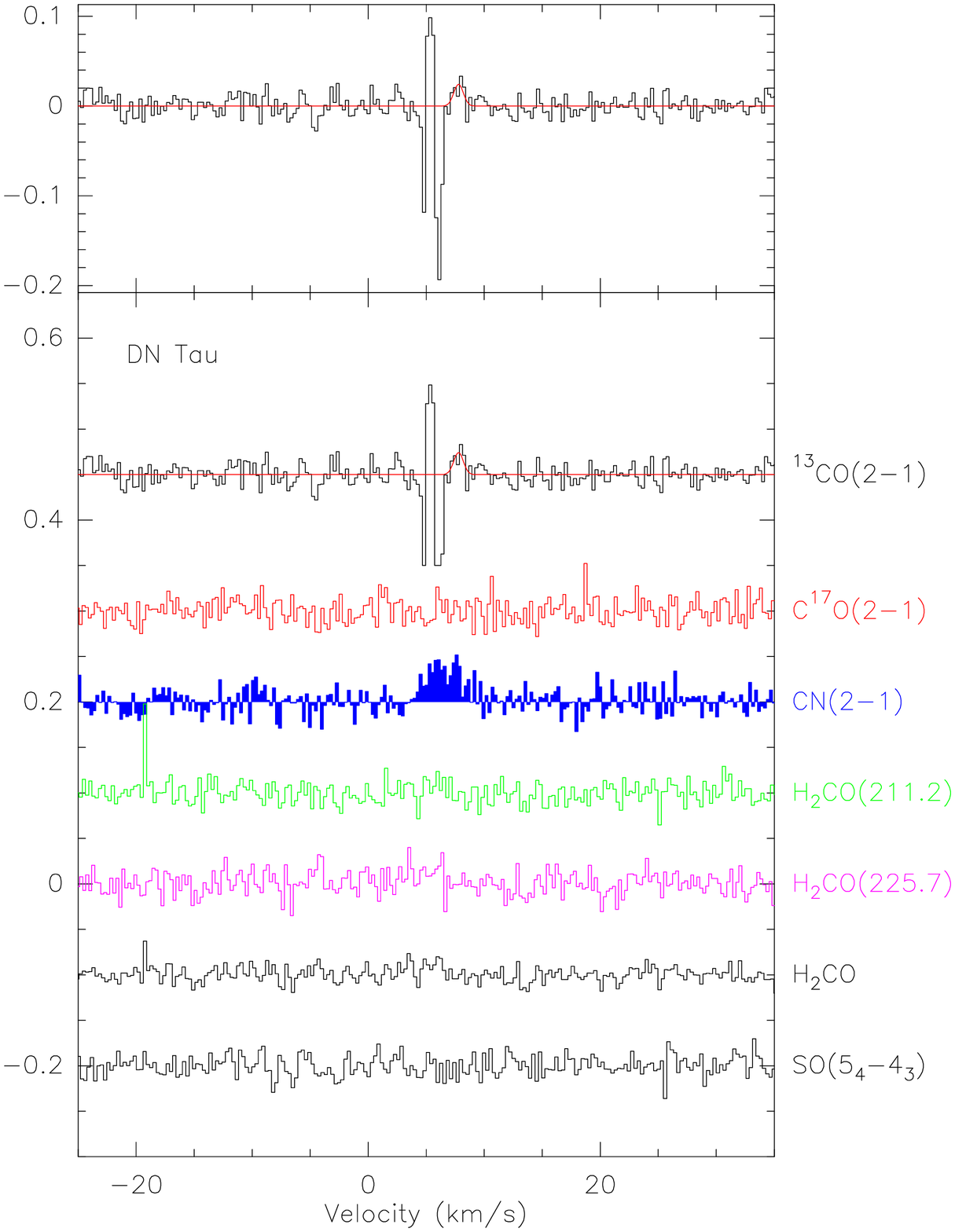}
\caption{Spectra of the observed transitions towards DN Tau}
\label{fig:DN_TAU}
\end{figure}
\clearpage
\begin{figure}
\includegraphics[height=11.5cm]{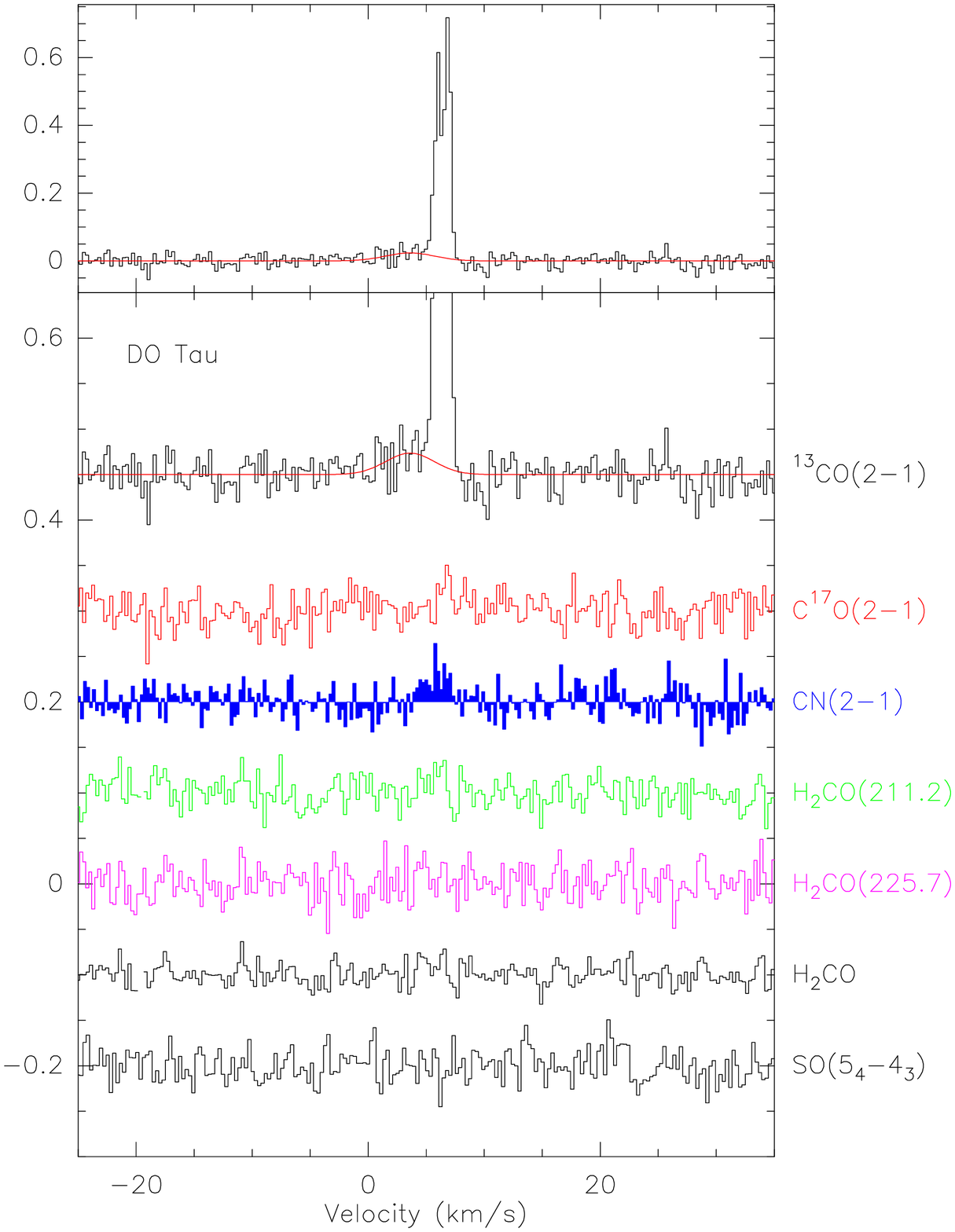}
\caption{Spectra of the observed transitions towards DO Tau}
\label{fig:DO_TAU}
\end{figure}
\begin{figure}
\includegraphics[height=11.5cm]{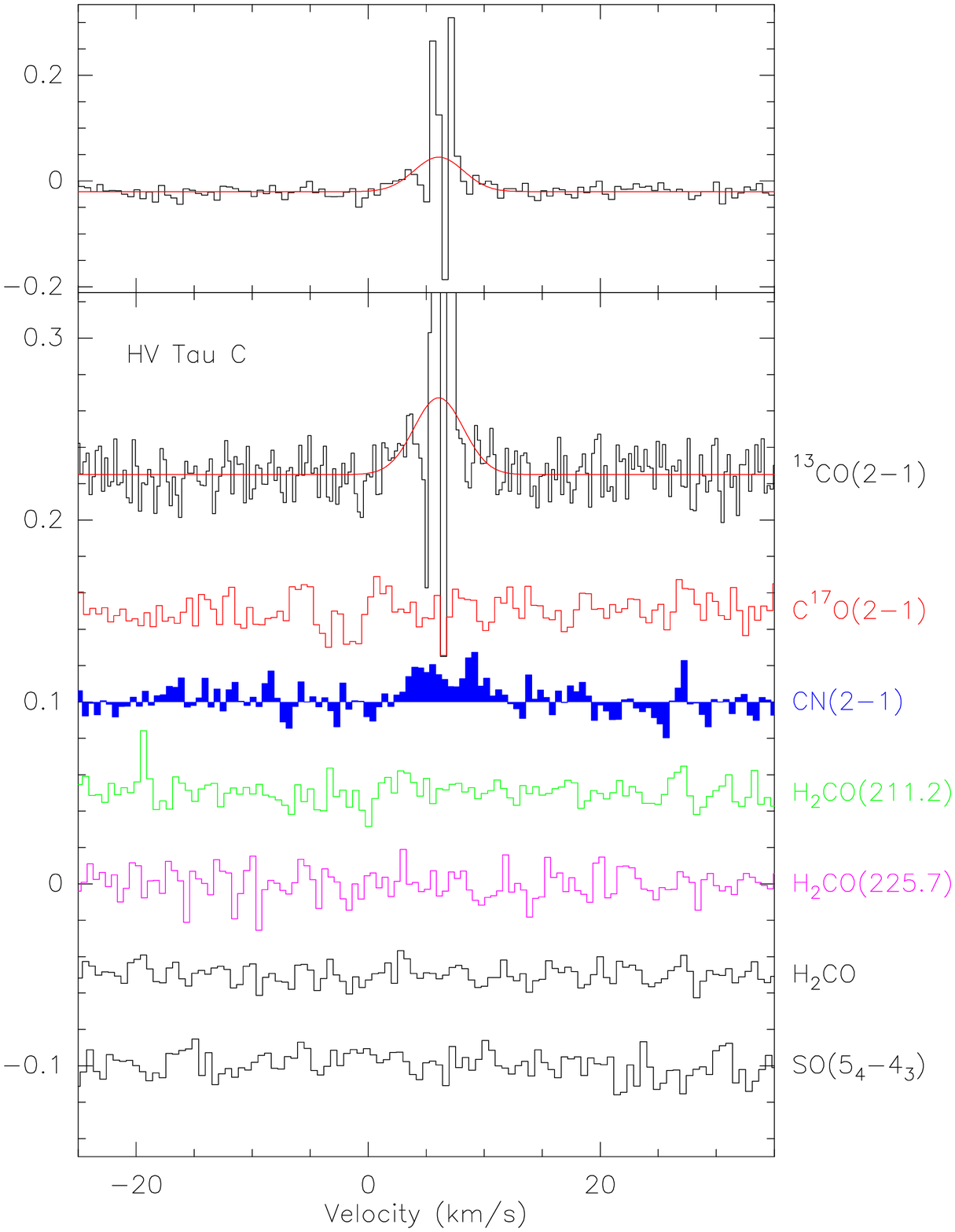}
\caption{Spectra of the observed transitions towards HV Tau C}
\label{fig:HV_TAU-C}
\end{figure}
\begin{figure}
\includegraphics[height=11.5cm]{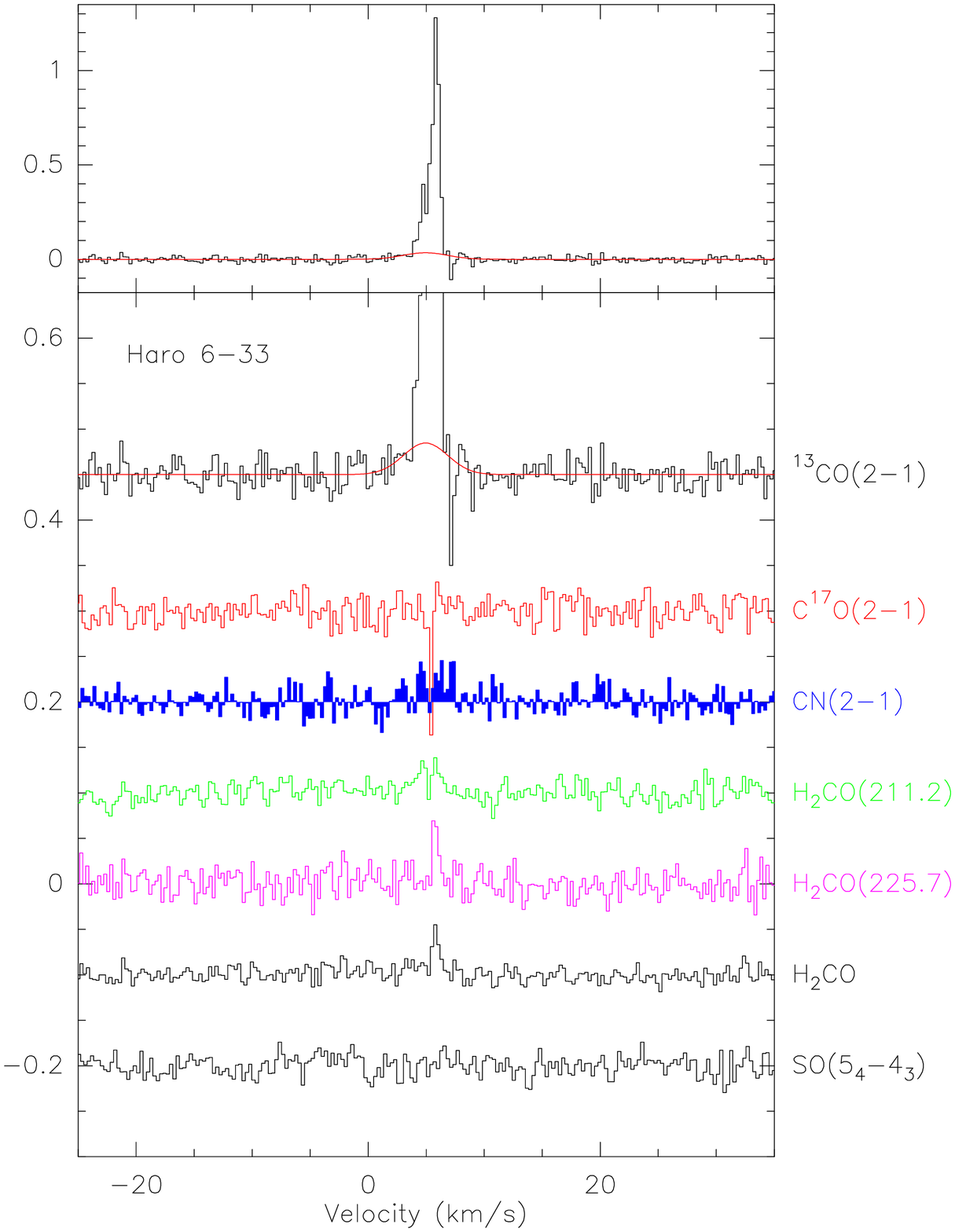}
\caption{Spectra of the observed transitions towards Haro 6-33}
\label{fig:HARO6-33}
\end{figure}
\begin{figure}
\includegraphics[height=11.5cm]{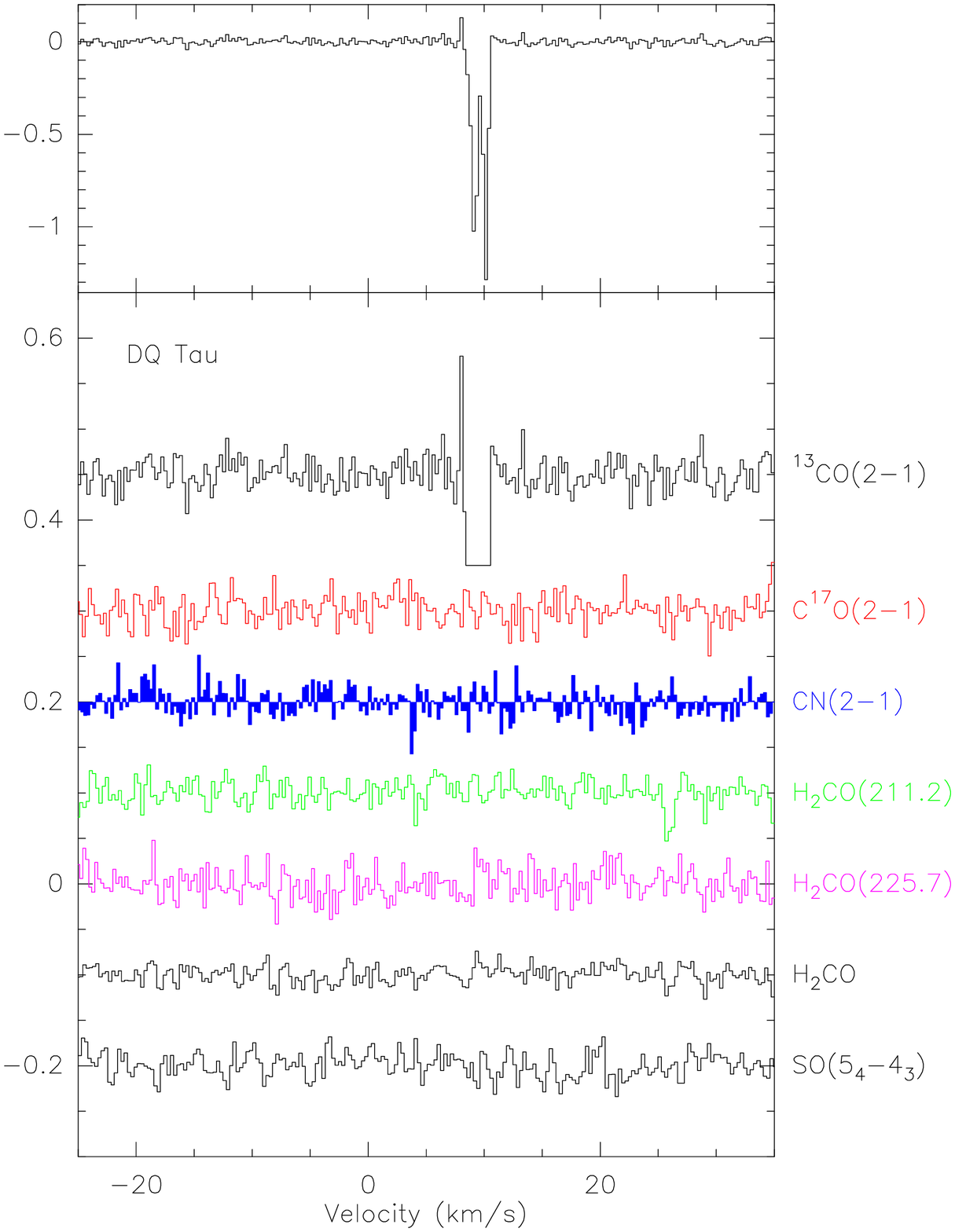}
\caption{Spectra of the observed transitions towards DQ Tau}
\label{fig:DQ_TAU}
\end{figure}
\clearpage
\begin{figure}
\includegraphics[height=11.5cm]{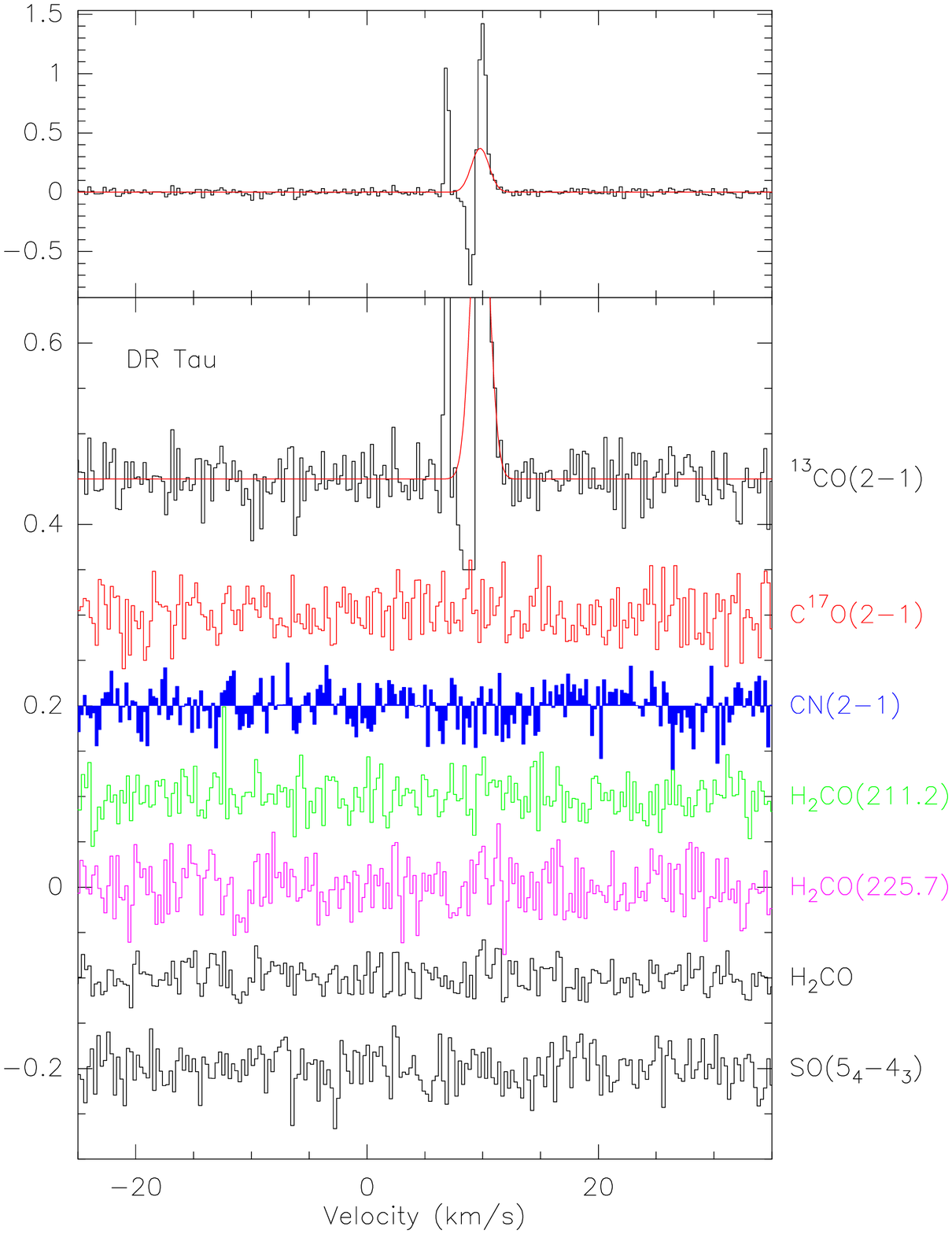}
\caption{Spectra of the observed transitions towards DR Tau}
\label{fig:DR_TAU}
\end{figure}
\begin{figure}
\includegraphics[height=11.5cm]{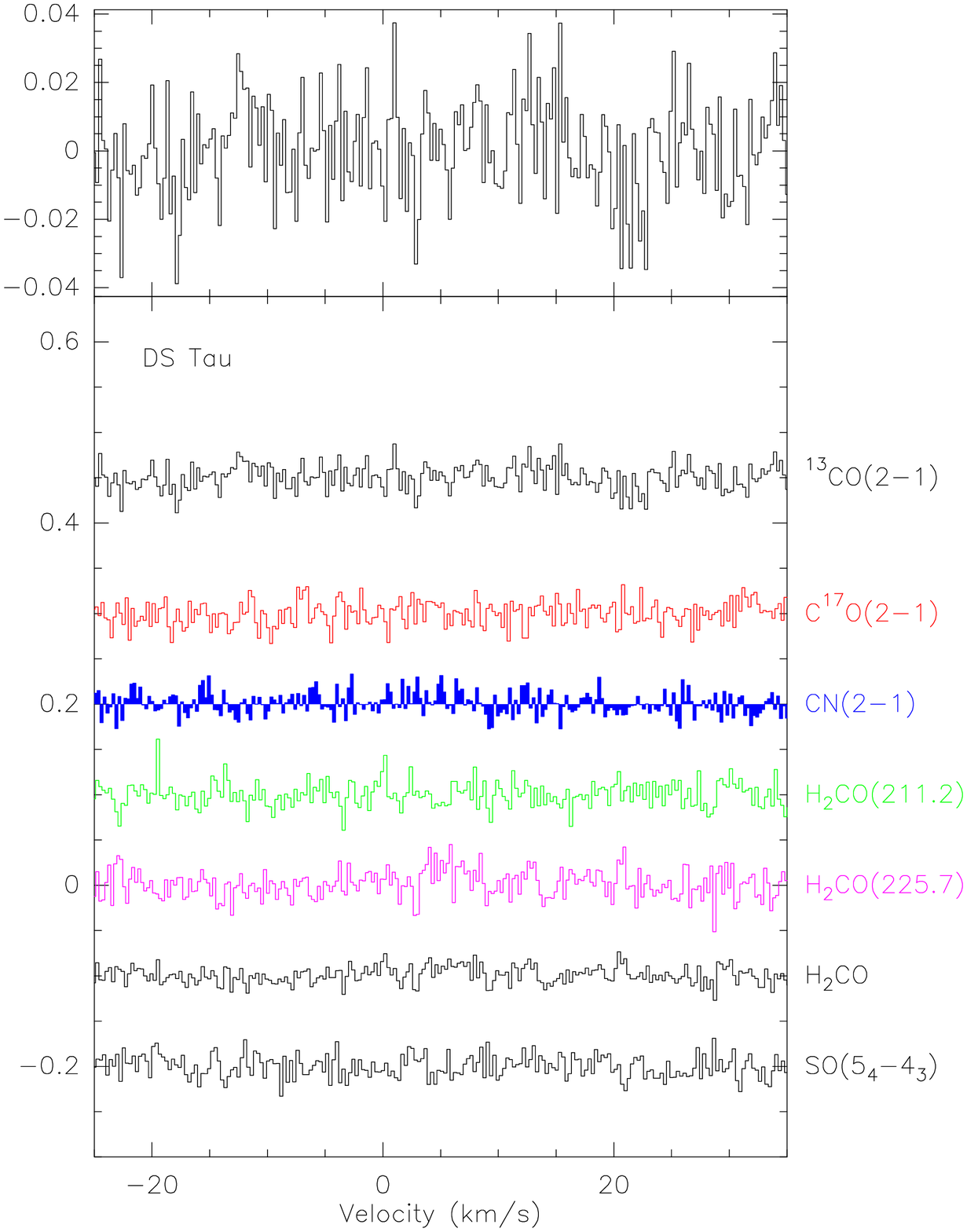}
\caption{Spectra of the observed transitions towards DS Tau}
\label{fig:DS_TAU}
\end{figure}
\begin{figure}
\includegraphics[height=11.5cm]{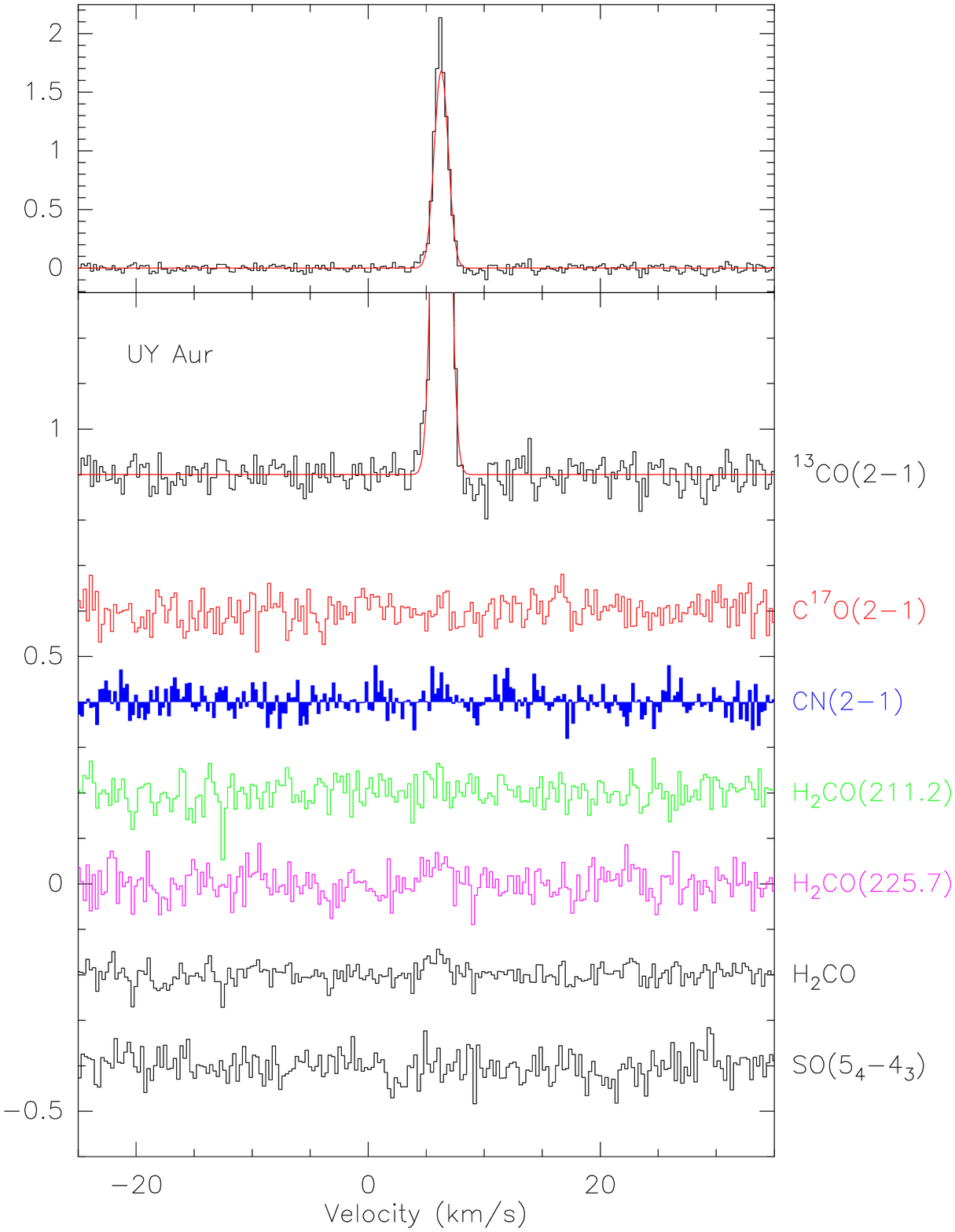}
\caption{Spectra of the observed transitions towards UY Aur}
\label{fig:UY_AUR}
\end{figure}
\begin{figure}
\includegraphics[height=11.5cm]{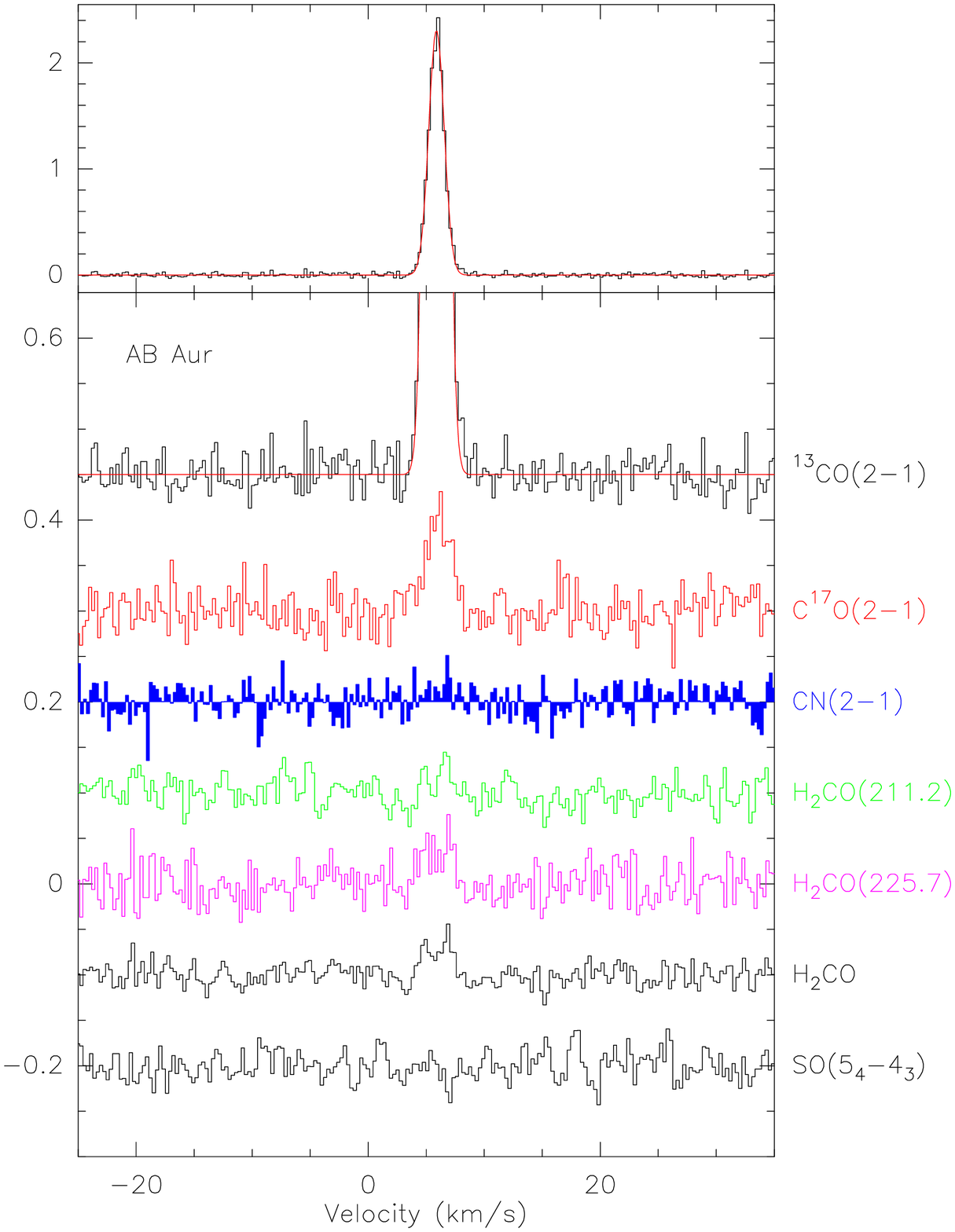}
\caption{Spectra of the observed transitions towards AB Aur}
\label{fig:AB_AUR}
\end{figure}
\clearpage
\begin{figure}
\includegraphics[height=11.5cm]{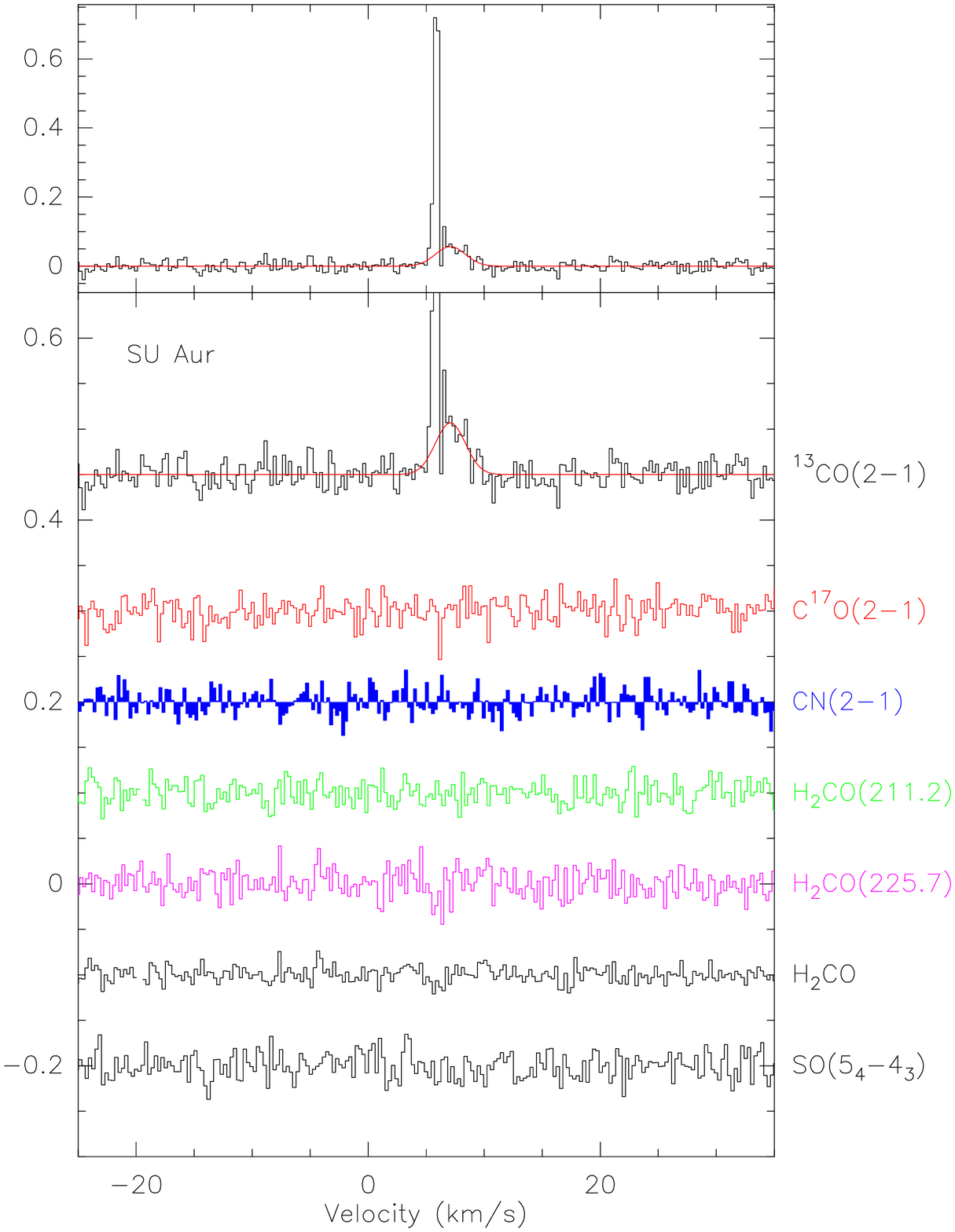}
\caption{Spectra of the observed transitions towards SU Aur}
\label{fig:SU_AUR}
\end{figure}
\begin{figure}
\includegraphics[height=11.5cm]{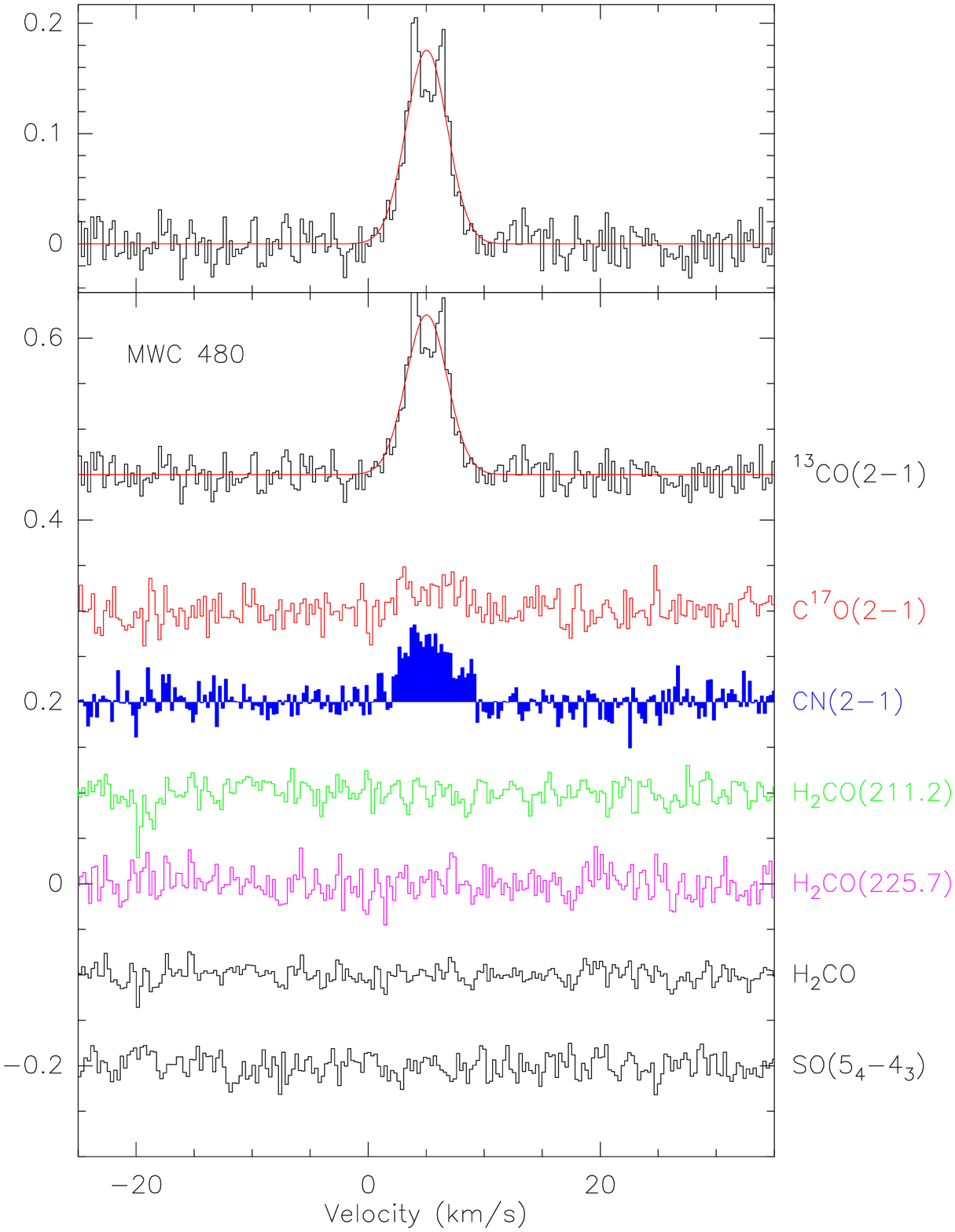}
\caption{Spectra of the observed transitions towards MWC 480}
\label{fig:MWC_480}
\end{figure}
\begin{figure}
\includegraphics[height=11.5cm]{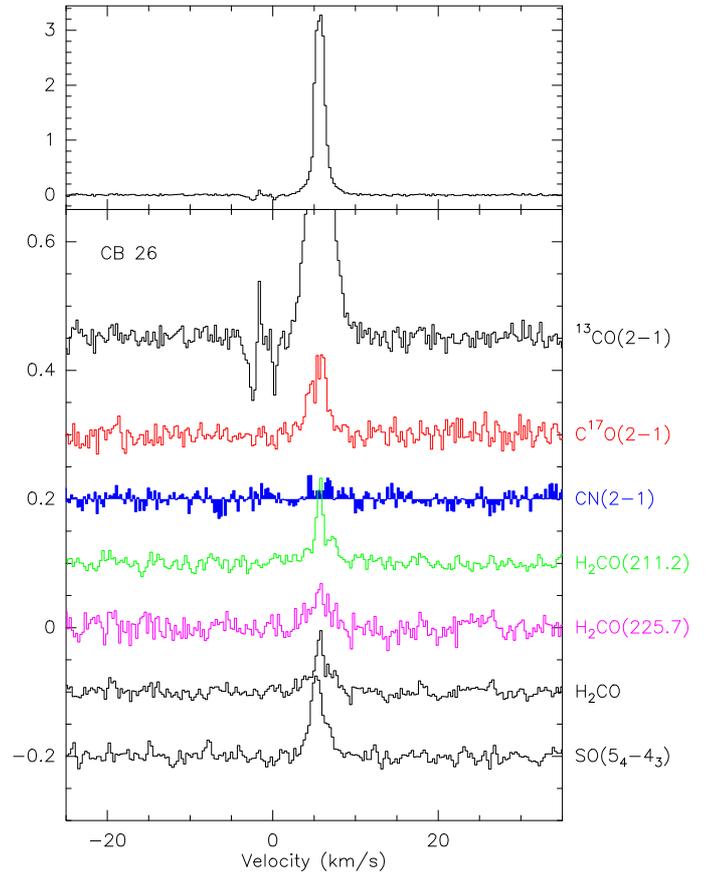}
\caption{Spectra of the observed transitions towards CB 26}
\label{fig:CB_26}
\end{figure}
\begin{figure}
\includegraphics[height=11.5cm]{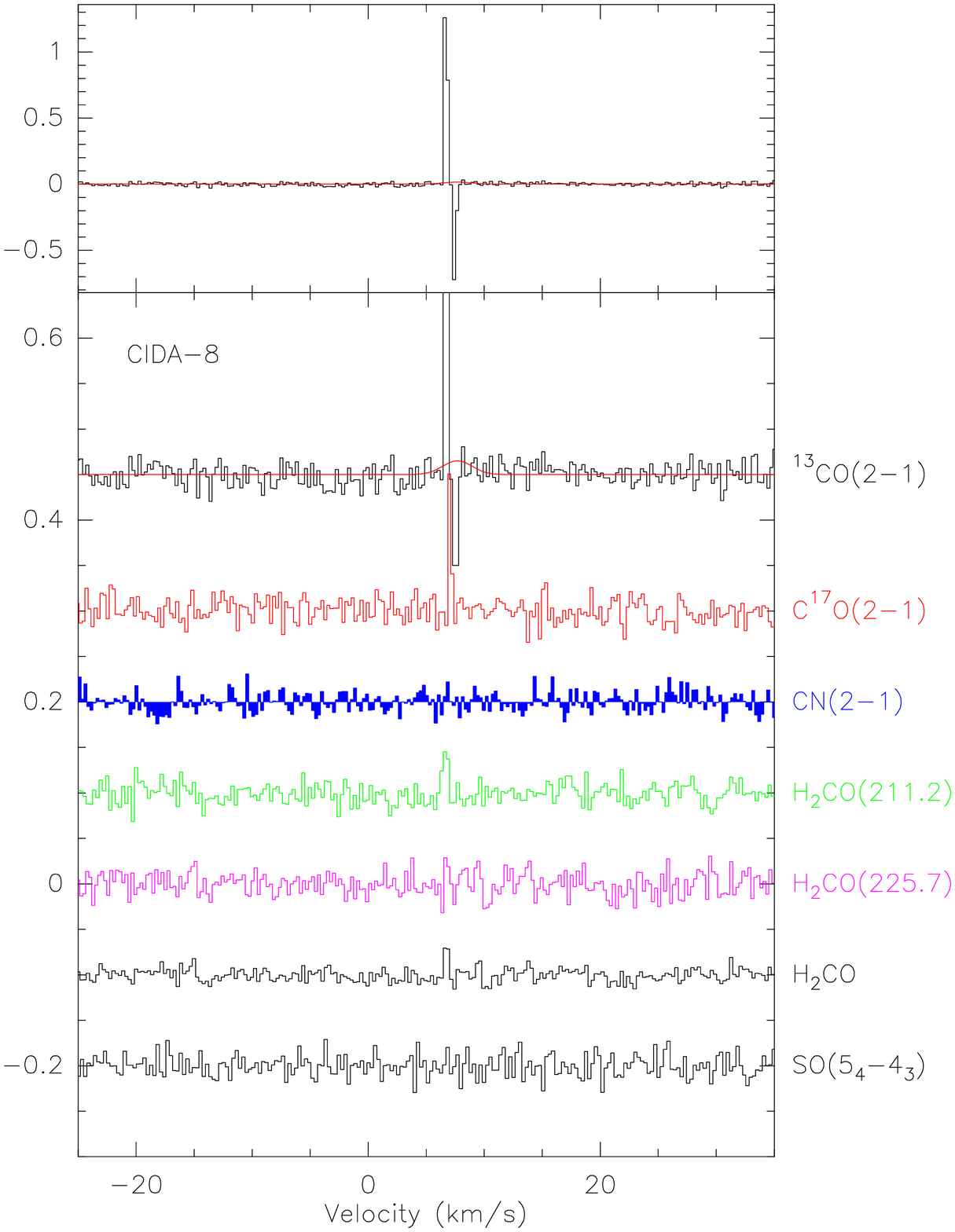}
\caption{Spectra of the observed transitions towards CIDA-8}
\label{fig:CIDA-8}
\end{figure}
\clearpage
\begin{figure}
\includegraphics[height=11.5cm]{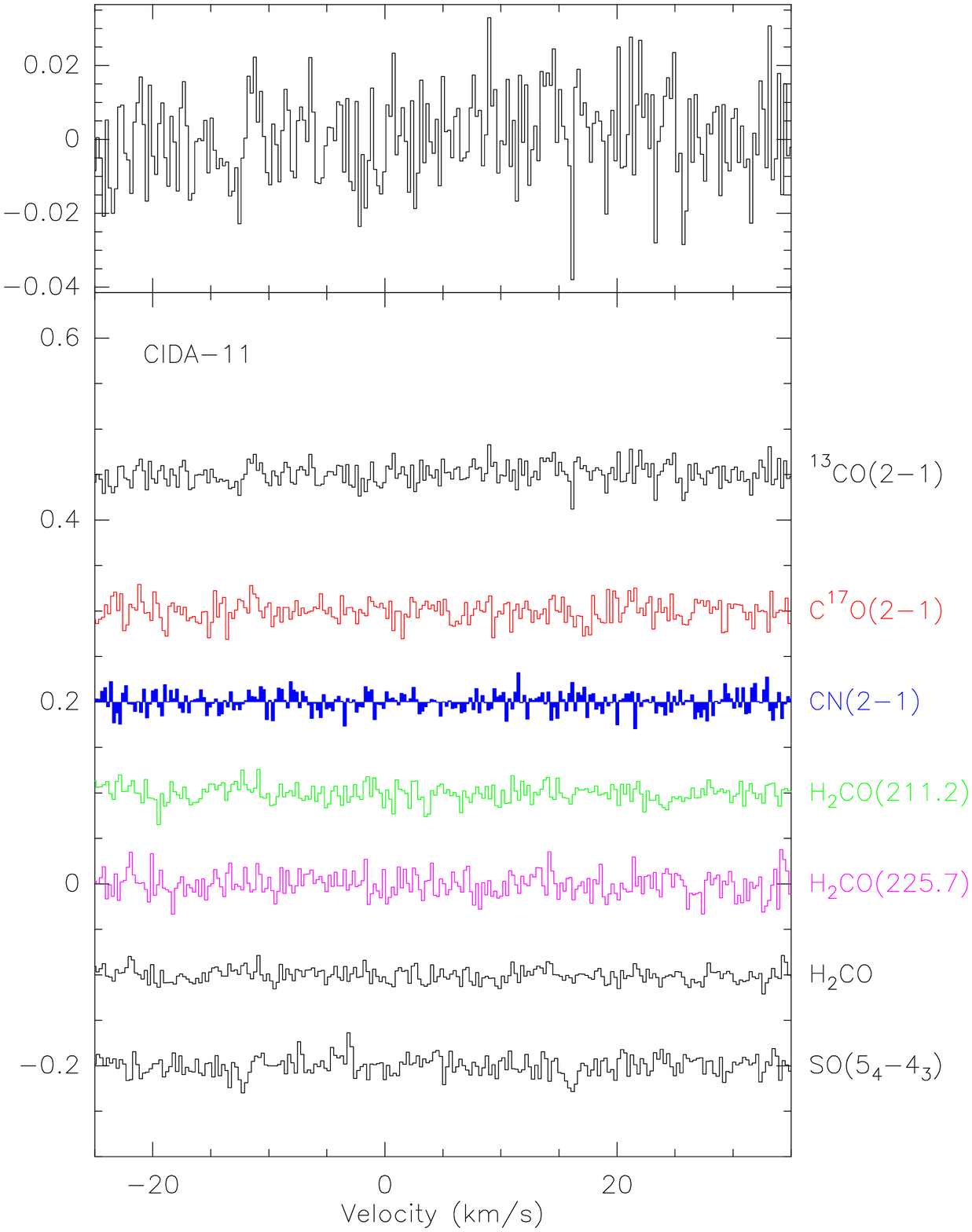}
\caption{Spectra of the observed transitions towards CIDA-11}
\label{fig:CIDA-11}
\end{figure}
\begin{figure}
\includegraphics[height=11.5cm]{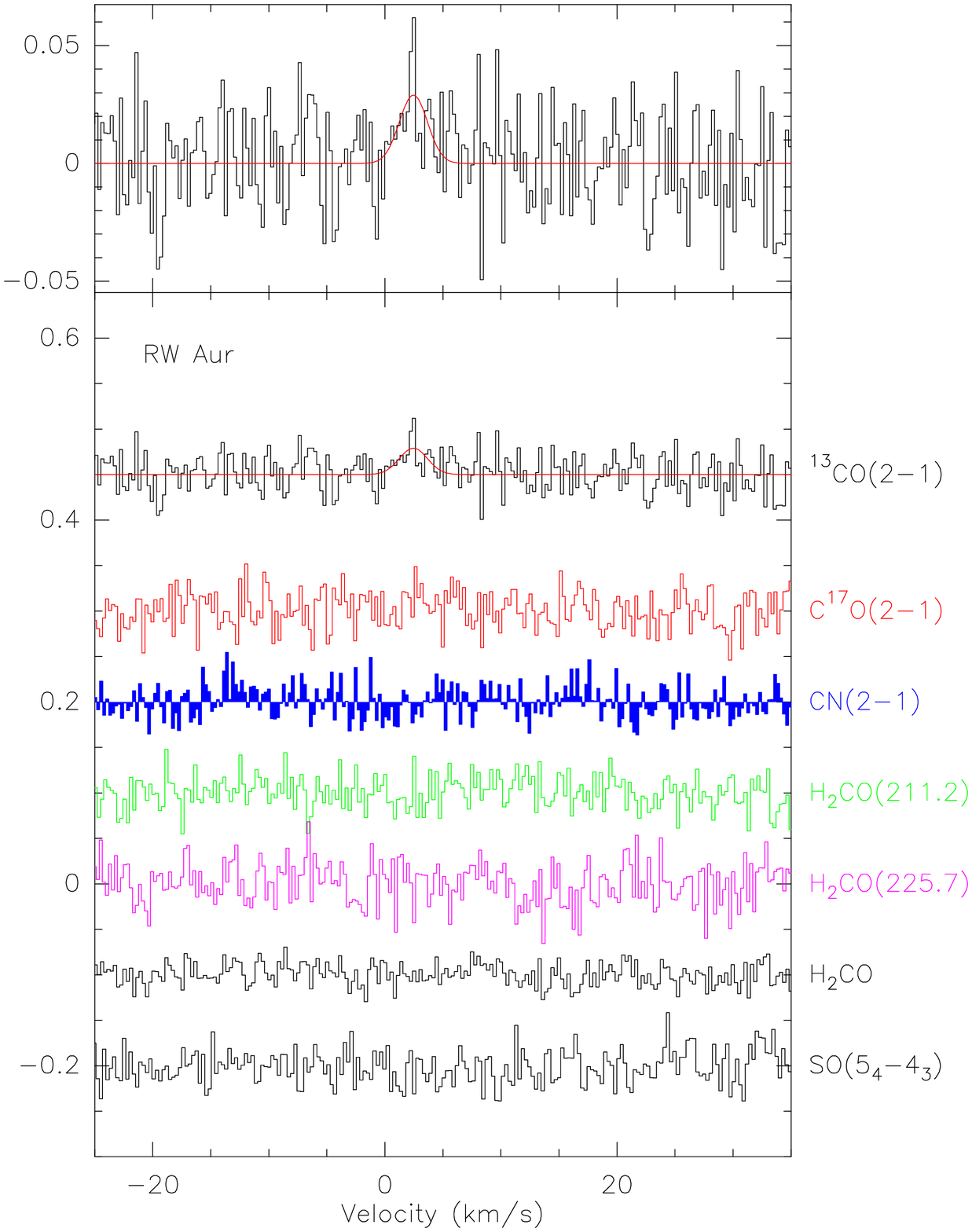}
\caption{Spectra of the observed transitions towards RW Aur}
\label{fig:RW_AUR}
\end{figure}
\begin{figure}
\includegraphics[height=11.5cm]{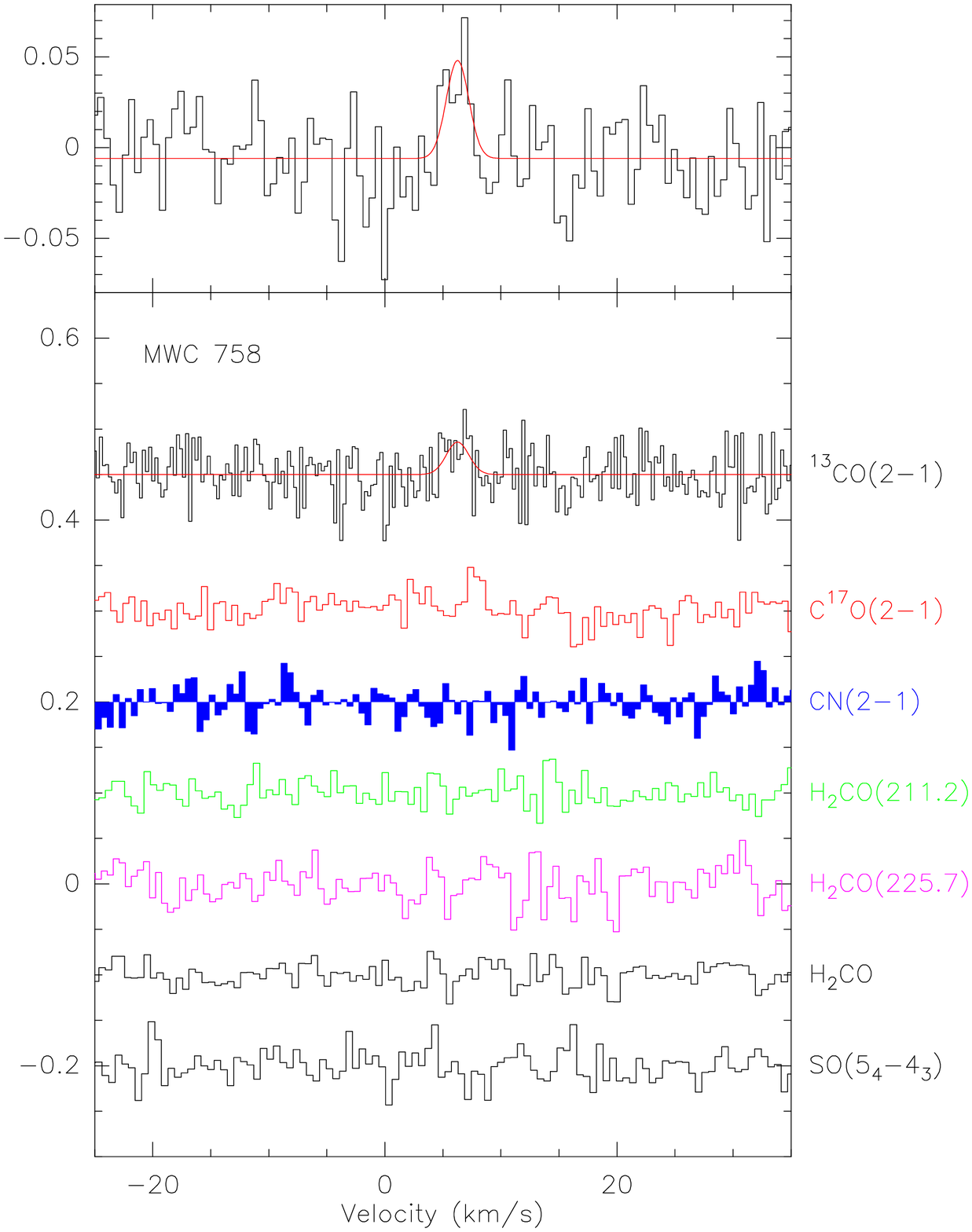}
\caption{Spectra of the observed transitions towards MWC 758}
\label{fig:MWC_758}
\end{figure}
\begin{figure}
\includegraphics[height=11.5cm]{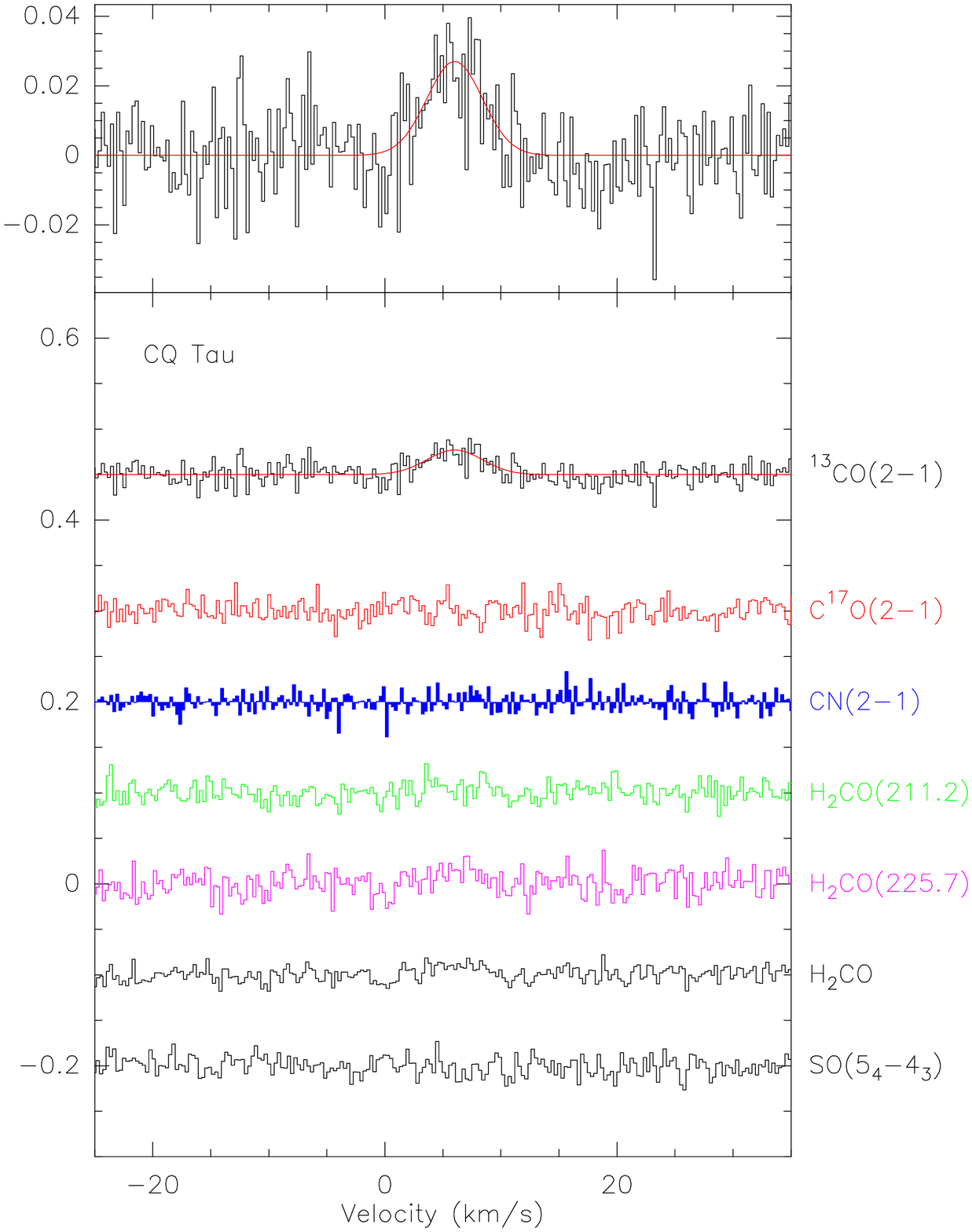}
\caption{Spectra of the observed transitions towards CQ Tau}
\label{fig:CQ_TAU}
\end{figure}
\clearpage
\begin{figure}[t]
\includegraphics[height=11.5cm]{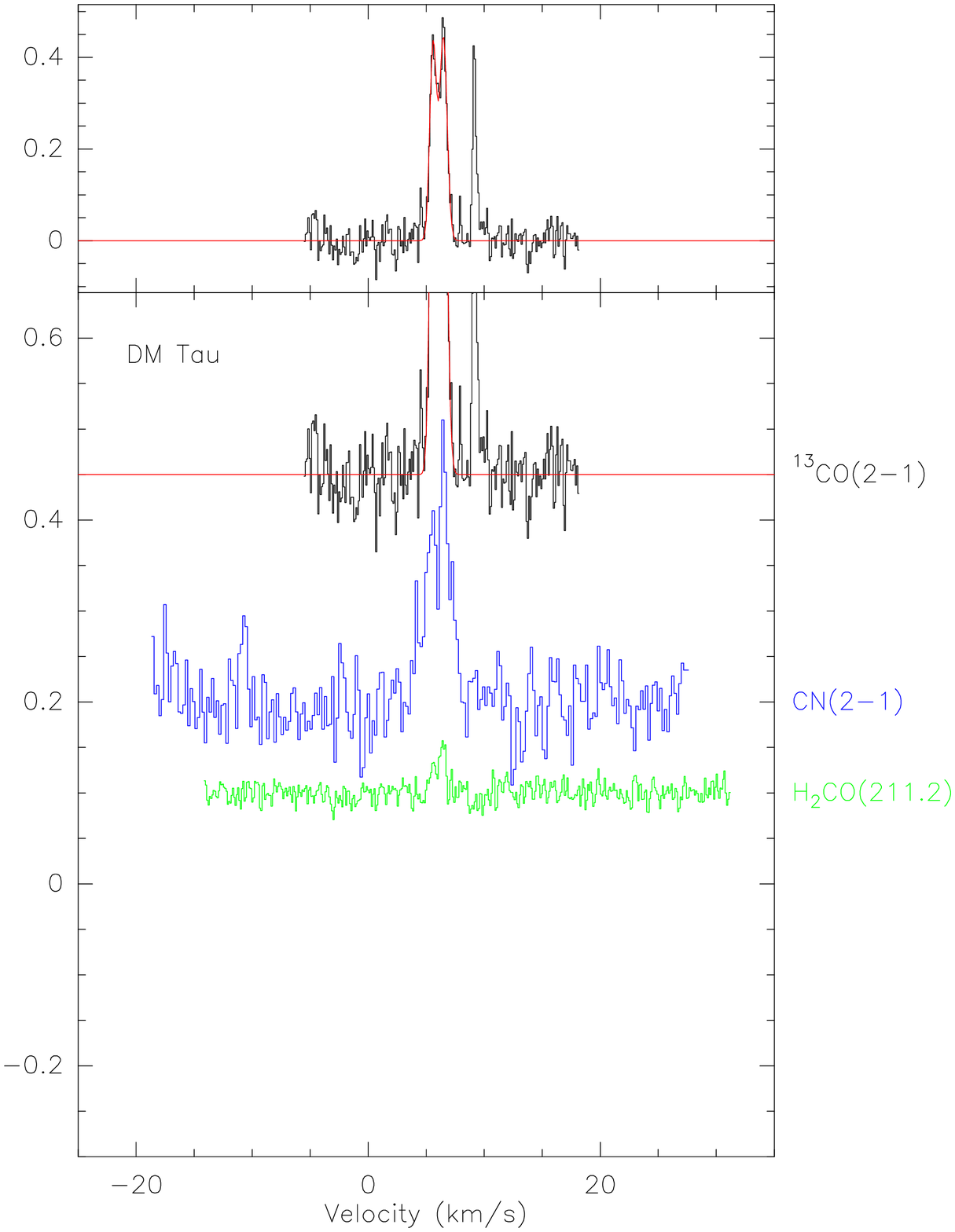}
\caption{Spectra of the observed transitions towards DM Tau}
\label{fig:DM_TAU}
\end{figure}
\begin{figure}[t]
\includegraphics[height=11.5cm]{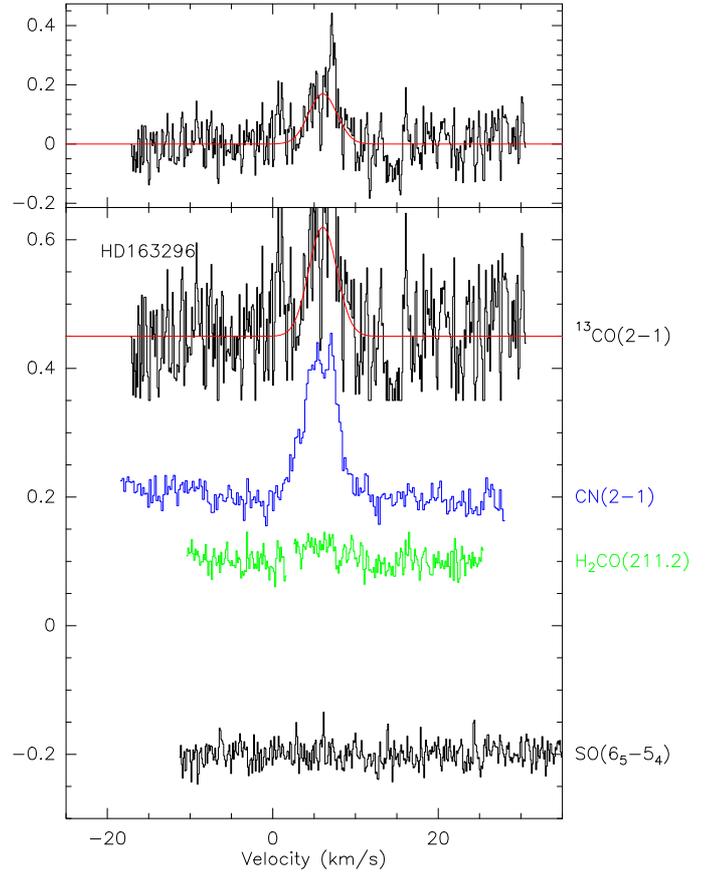}
\caption{Spectra of the observed transitions towards HD 163296.
The observed SO transition is not the same as in all other sources.}
\label{fig:HD163296}
\end{figure}
%
%
% \begin{figure}
% \includegraphics[height=11.5cm]{A-LKCA_15.eps}
% \caption{Spectra of the observed transitions towards LkCa 15}
% \label{fig:LKCA_15}
% \end{figure}
%
%end input

\section{The CW Tau / CIDA-1 line of sight}
\label{app:cloud}

\begin{figure}[!h]
\includegraphics[width=\columnwidth]{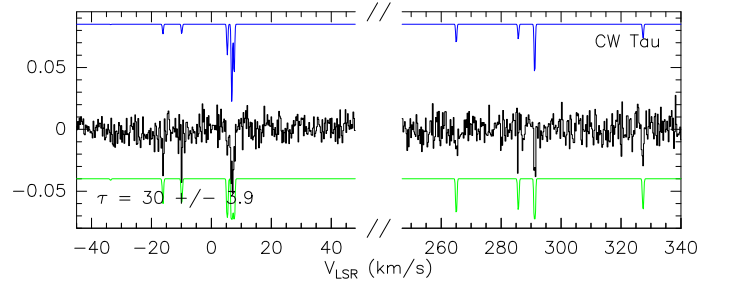}
\caption{Spectra of CN towards \object{CW Tau}. The best fit with the hyperfine
components is drawn below. The upper profile indicates the expected
profile for an optically thin line. Detection of even weak components suggests
high optical depth ($\sim 20-30$). The intensity scale is $T_A^*$ (K).}
\label{fig:cw_tau-cn}
\end{figure}

CW Tau and CIDA-1, which are only separated by 1 arcmin, exhibit strong confusion in $^{13}$CO, and are the only
sources in which confusion in CN is obvious. Confusion is also visible in H$_2$CO and even C$^{17}$O. In addition, in the
\citet{Salter+etal_2011} study, CW Tau is the source which has extended HCO$^+$ (3-2) emission.

In the beam-switched
observations, the detection of hyperfine components of CN (Fig.\ref{fig:cw_tau-cn}) suggests significant optical
depth.  Optically thick CN from clouds could introduce difficulties in using CN as a kinematical disk tracer,
as it could absorb the disk emission. This potential problem is mitigated by
several independent arguments. First, the cloud line width is small, so only a small fraction of disk kinematics
will be affected. Second, CN has several hyperfine components, of very different opacities: this would help in modeling the absorption layer. Furthermore,
anomalously high ($\tau \gg 1$) apparent opacities can be
the result of opacity gradients in a cloud with more moderate ($\tau > 1$) average opacity, as the strongest lines saturate
first. However, observations of the CW Tau line of sight in frequency switching shows  a flat-topped, 2 K line of $^{13}$CO, with a line width of $1.40 \kms$ and the same velocity than the CN signal, but no significant CN emission at a level of 50 mK ($3 \sigma$). Thus, in the beam-switched observations, the detected signal probably originates from emission in one or both of the reference beams,
and in the cloud, CN is most likely not much excited. Although high opacities cannot be excluded,
the unusual hyperfine line ratios may also be related to weak excitation-dependent anomalies, since CN collision rates display
significant dependencies on the hyperfine levels \citep{Kalugina+etal_2012}. Non negligible opacities,
low excitation and possible hyperfine anomalies have already been reported for CN in dark clouds by \citet{Crutcher+etal_1984}.

\section{Formation of Lines in Keplerian Disks}
\label{app:disk}

\begin{figure}[!t]
\begin{center}
\resizebox{6.0cm}{!}{\includegraphics[angle=270]{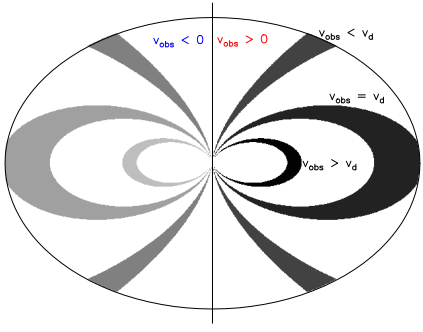}} \caption[Cinematique]{\label{fig:cinematique}
Regions of the disk which yield equal projected velocities $\mathrm{v}_\mathrm{obs}$. The ellipse is the
projection of the disk outer edge observed with an inclination of $45^\circ$.}
\end{center}
\end{figure}

Line formation in a Keplerian disk is strongly constrained by the velocity gradient
\citep{Horne+Marsh_1986}.
We use here this property to explore to what extent the detected CN lines trace the (putative)
circumstellar disks. Unless the disk is seen face on,
only a fraction of the disk projects at any given velocity. This fraction can be estimated with simple
reasoning: we follow here the derivation presented by \citet{Guilloteau+etal_2006}.
$r,\theta$ being the cylindrical coordinates in the disk plane, the line of sight velocity
(in the system rest frame) is
\begin{equation}
V_\mathrm{obs}(r,\theta) = \sqrt{GM_{*}/r} \sin{i}\cos{\theta}
\end{equation}
The locii of isovelocity are given by
\begin{eqnarray}
r(\theta) &=& ( GM_{*}/V_\mathrm{obs}^2 ) \sin^2{i}\cos^2{\theta}
\end{eqnarray}
With a finite local line width $\Delta \mathrm{v}$ (assuming rectangular line shape for simplification), the line at
a given velocity $V_\mathrm{obs}$ originates from a region included between $r_i(\theta)$ and $r_s(\theta)$ :
\begin{eqnarray}
r_i(\theta) &=& \frac{GM_{*}}{(V_\mathrm{obs}+\Delta \mathrm{v}/2)^2}\sin^2{i}\cos^2{\theta} \\
r_s(\theta) &=& \min\left[R_\mathrm{out},\frac{GM_{*}}{(V_\mathrm{obs}-\Delta \mathrm{v}/2)^2}
\sin^2{i}\cos^2{\theta}\right]
\end{eqnarray}
Figure \ref{fig:cinematique} indicates the regions of equal projected velocities for 6 different values:
$V_\mathrm{obs}>\mathrm{v}_d$, $V_\mathrm{obs} = \mathrm{v}_d$, $V_\mathrm{obs}<\mathrm{v}_d$, and their symmetric counterpart at negative
velocities, where $\mathrm{v}_d$ :
\begin{eqnarray}
\mathrm{v}_d &=&  \sqrt{GM_{*} / R_\mathrm{out}} \sin{i} \label{eq:vd}
\end{eqnarray}
is the projected velocity at the outer disk radius $R_\mathrm{out}$.
Figure \ref{fig:cinematique} shows that the fraction of the disk covered by the gas at velocities
$V_\mathrm{obs} \simeq \mathrm{v}_d$ is of order $\Delta \mathrm{v} / \mathrm{v}_d$, and drops very rapidly for larger velocities.
The larger area covered for $V_\mathrm{obs} \simeq \mathrm{v}_d$ explains the classical double peaked line profiles emerging from Keplerian disks.

For moderately inclined ($i < 70^\circ$) disks in Keplerian rotation, the integrated line flux is given by
\begin{equation}
\int S_{\nu} d\mathrm{v} = B_{\nu}(T_0) (\rho \Delta V) \pi R_{\mathrm{out}}^2 / D^2 {\rm cos}(i)
\label{eq:snu}
\end{equation}
where $T_0$ is the disk temperature (assumed uniform for simplicity), $\Delta V$ the local
linewidth, $R_\mathrm{out}$ the disk outer radius, $D$ is the source distance and $i$
the disk inclination.  $\rho$ is a factor of order of the
line of sight $\tau_l$ for optically thin lines, saturating as $\propto \log(\tau_l)$ for large optical depths
\citep[see][their Fig.4]{Guilloteau+Dutrey_1998}.  The optically thick limit is obtained because of the Keplerian
shears which, as demonstrated above, limits the maximum fraction of the disk
covered at any velocity to $\delta V /\mathrm{v}_d$.
$\delta V$ is the local linewidth including opacity broadening, so $\delta V \sim \log(\tau_l) \Delta V$ for large opacities.

Note that as $\tau_l \simeq \tau_p / \cos(i)$, where $\tau_p$ is the
line opacity perpendicular to the disk plane, and since $\tau_p \propto \Sigma/\Delta V$,
where $\Sigma$ is the molecule surface density,
the (apparent) dependence on inclination and line width disappears
for optically thin lines: the integrated line flux scales as the number of molecules.

From Eq.\ref{eq:snu}, we can derive the outer radius of the molecule distribution provided
with have an estimate of $\rho$
\begin{equation}
R_\mathrm{out}  =  D \left( \frac{\int S_{\nu} d\mathrm{v} }{B_{\nu}(T_0) (\rho \Delta V) \pi {\rm cos}(i)} \right) ^{1/2}
\label{eq:rout}
\end{equation}
Equation \ref{eq:rout} is valid for single lines. For molecules with hyperfine structure like CN,
line blending must be accounted for.  We do so in taking into account the relative
weight of the hyperfine components: the integrated flux of the J=5/2-3/2 fine structure
group presented in Table \ref{tab:cn} is multiplied by 0.45 to obtain the flux in the strongest
hyperfine component, which then used in Eq.\ref{eq:rout} to derive the outer radius.
The fitted opacity, if available, is used to provide
an estimate of $\rho$ ($\rho =\tau$, $\tau$ being the sum of the opacities of all hyperfine
components), which is otherwise set to 0.3.

\end{document}